\documentclass[a4paper,11pt]{article}
\pdfoutput=1 

\usepackage{jheppub} 
\usepackage{lineno}
\usepackage{mathrsfs}
\usepackage{MnSymbol}
\usepackage{color}
\usepackage[dvipsnames]{xcolor}
\usepackage[many]{tcolorbox}
\usepackage{enumitem}
\usepackage{dsfont} 
\usepackage{subcaption}

\usetikzlibrary{decorations.markings,arrows.meta}

\tikzset{
  MisalignedU/.style={
    -, double, thick,
    postaction={
      decorate,
      decoration={
        markings,
        mark=between positions 0.1 and 1 step 6mm with {
        \begin{scope}
            \pgftransformshift{\pgfpointorigin}   
            \draw(-5pt,5pt) -- (0pt,1pt);
            \draw(3pt,-1pt) -- (8pt,-5pt);
        \end{scope}
        }
      }
    }
  }
}

\tikzset{
  MisalignedD/.style={
    -, double, thick,
    postaction={
      decorate,
      decoration={
        markings,
        mark=between positions 0.1 and 1 step 6mm with {
        \begin{scope}
            \pgftransformshift{\pgfpointorigin}   
            \draw(0pt,5pt) -- (-5pt,1pt);
            \draw(8pt,-1pt) -- (3pt,-5pt);
        \end{scope}
        }
      }
    }
  }
}

\tikzset{
  Aligned/.style={
    -, double, thick,
    postaction={
      decorate,
      decoration={
        markings,
        mark=between positions 0.1 and 1 step 6mm with {
        \begin{scope}
            \pgftransformshift{\pgfpointorigin}   
            \draw(-5pt,5pt) -- (0pt,1pt);
            \draw(-5pt,-5pt) -- (0pt,-1pt);
        \end{scope}
        }
      }
    }
  }
}

\tikzset{
  SingleU/.style={
    -, thick,
    postaction={
      decorate,
      decoration={
        markings,
        mark=between positions 0.1 and 1 step 6mm with { \begin{scope}
            \pgftransformshift{\pgfpointorigin}   
          \draw(-5pt,4pt) -- (0pt,0pt);   
        \end{scope}
        }   
      }
    }
  }
}

\tikzset{
  SingleD/.style={
    -, thick,
    postaction={
      decorate,
      decoration={
        markings,
        mark=between positions 0.1 and 1 step 6mm with { \begin{scope}
            \pgftransformshift{\pgfpointorigin}   
          \draw(-5pt,-4pt) -- (0pt,0pt);   
        \end{scope}
        }   
      }
    }
  }
}

\tikzstyle{Orange Dot}=[fill={rgb,255: red,241; green,143; blue,31}, draw=black, shape=circle]
\tikzstyle{Green Dot}=[fill={rgb,255: red,120; green,151; blue,95}, draw=black, shape=circle]
\tikzstyle{Cyan Dot}=[fill={rgb,255: red,0; green,159; blue,223}, draw=black, shape=circle]
\tikzstyle{White Dot}=[draw=black, shape=circle]

\setlist[description]{leftmargin=0.5 cm}

\newtcolorbox{empheqboxed}{colback=gray!30, 
 colframe=white,
 width=\textwidth,
 sharpish corners,
 top=-2mm, 
 bottom=0mm
}

\def\tst{t}

\def\SymN{\textrm{Sym}^{N}(\mathcal{M})}

\tikzstyle{Orange Dot}=[fill={rgb,255: red,241; green,143; blue,31}, draw=black, shape=circle]
\tikzstyle{Green Dot}=[fill={rgb,255: red,120; green,151; blue,95}, draw=black, shape=circle]
\tikzstyle{Cyan Dot}=[fill={rgb,255: red,0; green,159; blue,223}, draw=black, shape=circle]
\tikzstyle{none}=[]
\tikzstyle{Dashed}=[-, dashed]
\tikzstyle{Arrowright}=[->]
\tikzstyle{Average}=[-, dashed, draw = {rgb,255: red,120; green,151; blue,95}, thick]
\tikzstyle{Orange}=[-, draw = {rgb,255: red,241; green,143; blue,31}, thick]
\tikzstyle{Cyan}=[-, draw = {rgb,255: red,0; green,159; blue,223},thick]

\usetikzlibrary{decorations.pathmorphing}

\arxivnumber{} 
\preprint{YITP-25-170}

\title{Interface Correlators in Symmetric Product Orbifolds}

\author[a,b]{Sebastian Harris,}
\author[a,b,c]{Volker Schomerus,}
\author[d]{and Takashi Tsuda}

\affiliation[a]{Deutsches Elektronen-Synchrotron DESY, Notkestr. 85, D-22607 Hamburg, Germany}
\affiliation[b]{Zentrum f\"ur Mathematische Physik, Bundesstrasse 55, D-20146 Hamburg, Germany}
\affiliation[c]{II. Institut f\"ur Theoretische Physik, Universit\"at Hamburg, Luruper Chaussee 149,\\ D-22761 Hamburg, Germany}
\affiliation[d]{Center for Gravitational Physics and Quantum Information, Yukawa Institute for Theoretical Physics, Kyoto University, Kitashirakawa Oiwakecho, Sakyo-ku, Kyoto 606-8502, Japan}

\emailAdd{sebastian.harris@desy.de}
\emailAdd{volker.schomerus@desy.de}
\emailAdd{takashi.tsuda@yukawa.kyoto-u.ac.jp}

\abstract{
    Symmetric product orbifolds provide a controlled environment to explore generic features of gauge theory and holography. 
    The tractability of these theories lies in the complete characterisation of their gauge structure through holomorphic covering maps.
    In this paper, we introduce a novel class of generalised covering maps, which define a universal family of interfaces between symmetric product orbifolds.
    These interfaces coincide with the holographic interfaces that were recently proposed as duals to AdS$_2$ branes in pure NSNS AdS$_3$ backgrounds.
    The new covering-map description enables efficient evaluation of interface correlators via a generalisation of the Lunin–Mathur method. To organise these computations, we derive a generalised Riemann–Hurwitz formula for interface coverings and introduce novel diagrammatic rules that systematically classify these maps.
    The new framework allows us to define a concrete grand-canonical ensemble that has the correct properties to compute correlation functions dual to open string scattering amplitudes. 
    Using the generalised Riemann-Hurwitz formula, we explicitly show that the correlators of the ensemble structurally match string perturbation theory to all orders in the string coupling. 
}

\begin{document}
\maketitle
\flushbottom

\section{Introduction}

    In many respects, symmetric product orbifolds are close cousins of higher dimensional weakly coupled gauge 
    theories and can serve as a very controlled laboratory to study features of perturbative expansions and most 
    importantly of holography. In particular, correlation functions of symmetric product orbifolds $\SymN$ of 
    some seed CFT $\mathcal {M}$ are known to possess a Feynman-like expansion. The role of the Feynman diagrams 
    is played by covering maps $\Gamma$ of the 2-dimensional surface $X$ the symmetric product orbifold 
    is defined on. These covering maps even possess a diagrammatic description that is quite reminiscent of 
    the usual double line networks that are used in gauge theory \cite{Pakman:2009zz}. More importantly, the 
    large $N$ expansion is similar to the 't Hooft expansion in gauge theory, suggesting a holographic 
    description through some dual string theory in $AdS_3$. And indeed, an exact dual string theory for the 
    symmetric product orbifold of a supersymmetric 4-torus has been established in \cite{Eberhardt:2018ouy,Eberhardt:2019ywk}. It 
    involves strings on an AdS$_3 \times $S$^3\times \mathbb{T}^4$ geometry with one unit of 
    NSNS- and no RR-flux turned on. This correspondence is quite remarkable. Tensionless strings are deeply 
    quantum and hence dual to a weakly coupled theory. Finding similar dual pairs between weakly coupled
    gauge theories in higher dimensions and (non-geometric) dual string theories is an important challenge
    that has seen limited progress, see however \cite{Berkovits:2025xok} for some recent proposal and further 
    references. On the other hand, many aspects of holography are actually still better understood in
    higher dimensional setups and in particular for the paradigmatic example of the duality between 
    $\mathcal{N}=4$ SYM theory and string theory in AdS$_5 \times $S$^5$. This applies in particular to the 
    emergence of a geometric supergravity regime which is well established in some higher dimensional 
    versions of the correspondence. In addition, the higher dimensional theories have been probed by 
    a wide range of different observable, including non-local ones such as line and surface defects  
    and interfaces. These have received much less attention in the context of symmetric product 
    orbifolds. Karch-Randal like interfaces that connect two symmetric product orbifolds with 
    different degree of the orbifold group, for instance, have only been constructed a few months 
    ago in \cite{Harris:2025wak}. In this paper, we continue our analysis of these 
    interfaces by providing the necessary tools to efficiently compute their local correlation 
    functions, especially with the aim to establish universal holographic features.
    
    The interfaces under consideration connect symmetric product orbifolds with orbifold groups $S_{N_\pm}$ 
    of degrees $N_\pm$. Two essential pieces of data enter their construction: A pair of boundary states 
    $|a_\pm\rangle$ of the seed theory $\mathcal{M}$ and an integer $P \leq \textrm{min}(N_\pm)$. While the 
    former controls the precise way in which bulk excitations in the symmetric product orbifold are reflected 
    off the interface, the integer $P$ determines the \textit{degree of transmissivity}, with large values of 
    $P$ corresponding to a maximally transparent setup. Note that, for $N_+ \neq N_-$, reflection processes 
    are unavoidable. It is conjectured that these interfaces provide a holographic dual for AdS$_2$-branes 
    in minimal pure NSNS flux limits of string theory on AdS$_3$ backgrounds. Concretely, this proposal was 
    brought forward in \cite{Harris:2025wak} for strings on AdS$_3\times$ S$^3 \times \mathbb{T}^4$, and 
    recently also extended to AdS$_3 \times$S$^3\times$S$^3\times$S$^1$ \cite{Belleri:2025eun}. Using the 
    boundary states of the folding trick description of the interfaces, Reference \cite{Harris:2025wak} 
    computed the torus partition function for a pair of interfaces. This CFT partition function was then 
    compared with the annulus partition function of open strings on the relevant AdS$_2$ branes and 
    both quantities were found to agree. In this sense, the interfaces defined in \cite{Harris:2025wak} 
    appear as natural low-dimensional analogues of Wilson lines in four-dimensional gauge theories. They 
    can also be viewed as two-dimensional counterparts of Karch-Randall interfaces \cite{Karch:2000gx}. 
    
    As convincing as the agreement of partition functions may seem, a complete derivation of the 
    holographic correspondence does require to also match CFT correlation functions of 
    bulk and interface operator with scattering amplitudes of tensionless closed 
    and open strings on AdS$_2$-branes in AdS$_3$. In order to approach such a comparison, we need to be able to compute the relevant CFT correlators. 
    This is the problem we address in the current work. In order to achieve this 
    goal, we extend the Lunin-Mathur covering space techniques to interface CFT. 
    This involves two steps. In a first step, the correlation functions in the 
    interface symmetric product orbifold CFT are represented as a sum over an 
    appropriate set of interface branched coverings $\Gamma$, see Subsection 
    \ref{sec:interface_coverings} and eq.~\eqref{eq:Interfaces_unnormalised_corr}. 
    The summands involve correlation functions in the seed CFT $\mathcal{M}$ on 
    surfaces with boundaries. These are weighted with some coefficients 
    $\alpha[\Gamma]$ that do not depend on the seed CFT. Following the ideas of 
    Lunin-Mathur \cite{Lunin:2000yv}, the correlation functions in the seed CFT 
    can then be calculated with the help of (boundary) Liouville field theory. 
    For pure twist field insertions, the dependence on the seed CFT is restricted 
    to the bulk and boundary entropies, i.e.~the central charge $c$ and the 
    $g$-factors $g_\pm$ of the two boundary conditions $a_\pm$, 
    see for example eq.~\eqref{eq:Z_1_Sigma_Interface_master}. 
    
    In order to construct the sum over coverings $\Gamma$ and compute the weight factors 
    $\alpha[\Gamma]$, we develop a diagrammatic approach  that extends the Feynman-like 
    diagrams that were introduced for bulk symmetric product orbifolds by Pakman, Rastelli 
    and Razamat in \cite{Pakman:2009zz}. In our case, the diagrams are built from five 
    different types of propagator-lines. Three of them are represented by double lines. 
    We shall refer to them as \textit{aligned} and \textit{misaligned}$_\pm$. In addition 
    there are two single lines of opposite orientation. Bulk twist fields of order $w$ are 
    represented by vertices with $w$ misaligned double lines, see Figure 
    \ref{fig:bulkPRR_vertices_bulk_rammification}. Interface twist field of order $\omega$, 
    on the other hand, are represented by vertices with two single lines and $\omega-1$ 
    aligned double lines, see Subfigure a, b, c and d of Figure \ref{fig:IPRR_vertices}. 
    In addition to these vertices that represent field insertions there are also two 
    classes of trivalent vertices that are not associated with any field insertions. These additional trivalent vertices allow to split a misaligned double line into either a pair 
    of aligned double lines, see Figure \ref{fig:bulkPRR_vertices_bulk_splitting}, or into 
    a pair of single lines, see Subfigure e and f of Figure \ref{fig:IPRR_vertices}. In order 
    to describe a valid interface covering map $\Gamma$, a diagram built out of these basic 
    ingredients needs to satisfy some simple rules, see Section \ref{sec:Interface_PRR}. 
    Given a valid diagram it is then easy to read off the degrees $N_\pm$ of the associated 
    covering map, the degree of transmissivity $P$ of the interface and the symmetry factor 
    $\alpha[\Gamma]$. 
    
    If the diagram has no single lines, it actually describes an ordinary bulk covering map. 
    In the corresponding diagrams, it is possible to eliminate all the misaligned double lines 
    by splitting them into pairs of aligned ones. Hence, these diagrams can be drawn with a 
    single type of double line which agrees with the pictorial descriptions proposed by 
    Pakman-Rastelli-Razamat (PRR). After we split the misaligned lines, bulk twist field of 
    order $w$ are represented by a vertex with $2w$ aligned double lines rather than $w$ 
    misaligned ones, see Figure \ref{fig:PRR_fused_vertex}. Thereby, we recover the graphical 
    rules for PRR diagrams. Let us note that our representation of the aligned double lines 
    looks a bit different from the one in the work of Pakman et al.~who used double lines 
    with one dashed and one solid edge. 
    \medskip 
    
    Another central result that we obtain from our new understanding of interface covering 
    maps concerns the structure of the large $N_\pm$ and large $P$ expansion that is shown 
    to match the string genus expansion for scattering amplitudes of closed and open strings 
    to all orders in the string coupling. In the bulk case it has been well understood 
    that the comparison of the two expansions requires passing to a grand-canonical 
    ensemble in which the parameter $N$ is controlled by a dual fugacity $\mu$. The 
    latter scales with the string coupling $g_s$ as $\mu \sim g_s^{2G-2}$ where $G$ 
    denotes the genus of the holographic boundary $X$. In the grand-canonical ensemble, 
    $N$ is not a number but one can check that the expectation value of $N$ scales as 
    $N \sim g_s^{-2}$. We extend these insights to the interface setup. In particular, we 
    introduce a new grand-canonical ensemble in which $N_\pm$ and the degree of 
    transmissivity $P$ are controlled by dual fugacities, see eq. \eqref{eq:ZXpmJ}. 
    With some appropriate scaling behaviour of these three fugacities, see 
    eq.~\eqref{eq:munug}, the expectation values of $N_\pm$ and $P$ scale as $g_s^{-2}$ 
    while the expectation value of $N_\pm -P$ turns out to scale as $g_s^{-1}$. 
    \medskip

    The plan for this work is as follows. The next section is devoted to covering maps and 
    their role in computing correlation functions in symmetric product orbifolds. We begin 
    by reviewing the well known notion of branched coverings in Section \ref{sec:branched_cover} 
    before we recall, in Section \ref{sec:symm_orb}, how correlation functions on symmetric 
    product orbifolds on a closed surface $X$ can be expressed as a sum over branched 
    coverings of $X$. The individual summands involve correlation functions of the seed 
    theory on the covering surface $\Sigma$.  We then extend this discussion to the case of 
    interfaces. In Subsection \ref{sec:interface_coverings}, we introduce the relevant notion 
    of generalised covering maps before we spell out how to compute correlation functions 
    involving interfaces as a sum over these covering maps, see Subsection \ref{sec:interfaces}
    and in particular eq.~\eqref{eq:Interfaces_unnormalised_corr}. We evaluate the right hand 
    side of this equation for some special setup in which we just insert two parallel interfaces 
    along some cycle of the torus, without any further local insertions of either bulk or 
    interface fields. The resulting formula coincides with the interface partition functions 
    we calculated in \cite{Harris:2025wak} from an algebraic construction that involves the 
    folding trick and explicit boundary states. This justifies a posteriori the rather natural 
    notion of generalised covering maps we propose in Subsection \ref{sec:interface_coverings}.
    
    Section \ref{sec:PRR} is devoted to the diagrammatic representation of interface covering 
    maps that we summarised above. After a short review of the PRR diagrams for correlation 
    functions in symmetric product orbifolds in Section \ref{sec:PRR_diagrams}, we describe 
    the corresponding interface diagrams in Section \ref{sec:Interface_PRR}. This is followed 
    by a brief discussion of the symmetries of interface PRR diagrams in Sections 
    \ref{sec:symmetry_factors} and \ref{sec:GaugeFixing}. These symmetries allow to read off 
    the group of deck transformations of the associated branched covering. Finally, we 
    determine the large $N$ behaviour of the expansion in Section \ref{sec:CanonicalLargeN}. 
    
    With all this control over the interface PRR expansion of correlation functions in 
    symmetric product orbifolds with holographic interface insertions, it remains to evaluate 
    its individual diagrams, i.e.~the relevant correlation functions of the seed theory on the 
    covering surfaces $\Sigma$. This is addressed in Section \ref{sec:Lunin_Mathur} where we 
    extend the ideas proposed by Lunin and Mathur for symmetric product orbifolds on a closed 
    surface to the interface setup. Once again, we start with a short review of the bulk analysis 
    before we discuss its extension to interfaces and several concrete examples. Section 
    \ref{sec:string_theory_&_GCE} is about grand-canonical partition functions. After a brief 
    review of the bulk case in Section \ref{sec:bulkgce}, we introduce grand-canonical interface 
    partition functions is Section \ref{sec:interfacegce} and show that these indeed have the 
    same structure as partition functions of closed and open strings with the scalings we 
    described above. In Section \ref{sec:conclusion}, we conclude by listing a number of 
    interesting open problems and directions for future research. Some more technical 
    explanations and additional material are collected in five appendices. 

\section{Interfaces in symmetric product orbifolds}\label{sec:symmetric_orbifold_boundaries_interfaces}
    In this section, we discuss a family of universal interfaces of
    symmetric product orbifolds that were introduced in Reference
    \cite{Harris:2025wak}. Using the folding trick, Section 2 of that 
    work gave a fully explicit construction of these interfaces in terms 
    of boundary states, from which in particular a closed formula for the 
    grand-canonical torus partition function with two interface insertions 
    was computed. Some very special cases of the interfaces had been considered previously, for instance in 
    References \cite{Gutperle:2024vyp,Belin:2021nck}.
    
    However, all these previous discussions, including Reference
    \cite{Harris:2025wak}, took a fairly algebraic 
    perspective on symmetric product orbifolds. In this section, we would 
    like to complement the existing literature on the topic by an analytic 
    perspective that focuses on an interpretation of objects in symmetric 
    product orbifolds in terms of branched covering maps, i.e.~we 
    represent interface correlation functions in symmetric product orbifolds
    through a sum over appropriately defined ``interface covering maps''. 

    In order to set the stage, we first review the corresponding description 
    for correlation functions of local fields in symmetric product orbifolds, see 
    Section~\ref{sec:symm_orb}. To make this work self-contained, the relevant 
    mathematical notions concerning branched coverings of Riemann surfaces are collected 
    in Section~\ref{sec:branched_cover}.  
    Analogously, our discussion 
    of interface correlators starts by introducing a class of generalised covering 
    maps in Section \ref{sec:interface_coverings} and proving that they satisfy a modified 
    Riemann-Hurwitz formula. 
    Equipped with this mathematical background, the final Section \ref{sec:interfaces} provides a new 
    description of the interfaces whose correlation functions are explored in the 
    rest of this work. As a first application of the definition in terms of a sum over generalised 
    covering maps, we re-calculate the interface partition functions 
    that were first computed in \cite{Harris:2025wak}. 
    Crucially, the new calculation turns 
    out to be considerably easier than the original one, which allows us to study much more general partition functions and correlators in this paper, than those accessible with the methods of Reference \cite{Harris:2025wak}. 
    
    \subsection{Branched coverings}\label{sec:branched_cover}
        To prepare for our discussion of correlations of local fields in symmetric
        product orbifolds, we start with a lightning 
        review of branched coverings. The definition of unramified and branched 
        coverings is stated in the next two paragraphs. This is followed by three 
        paragraphs in which we introduce the notation of monodromy and deck 
        transformations of branched coverings. The final paragraph reviews 
        a classical result of the theory of covering maps: The Riemann-Hurwitz formula. 
        
        \paragraph{Unramified coverings.} Throughout this section, let $\Sigma$ and $X$ 
        be Riemann surfaces. Unless stated otherwise,
        neither $\Sigma$ nor $X$ are assumed to be closed. Furthermore, let $\Gamma:\Sigma 
        \rightarrow X$ be a holomorphic surjection. If $\Gamma$ 
        is locally invertible and if every point $x \in X$ has $|\Gamma^{-1}(x)|=N$ preimages, 
        then $\Gamma$ is called a degree $N$ \emph{unramified covering} of $X$ by $\Sigma$. 
        
        \paragraph{Branched coverings.} 
        Let $B$ be a finite subset of $X$ and 
        $R := \Gamma^{-1}(B)$. If $R$ is a finite set and the restriction of $\Gamma$ to $\Sigma 
        \setminus R$ is a holomorphic covering of $X \setminus B$ then $\Gamma$ is 
        a \emph{branched covering} of $X$. In this case, we 
        refer to elements of $B$ and $R$ as \emph{branch points} and 
        \emph{ramification points} respectively. At each ramification point 
        $z \in \Sigma$, the covering map $\Gamma$ locally takes the form
        \begin{align}\label{eq:bulk_ramm}
            \Gamma(z+\delta z)  = \Gamma(z) + O(\delta z^w).
        \end{align}
        The integer $w$ is the \emph{ramification index} of $\Gamma$ in $z$. 
        For $z \notin R$, the index is one.
        
        \paragraph{Monodromy.} For a connected surface $X$, consider a closed curve $\gamma:[0,1] 
        \rightarrow X\setminus B$ based in the point $x \in X$, i.e.~a curve satisfying $\gamma(0) 
        = \gamma(1) = x$ that does not pass through any of the branch points of the covering map 
        $\Gamma$. 
        For each element $z$ of the set $\Gamma^{-1}(x)$ there is a unique lift 
        $\Gamma^*\gamma_z$ of $\gamma$ to $\Sigma$ such that 
        \begin{align}
            \Gamma \circ \Gamma^*\gamma_z = \gamma \quad \text{and} \quad \Gamma^*\gamma_z(0)=z \, .
        \end{align} 
        The map $z \mapsto \Gamma^*\gamma_z(1)$ is a bijection of the set $\Gamma^{-1}(x)$ onto 
        itself i.e.~an element of $\text{Aut}[\Gamma^{-1}(x)]$. It only depends on the homotopy 
        class $[\gamma]$ of the curve $\gamma$ and therefore defines a group homomorphism
        \begin{align}
            M_\Gamma: \pi_1(X,x) \rightarrow \text{Aut}[\Gamma^{-1}(x)] 
        \end{align}
        which sends $\gamma$ to an automorphism $\alpha_\gamma$ of the set $\Gamma^{-1}(x)
        \subseteq \Sigma$. The map $M_\Gamma$ is the \emph{monodromy action} of the covering 
        map $\Gamma$. Its image is the \emph{monodromy group} of $\Gamma$. 

        \paragraph{Deck transformations.} A \emph{Deck transformation} of the covering $\Gamma$ 
        is an automorphism $D$ of $\Sigma$ with the property $\Gamma \circ D = \Gamma$. 
        If $\Sigma$ is connected, then the image $D(z)$ of an arbitrary unramified point $z \in
        \Sigma$ already fixes $D$ completely. In this sense, the group Deck[$\Gamma$] of deck
        transformations can be identified with a subgroup of Aut$[\Gamma^{-1}(x)]$ for a generic 
        $x \in X$ in this case. Since deck transformations are precisely the maps that 
        preserve $\Gamma$, this subgroup is the stabiliser of the monodromy group of $\Gamma$ 
        under the conjugation of Aut$[\Gamma^{-1}(x)]$. 
    
        \paragraph{Local gauge choice.} Given a point $x \in X$ and a set $C$ of degree $N$ coverings of $X$, for which $x$ is not a branch point, we refer to a choice of labelling the $N$ points in the preimage $\Gamma^{-1}(x)$ of $x$ w.r.t.~every covering map $\Gamma \in C$ by the numbers $1,\dots, N$ as a \emph{local gauge choice\footnote{This terminology is justified in Appendix \ref{app:usual_gauge_theory}, where symmetric product orbifolds are discussed in the language of gauge theory, and the local gauge choices we define here are shown to correspond to local trivialisations of the gauge bundle.} for $C$ at $x \in X$}. 
        In particular, a local gauge choice for $C=\{\Gamma\}$ is a bijection
        \begin{align}
            \psi: \underline N:= \{ 1, 2, \dots, N\} \rightarrow \Gamma^{-1}(x),
        \end{align}
        that allows to identify elements $\alpha$ of 
        $\text{Aut}[\Gamma^{-1}(x)]$ with elements $\alpha^\psi :=  \psi^{-1} \circ \alpha \circ 
        \psi$ of the symmetric group $S_N$.
        Thus, local gauge choices lead to an identification of the monodromy group and deck transformations with subgroups of $S_N$. 
        As an application, we can for example use a local gauge choice to define a map
        \begin{align}\label{eq:M_Gamma_R}
            M_\Gamma^R: \pi_1(X,x) \rightarrow \mathbb{C}, \quad [\gamma] \mapsto \chi_R(M_\Gamma[\gamma]):=\chi_R(M_\Gamma^\psi[\gamma]),
        \end{align}
        given a representation $R$ of $S_N$ with character $\chi_R$. Since characters are invariant under the inner automorphisms of $S_N$, the value of $\chi_R(M_\Gamma[\gamma])$ is independent of the choice of $\psi$.
        
        \paragraph{Riemann-Hurtwitz formula.} In this paragraph, we 
        assume that $\Sigma$ and $X$ are closed. If $\Gamma$ is unramified, sufficiently fine
        triangulations of $X$ lift along $\Gamma$ to triangulations of $\Sigma$ with $N=
        \text{deg}[\Gamma]$ times as many vertices, edges and faces. Thus, the Euler 
        characteristics of $\Sigma$ and $X$ are related by $\chi(\Sigma) = N \chi(X)$. 
        More generally, consider a triangulation of $X$ whose vertices include all branch 
        points of $\Gamma$. Then the lifted triangulation still has the expected number of 
        edges and faces. However, $w-1$ vertices on $\Sigma$ are lost for each ramification 
        point with index $w$. Therefore, the Euler characteristic $\chi(\Sigma)$ of 
        $\Sigma$ is given by
        \begin{align}\label{eq:RH}
            \chi(\Sigma) = N \chi(X) - \sum\limits_{i=1}^n (w_i-1),
        \end{align}
        where $w_1, \dots ,w_n$ are the ramification indices of $\Gamma$. This is the 
        \emph{Riemann-Hurwitz formula}.

    \subsection{Symmetric product orbifolds}\label{sec:symm_orb}
        This section introduces symmetric product orbifolds as two-dimensional CFTs 
        defined by lifting correlation functions of a seed theory theory $\mathcal{M}$ 
        on a closed Riemann surface $X$ to various covering spaces. It is divided into 
        several paragraphs, each devoted to one aspect of the lift. The first of these 
        provides a definition. The definition is then applied to the computation 
        of vacuum partition functions. These allow us to determine the spectrum of the 
        symmetric orbifold and its local correlation functions. 
        
        \paragraph{Covering space lift.} 
        Let us denote by $\mathcal{A}_X[\mathcal{T}]$ the 
        set of observables $\mathcal{O}$ of a CFT $\mathcal{T}$ 
        on the surface $X$, i.e.~the set of all objects that can be inserted in its 
        correlation functions. 
        The most basic elements of $\mathcal{A}_X[\mathcal{T}]$ are products $\mathcal{O}_S = 
        \mathcal{O}_1(x_1)\mathcal{O}_2(x_2) \cdots \mathcal{O}_n(x_n)$ of local operators 
        $\mathcal{O}_i$ that are inserted at some finite set of points $x_i \in S \subseteq 
        X$. Beyond those, $\mathcal{A}_X$ also includes non local observables such as line
        operators. 
        
        We define the symmetric product orbifold $\text{Sym}^N(\mathcal{M})$ of the 
        seed theory $\mathcal{M}$ by a prescription that expresses the correlation 
        functions of its observables in terms of correlators of $\mathcal{M}$. This 
        is achieved through the following two steps. First, we associate a set
        $C_N[\mathcal{O}]$ of $N$-sheeted branched coverings\footnote{We consider 
        coverings that are related by an isomorphism that identifies their covering 
        spaces as identical. A more precise statement would be that $C_N[\mathcal{O}]$ 
        is a set of equivalence classes of covering maps.} of $X$ with each observable 
        $\mathcal{O} \in \mathcal{A}_X[\text{Sym}^N(\mathcal{M})]$.
       
        Second, we associate an observable $\mathcal{O}^\Gamma$ of the seed theory to 
        any pair $(\mathcal{O},\Gamma)$ of an observable $\mathcal{O}$ of the symmetric 
        product orbifold and a covering map $\Gamma \in C_N[\mathcal{O}]$. 
        For the multipoint observables $\mathcal{O}_S$ we described above, for example, 
        the associated observable in the seed theory is essentially given by some 
        $\mathcal{O}_{\Gamma^{-1}(S)}$. 
        Given these two ingredients, i.e.~the space $C_N[\mathcal{O}]$ of covering maps 
        associated with $\mathcal{O}$ and the map  
        \begin{align}
            \mathcal{O} \mapsto \mathcal{O}^\Gamma \in \mathcal{A}_{\Sigma}[\mathcal{M}],
        \end{align}
        that sends $\mathcal{O} \in \mathcal{A}_X[\text{Sym}^N(\mathcal{M})]$ to the 
        observable $\mathcal{O}^\Gamma \in \mathcal{A}_\Sigma[\mathcal{M}]$, we lift expectation values in the symmetric product orbifolds to expectation values 
        in the seed theory as, 
        \begin{align}\label{eq:unnorm}
            \langle \mathcal{O} \rangle_{\text{Sym}^N(\mathcal{M})}^{\text{unnormalised}} 
            = \frac{1}{N!} \sum\limits_{\Gamma \in C_N[\mathcal{O}]}\langle \mathcal{O}^\Gamma 
            \rangle_\mathcal{M},
        \end{align}
        where $\langle \mathcal{O}^\Gamma \rangle_\mathcal{M}$ is a correlator of 
        $\mathcal{M}$ on $\Sigma$ with a metric that is obtained by pulling back the 
        metric of $X$ along $\Gamma$. We have placed a subscript `unnormalised' on this 
        correlator since it does not necessarily yield $1$ for the one-point functions of the identity operator $\mathds{1}$. We define its normalised analogue as the quotient
        \begin{align}\label{eq:norm_vs_unnorm}
            \langle \mathcal{O} \rangle_{\text{Sym}^N(\mathcal{M})} = 
            \frac{\langle \mathcal{O} \rangle_{\text{Sym}^N(\mathcal{M})}^{\text{unnormalised}} }
            {\langle \mathds{1} \rangle_{\text{Sym}^N(\mathcal{M})}^{\text{unnormalised}} } .
        \end{align}
    
        \paragraph{The vacuum.}
        As a first example, let us discuss the concrete realisation of the general equation \eqref{eq:unnorm} for the case $\mathcal{O}=\mathds{1}$ of the identity, or vacuum operator.
        The set $C_N[\mathds{1}]$ is 
        \begin{align}\label{eq:C[1]}
            C_N[\mathds{1}] := \{N \text{ sheeted unramified coverings of } X\}\ . 
        \end{align}
        Given some $\Gamma \in C_N[\mathds{1}]$, the observable $\mathds{1}^\Gamma$ of the 
        seed theory on the covering surface $\Sigma$ is the identity operator, up to some 
        normalising prefactor $\alpha[\Gamma]$, i.e.
        \begin{align}\label{eq:1^Gamma}
            \mathds{1}^\Gamma = \alpha[\Gamma] \cdot \mathds{1} \quad \textrm{ where }
            \quad
           \alpha[\Gamma] := \frac{\text{deg}[\Gamma]!}{|\text{Deck}[\Gamma]|}\,.
        \end{align}
        By definition, the vacuum partition function $Z_N^X$ of  Sym$^N(\mathcal{M})$ on the surface $X$ is 
        \begin{align}
            Z_N^X :=\langle \mathds{1}\rangle_{\text{Sym}^N
          (\mathcal{M})}^{\text{unnormalised}}.
        \end{align}
        Let us compute $Z_N^X$ in the simplest case $X=\mathbb{S}^2$, where the set $C_N[\mathds{1}]$ has only a single element $\Gamma$, which covers the sphere by $N$ disconnected spheres.
        The group $\text{Deck}[\Gamma]$ consists of the $N!$ permutations of these spheres. 
        Therefore, we obtain $Z_N^{\mathbb{S}^2} = 1/N!$ in a normalisation where the sphere partition function of the seed theory is set to one. 
        A much greater amount of information is encoded in the partition function of higher genus surfaces $X$.
        In particular, the torus partition function contains the entire spectral data 
        of local operators and the genus two partition function yields all structure 
        constants.

       \paragraph{Torus partition function.} 
        Let us compute the torus partition function from def.~\eqref{eq:C[1]}. 
        Using the Riemann-Hurwitz formula, one can easily conclude that the covering 
        space of an unramified covering of $\mathbb{T}^2$ must be a disjoint union of 
        tori. More precisely, all connected unramified coverings $\Gamma_{w,\ell}^k$ 
        of the torus $\mathbb{C}/(\mathbb{Z} + \tau \mathbb{Z})$ arise from linear maps
        \begin{align}
            \Gamma :  \mathbb{C} \rightarrow \mathbb{C}, \quad z \mapsto (m\tau +w) 
            z \quad \text{with} \quad m,w\in\mathbb{Z} 
        \end{align}
        by quotienting out the lattices $\mathbb{Z} 
        + t \mathbb{Z}$ and $\mathbb{Z} 
        + \tau \mathbb{Z}$ of the domain and image of $\Gamma$ respectively. 
        The modular parameter $t$ of the covering space takes the form $t = \frac{\ell \tau + k}{m \tau +w}$ with $\ell,k \in 
        \mathbb{Z}$ and
        modular 
        transformations allow us to fix $m=0$, $w,\ell>0$ and $1 
        \leq k \leq w$.
        As anticipated, the coverings are thus labelled by three integers. Since 
        $\text{Deg}[\Gamma^k_{w,\ell}]= w\ell$, we find 
        \begin{align}\label{eq:torus_partition}
            Z_N^{\mathbb{T}^2} = \frac{1}{N!} \sum\limits_{\Gamma \in C_N[\mathds{1}]} 
            \alpha[\Gamma] \langle \mathds{1} \rangle_\mathcal{M} = \sum\limits_{\sum w_i \ell_i n_i=N}\sum
            \limits_{k_i=1}^{w_i}
            \prod\limits_{i}\frac{1}{n_i!} \left(\frac{Z(\tfrac{\ell_i \tau + k_i}{w_i})}{|\text{Deck}[\Gamma_{w_i,\ell_i}^{k_i}]|}
            \right)^{n_i}\,.
        \end{align}
        $\text{Deck}[\Gamma_{w,\ell}^k]$ is the group of translations $z \mapsto z +  
        \frac{a\tau+b}{w}$ with integer $a$ and $b$, which has size $|\text{Deck}
        [\Gamma_{w,\ell}^k]| = w \ell$. In the grand-canonical ensemble, the constraint 
        $\sum w_i \ell_i n_i=N$ is lifted and we obtain the simple formula
        \begin{align}
            \sum\limits_{N=0}^\infty Z_N^{\mathbb{T}^2} \mu^N 
            = \exp\left( \sum\limits_{w,\ell=0}^\infty \sum\limits_{k=1}^{w}
            \frac{\mu^{w\ell}}{w \ell} Z\left(\frac{\ell \tau + k}{w}\right) \right),
        \end{align}
        from which the spectrum of the orbifold can be read off. This completes our short computation of the torus partition function in symmetric product 
        orbifolds. 

        \paragraph{Correlation functions of twist fields.}

        A universal (i.e.~seed theory independent) feature of all symmetric product orbifolds is 
        the existence of twist fields -- local operators $\sigma_{\mathbf{w}}$ labelled by integer partitions $\mathbf{w} = [w_1, \dots ,w_\ell]$ whose conformal weights are
        \begin{align}
        \label{eq:TwistPartition}
            h_\mathbf{w} = \bar h_\mathbf{w} = 
            \frac{c_\mathcal{M}}{24}\sum\limits_{i=1}^\ell \left(w_i - \frac{1}{w_i}
            \right).
        \end{align}
        The state corresponding to $\sigma_{\mathbf{w}}$ contributes to the torus partition 
        function \eqref{eq:torus_partition} through a particular set of covering maps. These 
        are the maps that, upon cutting the torus along the real cycle, yield coverings of a 
        cylinder $X$ by $\ell$ disconnected cylinders $\Sigma_i$, such that $\Sigma_i$ winds 
        $X$ exactly $w_i$ times. Therefore, 
        \begin{equation}\label{eq:twist_field_C}
            \hspace{-5 pt }C\left[\prod\limits_{i=1}^n\sigma_{\mathbf{w}_i}(x_i)\right] =
            \Big\{ \; \parbox{25em}{$N$ sheeted coverings of $X$ with branchpts.~at $x_i$ 
            lifting to $\ell_i$ ramification pts.~with ramification indices $w_1, \dots, 
            w_{\ell_i}$. } \; \Big\}.
        \end{equation}
        Observables made up of pure twist fields $\sigma_\mathbf{w}$ in the symmetric product 
        orbifolds are mapped to identity fields in the seed theory. More precisely, the 
        prescription is 
        \begin{align}\label{eq:twist_field_Gamma}
            \left[\prod\limits_{i=1}^n\sigma_{\mathbf{w}_i}(z_i)\right]^{\Gamma} 
            = \alpha[\Gamma] \mathds{1} \quad \text{where} \quad \alpha[\Gamma]= 
            \frac{\text{deg}[\Gamma]!}{\left|\text{Deck}[\Gamma]\right|}, 
        \end{align}
        as for the identity field. Partitions $\mathbf{w} = [w,1,1,\dots,1]$ with only a single $w>1$ play 
        a special role. They are referred to as single cycle twist fields and we denote 
        them simply as $\sigma_w$, labelled by the single integer $w$. 
        
        \paragraph{Connected sphere correlators.} As an important example, let us discuss 
        correlators of twist fields on the sphere $X = \mathbb{S}^2$ in more detail. Concretely, let us focus on contributions to these correlators that stem 
        from covering maps 
        \begin{align}
            \Gamma : \Sigma \sqcup [\mathbb{S}^2]^{\sqcup(N - n_c)} \rightarrow X,
        \end{align}
        where $\Sigma$ is \textit{connected}, $\Gamma|_{\Sigma}$ has degree $n_c$ and 
        $\Gamma$ is the identity map when restricted to the sphere factors. We denote 
        this subset of covering maps by $C_{n_c}^\text{conn}[\mathcal{O}]$. The resulting 
        partial sum of expectation values in the seed theory is referred to as the 
        \textit{connected part} of the correlator. For general observables
        $\mathcal{O}$, the connected expectation value is defined as 
        \begin{align} 
            \langle \mathcal{O}\rangle^\text{conn,unnormalised}_{\text{Sym}^N(\mathcal{M})} = 
             \frac{1}{N!} \sum_{n_c=1}^N \ \sum_{\Gamma \in C_{n_c}^{\text{conn}}[\mathcal{O}]}
             \langle \mathcal{O}^{\Gamma}\rangle_\mathcal{M}\ . 
        \end{align} 
        In order to compute a connected correlator 
        \begin{align}
            \langle \sigma_{w_1}(x_1) \dots \sigma_{w_n}(x_n)
            \rangle^\text{conn,unnormalised}_{\text{Sym}^N(\mathcal{M})} 
        \end{align}
        of a product of single cycle twist 
        fields, one has to determine the factor $\alpha[\Gamma]$ of connected covering 
        maps. In order to do so, pick some unramified point  $z \in 
        \Sigma$ and choose $n$ closed curves $\gamma_i$ based in $z$ such that $\gamma_i$ 
        winds around the ramification point $z_i$ associated with $x_i$ exactly once and 
        has vanishing winding number with respect to all the other ramification points.
        Furthermore, let $x := \Gamma(z)$. The monodromy action $M_\Gamma$ 
        then associates a cyclic permutation $g_i:=M_\Gamma[\gamma_i] \in 
        \text{Aut}[\Gamma^{-1}(x)]$ of order $|g_i| = w_i$ to each of the twist fields. 
      
        From the collection $g_1, \dots, g_n$ of permutations, we can calculate the 
        symmetry factor $\alpha[\Gamma]$ that appears in the prescription
        \eqref{eq:twist_field_Gamma} as 
        \begin{align}\label{eq:conjugacy_class}
            \alpha[\Gamma] = |[g_1, \dots, g_n]| && \text{with} && [g_1, \dots, g_n] = 
            \{(g g_1 g^{-1}, \dots, g g_n g^{-1})| g \in \text{Aut}[\Gamma^{-1}(x)]\}
        \end{align}
        because $\text{Deck}[\Gamma]$ is precisely the stabiliser of $(g_1, \dots, g_n)$ 
        under conjugation. Let us stress that this formula for $\alpha[\Gamma]$ only holds 
        for connected covering maps $\Gamma$. 

        \paragraph{Wilson loops.} Let us now introduce a first family of line operators, 
        namely the \emph{Wilson loop} $\mathcal{W}_R[\gamma] \in \mathcal{A}_X [\text{Sym}^N
        (\mathcal{M})]$. Here, $\gamma$ is a closed curve on $X$ that starts and ends at some
        point $x \in X$ and $R$ labels a representation of the symmetric group $S_N$. Inserting a
        Wilson loop $\mathcal{W}_R[\gamma]$ into a correlator does not alter the 
        set $C_N[\mathcal{O}]$ of coverings we sum over, i.e. 
        \begin{align} 
            C_N[\mathcal{O} \mathcal{W}_R[\gamma]] = C_N[\mathcal{O}]. 
        \end{align} 
        What changes is only the sum on the right hand side of our defining relation \eqref{eq:unnorm}. 
        After insertion of the Wilson line $\mathcal{W}_R[\gamma]$ the formulas reads 
        \begin{align}
            \langle \mathcal{O} \mathcal{W}_R[\gamma] 
            \rangle_{\text{Sym}^N(\mathcal{M})}^{\text{unnormalised}} 
            = \frac{1}{N!} \sum\limits_{\Gamma \in C_N[\mathcal{O}]}\, M^R_\Gamma[\gamma] 
            \ \langle \mathcal{O}^\Gamma \rangle_\mathcal{M}. 
        \end{align}
        Here, we have used the map $M^R_\Gamma$ defined in eq.~\eqref{eq:M_Gamma_R} as 
        the evaluation of the character of $R$ on the monodromy action of $\Gamma$. 
        The only effect of a the Wilson loop insertion into the expectation 
        value of some observable $\mathcal{O}$ of the $N$-fold symmetric product 
        orbifold is therefore to multiply the individual summands in \eqref{eq:unnorm} 
        by some complex numbers. For a discussion of open Wilson loops i.e.~Wilson lines 
        as well as dual 't Hooft lines, which change the structure of covering maps 
        contributing to correlation functions, see App.~\ref{App:Wilson_tHooft}.

    \subsection{Interface covering maps}\label{sec:interface_coverings}
        In this section, we generalise the notion of (un)ramified coverings to a broader 
        class of maps, which we call \emph{interface coverings}. As the name suggests, 
        these maps compute correlators in the presence of the interfaces, which we 
        introduce in Sec.~\ref{sec:interfaces}. An important result of this section 
        is the generalised Riemann-Hurwitz formula \eqref{eq:interface_RH} 
        satisfied by interface coverings.
        
        \paragraph{Unramified interface coverings.}
        Let us consider a connected Riemann surface $X$ without boundary and a based loop $\gamma:[0,1]
        \mapsto X$ with $\gamma(0) = x = \gamma(1)$ therein. We denote the image $\gamma([0,1])$ by $s_\gamma$ and assume that the complement $X\setminus s_\gamma$ of $s_\gamma$
        in $X$ splits into two connected components $X^o_\pm$.
        Let us denote their closure in $X$ by $X_\pm = X^o_\pm \cup s_\gamma$. By construction the two sets $X_+$ and $X_-$
        possess the same boundary $\partial X_\pm = s_\gamma$. 

        An \emph{interface covering map} (for an interface localised along the curve $\gamma$)
        is a surjective holomorphic map $\Gamma : \Sigma \rightarrow X$ whose restrictions $\Gamma_\pm$ 
        to the subspaces $\Sigma^o_\pm:= \Gamma^{-1}(X^o_\pm)$ are covering maps in the sense of 
        Sec.~\ref{sec:branched_cover}. The degrees $N_\pm$ of these covering maps may differ, 
        i.e.~the size $N_+$ of the set $\Gamma^{-1}(x)$ for an element $x$ of $X^o_+$ may not be the same as the number $N_-$ of points in $\Gamma^{-1}(x)$ for $x\in X^o_-$.\footnote{Here it is important that 
        $X\setminus s_\gamma$ is not connected. If it was, the degree of $\Gamma$ could not 
        jump across $\gamma$.} We define the \emph{degree of transmissivity} $P$ of $\Gamma$
        on $\gamma$ as
        \begin{align} 
            P := |  \Gamma^{-1}(\gamma(t)) \cap \Sigma^o |\   
        \end{align}
        for any $t \in [0,1]$. Let us stress that the boundary $\partial \Sigma$ of the covering space is allowed to be non-empty. 
        Since $\Gamma(\partial \Sigma) \subseteq s_\gamma$, our definition 
        of the degree of transmissivity implies that $\partial \Sigma$ has at most $N_+ + N_- - 2 P$ connected 
        components. Let us also note that $P \leq \text{min}(N_-,N_+)$. 
        While we, for concreteness and ease of notation, restricted to the specific case of a single curve $\gamma$ splitting $X$ into two connected components here, the definition given above trivially generalises to an arbitrary number of curves splitting $X$ into arbitrarily many components.
        
        \paragraph{Ramification of interface coverings.}
        Let $B$ be a finite subset of $X$ and 
        $R := \Gamma^{-1}(B)$. If $R$ is a finite set and the restriction of $\Gamma:\Sigma \rightarrow X$ to $\Sigma 
        \setminus R$ is an unramified interface covering of $X \setminus B$ then $\Gamma$ is 
        a \emph{branched interface covering} of $X$.
        We refer to elements of $B$ and $R$ as branch points and ramification points respectively. 
        Branch points that lie on the curve $\gamma$ are called \emph{interface branch points} below. 
        The other branch points are referred to as \emph{bulk branch points}. 
        The ramification points associated with bulk branch points are simply ramification points of the restrictions $\Gamma_\pm$ of $\Gamma$ to $X_\pm^o$ in the sense of Section \ref{sec:branched_cover}.

        Ramification points of an interface covering map fall into two different 
        classes: They can either be in the interior of $\Sigma$, in which case we speak of \emph{interior ramification points}, or on the boundary $\partial 
        \Sigma$, in which case we speak of \emph{boundary ramification points}.
        The local behaviour of the covering map near an interior ramification point is captured by eq.~\eqref{eq:bulk_ramm}.
        All ramification points associated with bulk branch points are interior ramification points. 
        However, some interior ramification points may map to interface branch points.
        We will occasionally find it convenient to associate elements of this subset of interior ramification points with an \emph{interface ramification index} $\omega := 2w$, where $w$ is the index defined through eq.~\eqref{eq:bulk_ramm}.
        If all ramification points of an interface branch point are interior ramification points, then the degree of transmissivity $P$ of the interface covering map is the same on both sides of that branch point and bounds the value of the index as
        $\omega \leq 2P$.  

\begin{figure}
            \begin{subfigure}{.5\textwidth}
                \centering
                \includegraphics[width=0.7\columnwidth]{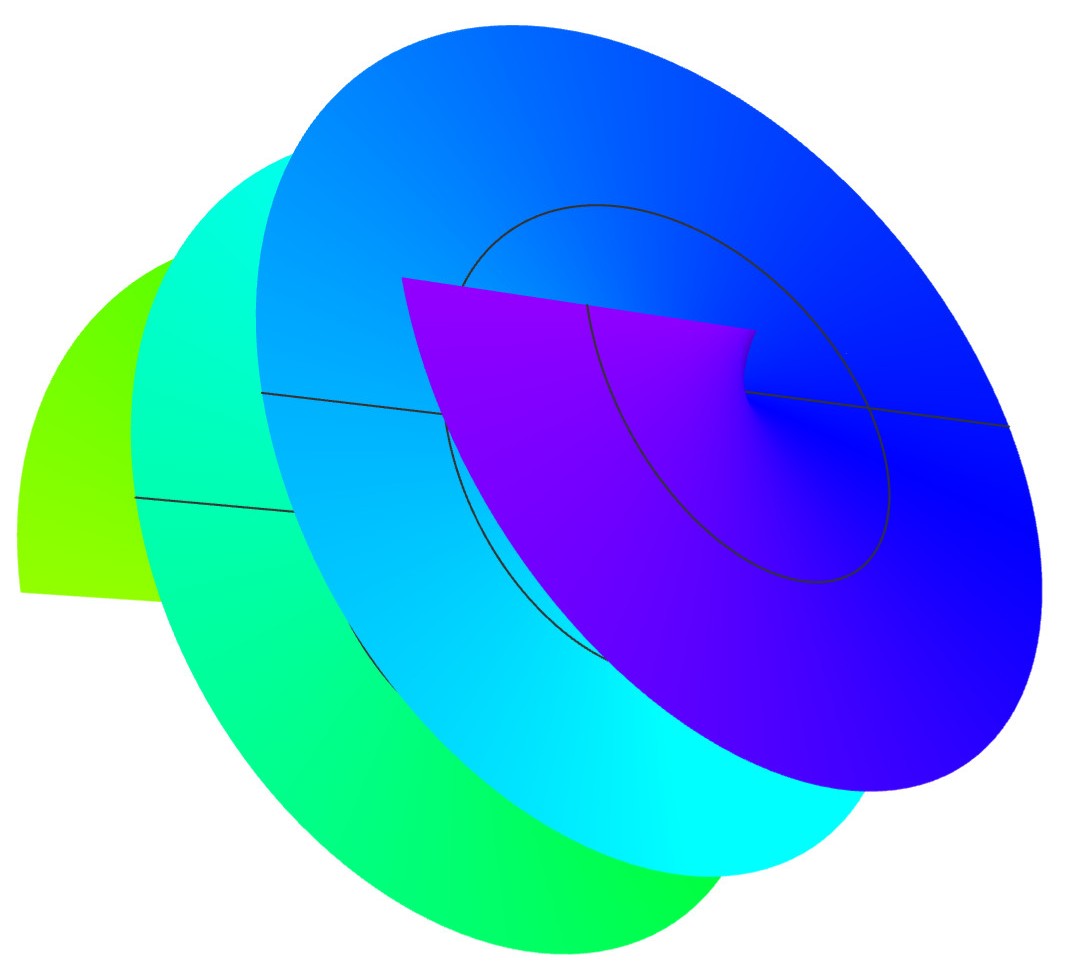}
                \caption{Type $6^{LL}$ ramification point of a $N_\pm = 3$ \\ covering. The degree of transmissivity jumps \\ from $2$ to $3$ at the ramification point.}
            \end{subfigure}
            \begin{subfigure}{.5\textwidth}
                \centering   
                \includegraphics[width=0.7\columnwidth]{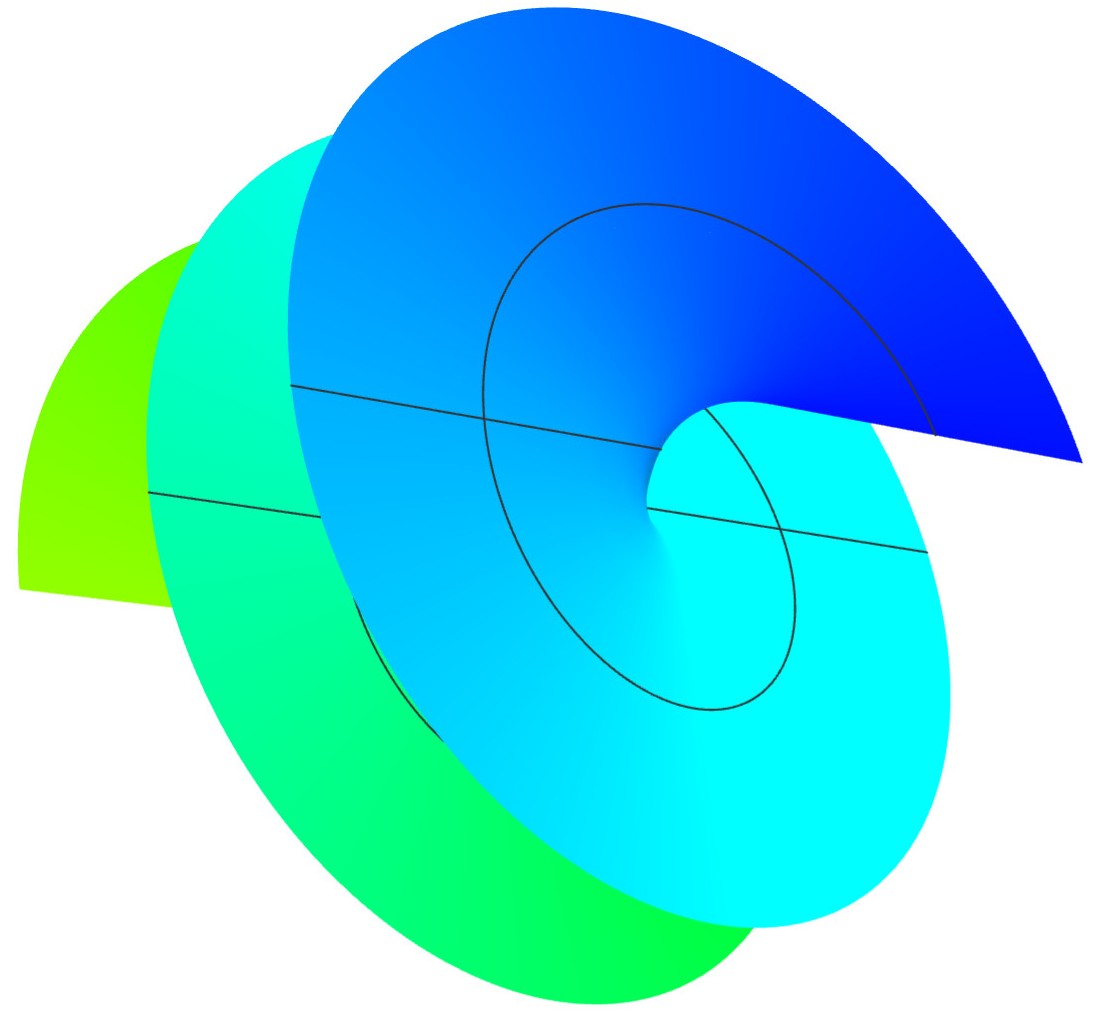}
                \caption{Type $5^{RL}$ ramification point of a $N_-=2$, $N_+=3$ covering. The degree of transmissivity is constantly $2$.}
            \end{subfigure}
            \caption{Visualisation of coverings in a neighbourhood of interface ramification points.
            The covering map is the vertical projection down the symmetry axis of the spiral. 
            The preimage of the interface is marked by horizontal black lines on the covering space.}
            \label{fig:IntTwistRamif}
        \end{figure}

        Let us now turn to boundary ramification points, of which we distinguish four different types. 
        To define these types, holomorphically identify a small neighbourhood of the ramification point $z$ with the upper 
        half complex plane, such that $z$ is identified with $0$. 
        Let $c$ be the curve in $\Sigma$ that, through the identification, corresponds to the curve 
        in the upper half plane which maps $\varphi\in[0,\pi]$ to $\varepsilon e^{i\varphi}$.
        If $\Gamma$ is unramified in $z$, then $c$ lies completely within $\Sigma_-$ or 
        $\Sigma_+$ and we say that $z$ has ramification index $1^{LR}$ or $1^{RL}$ respectively. 
        If however $\Gamma$ is ramified in $z$, there 
        are $\omega-1 \geq 1$ transitions from $\Sigma_\pm$ to $\Sigma_\mp$ along $c$. 
        In this case, we say $\Gamma$ has \emph{interface ramification index} $\omega$ in $z$.
        We furthermore classify $z$ as   
        \begin{align}
            & \text{type}\  \omega^{LL}\ \text{if} \ c(0) \in \Sigma_-,\, c(\pi)\in \Sigma_+, \quad  \quad 
            & \text{type}\  \omega^{RR} \ \text{if} \ c(0)\in \Sigma_+, \, c(\pi)\in \Sigma_-,\\[2mm] 
            & \text{type}\  \omega^{LR} \ \text{if} \ c(0),c(\pi)\in \Sigma_-, \quad  \quad  &\text{type}\  \omega^{RL}\ 
            \text{if} \ 
            c(0),c(\pi)\in \Sigma_+ . \quad \quad
        \end{align}
        See Figure \ref{fig:IntTwistRamif} for examples. Note that type $\omega^{LL}$ and $\omega^{RR}$ ramification points force the degree of 
        transmissivity to jump by one unit. 
            
        \paragraph{Interface Riemann-Hurwitz formula.}
        In this final paragraph, we prove that interface coverings of closed Riemann surfaces $\Sigma$ satisfy the modified Riemann-Hurwitz formula
        \begin{align}\label{eq:interface_RH}
            \chi(\Sigma) = N_- \chi(X_-) + N_+ \chi(X_+) - \sum\limits_{i=1}^{n_c} (w_i-1) - \frac{1}{2} \sum\limits_{i=1}^{n_o} (\omega_i-1),
        \end{align}
        where $w_1, \dots, w_{n_c}$ are the indices of the $n_c$ interior ramification points and 
        $\omega_1, \dots, \omega_{n_o}$ are the indices of the $n_o$ boundary ramification points.
        To simplify the discussion, we w.l.o.g.~assume that all interface branch points correspond to boundary ramification points\footnote{Covering maps with interior ramification points that are mapped to interface branch points can be obtained as limits of coverings without such points. Hence eq.~\eqref{eq:interface_RH} also directly follows for these maps from the arguments given in this paragraph.}. 
        To derive eq.~\eqref{eq:interface_RH}, consider a triangulation 
        of $X_-$ that contains all bulk and boundary branch points as vertices and is fine 
        enough that all edges and faces lift to exactly $N_-$ copies on $\Sigma_-$.
        As discussed in Section \ref{sec:branched_cover}, while generic vertices too lift to 
        $N_-$ copies on $\Sigma_-$, we loose $w_i^--1$ vertices on $\Sigma_-$ for each of 
        the $n_c^-$ interior ramification points.
        The loss of vertices at boundary ramification points of type $LL$, $RR$, $LR$ and $RL$ 
        can be determined completely analogously, leading to the formula
        \begin{align}
            \chi(\Sigma_-) = N_- \chi(X_-) - \hspace{-1mm}\sum\limits_{i=1}^{n_c^-}
            (w_i^--1) - \hspace{-2mm}\sum\limits_{i=1}^{n_o^{LR}}\hspace{-1mm}
            \left(\tfrac{\omega_i^{LR}+1}{2}-1\right)
            - \hspace{-2mm}\sum\limits_{i=1}^{n_o^{RL}}\hspace{-1mm}
            \left(\tfrac{\omega_i^{RL}-1}{2}-1\right)
            - \hspace{- 1mm}\sum\limits_{A}\hspace{-1mm}\sum\limits_{i=1}^{n_o^{AA}}
            \hspace{-1mm}\left(\tfrac{\omega_i^{AA}}{2}-1\right).
        \end{align}
        Combining 
        this result with the analogous formula for $\chi(\Sigma_+)$, we obtain
        \begin{align}
            \chi(\Sigma_-) + \chi(\Sigma_+) = N_- \chi(X_-) + N_+ \chi(X_+) 
            -\sum\limits_{i=1}^{n_c}(w_i-1) -\sum\limits_{i=1}^{n_o}(\omega_i-2).
        \end{align}
        Here $n_c = n^+_c + n^-_c$ and similarly $n_o$ counts the total number of boundary ramification points. We can view $\Sigma_- \cap \Sigma_+$ as a graph whose vertices are the boundary ramification points. The valency of each vertex is $\omega-1$, where $\omega$ is 
        the corresponding ramification index. Therefore, 
        \begin{align}
            \chi(\Sigma_- \cap \Sigma_+) = n_o - \frac{1}{2}\sum\limits_{i=1}^{n_o}(\omega_i-1).
        \end{align}
        The result \eqref{eq:interface_RH} follows directly from the standard identity 
        \begin{align}
            \chi(\Sigma) = \chi(\Sigma_-) + \chi(\Sigma_+) - \chi(\Sigma_- \cap \Sigma_+).
        \end{align}
        This concludes our proof of the interface Riemann-Hurwitz formula \eqref{eq:interface_RH}. 
        
        The central role that this result will play in the rest of this work is analogous to the significance that the standard Riemann-Hurwitz formula has for symmetric product orbifolds in the absence of interfaces.
        It will most prominently be applied to derive topological expansions of interface correlators in Sections \ref{sec:CanonicalLargeN} and \ref{sec:string_genus_gce}, but is more generally also a useful tool to decide which topologies are allowed to contribute to a given correlation function.

    \subsection{Holographic interfaces}\label{sec:interfaces}
        We now use the interface covering maps introduced in Section \ref{sec:interface_coverings} 
        to define the interfaces of symmetric product orbifolds that are studied in this work.
        As an application of the new description presented here, we use it to compute a grand-canonical 
        torus partition function with two interface insertions. In the new formalism, the full derivation 
        is rather simple and can conveniently be described in less than a page. 
        Compared to this, the derivation in \cite{Harris:2025wak} (which moreover only partially 
        established the result) relied on rather lengthy and complicated combinatorial arguments.
        This new computational power is one of the key improvements of this paper and ultimately, 
        in Section \ref{sec:string_theory_&_GCE}, allows us to compute much more general partition 
        functions that were out of reach with the approach of \cite{Harris:2025wak} . 
        
        \paragraph{Normalisation of interface correlators.} In the setup described in Section \ref{sec:interface_coverings}, consider $\text{Sym}^{N_\pm}(\mathcal{M})$ on $X_\pm$. To 
        obtain a well defined theory on $X$, we need to provide CFT glueing data 
        along each of the $n$ connected components $\gamma_1$, $\dots$, $\gamma_n$ of $\partial X_\pm$.\footnote{We allow for a straightforward generalisation of the discussion in the previous subsection in 
        which the boundary $\partial X_+ = \partial X_-$ can possess several disconnected components
        $\gamma_i$ so that the interface $\mathcal{I}$ is a product $\mathcal{I} = \mathcal{I}_1 \cdots 
        \mathcal{I}_n$.} In this paper, we only study interfaces that preserve the lifting property that 
        defines symmetric orbifolds. In the presence of such an interface $\mathcal{I}$, the set $C[\mathcal{I}
        \mathcal{O}]$ of covering maps is no longer restricted to consist of usual branched covers, but 
        generically also includes interface coverings of the type introduced in the previous subsection.
        With this in mind, the formula for unnormalised interface correlators has exactly the same form 
        \eqref{eq:norm_vs_unnorm} as in the case of local operator insertions, 
        \begin{align}\label{eq:Interfaces_unnormalised_corr}
            \langle \mathcal{I}\mathcal{O} \rangle_{\text{Sym}^{N_-|N_+}(\mathcal{M})}^{\text{unnormalised}} 
            = \frac{1}{N_-!N_+!} \sum\limits_{\Gamma \in C[\mathcal{I}\mathcal{O}]}\langle 
            \mathcal{I}\mathcal{O}^\Gamma \rangle_\mathcal{M}
        \end{align}
        Note, however, that  we use the convention that the factor $N!$ appearing in 
        eq.~\eqref{eq:norm_vs_unnorm} is replaced by $N_-!N_+!$. As before, we also define normalised 
        interface correlators 
        \begin{align}\label{eq:Interfaces_normalised_corr}
            \langle \mathcal{I}\mathcal{O} \rangle_{\text{Sym}^{N_-|N_+}(\mathcal{M})} 
            := \frac{\langle \mathcal{I}\mathcal{O} \rangle_{\text{Sym}^{N_-|N_+}
            (\mathcal{M})}^{\text{unnormalised}} }{\langle \mathcal{I} \rangle_{\text{Sym}^{N_-|N_+}(\mathcal{M})}^{\text{unnormalised}} }.
        \end{align}
        In contrast to \eqref{eq:norm_vs_unnorm}, the interface prescription \eqref{eq:Interfaces_normalised_corr} is to normalise by an interface 
        partition function that contains the insertion of $\mathcal{I}$ instead of a vacuum bulk partition function. 
        
        \paragraph{Definition of the interfaces.} 
        Let us define a specific class of interface operators $\mathcal{I}$ that can be inserted 
        along the connected components of $\partial X_\pm$ in the setup of the previous paragraph. The definition we give here will later be 
        shown to coincide with the interfaces $\mathcal{I}^{(P)}_{|a_\pm\rangle}$ of symmetric product 
        orbifolds that were introduced in \cite{Harris:2025wak}. The interfaces are parametrised by a 
        non-negative integer $P \leq \text{min}(N_-,N_+)$, a pair of boundary states $|a_-\rangle$ and 
        $|a_+\rangle$ of the seed theory $\mathcal{M}$, as well as characters $\chi_\pm$ of $S_{N_\pm - P}$ 
        and $\chi_P$ of $S_P$.\footnote{The dependence on the characters was suppressed in the notation of 
        \cite{Harris:2025wak} since we were mostly considering the case in which all three characters are 
        trivial.} To simplify notation, we shall denote the collection of these data by $\varpi = 
        (P,|a_-\rangle,|a_+\rangle;\chi_-,\chi_+, \chi_P)$. Correlation functions with $\mathcal{I}_\varpi[\gamma]$ inserted 
        along some curve $\gamma$ are computed by summing over all interface coverings $\Gamma$ whose degrees 
        on $X_\pm$ are $N_\pm$ and whose degree of transmissivity on $\gamma$ is $P$. The relevant correlation 
        functions in the seed theory $\mathcal{M}$ on the covering surfaces $\Sigma$ depend on the choice
        of boundary states $|a_\pm\rangle$, which describe the boundary condition imposed on the potentially non-trivial 
        boundary $\partial \Sigma$. This boundary splits into disjoint subsets
        \begin{align}\label{eq:def_Rpm}
            R^\gamma_\pm := \partial \Sigma_\pm \cap \Gamma^{-1}(\gamma) \cap \partial \Sigma \ .   
        \end{align}
        The intersection with $\Gamma^{-1}(\gamma)$ is redundant in case where $\partial X_\pm$ consists of a single connected 
        component $\gamma$. The intersection with $\partial \Sigma$ removes the part of $\Gamma^{-1}(\gamma)$ that runs 
        through the interior of $\Sigma$, i.e.~the subsets 
        \begin{align}
            T^\gamma := \Gamma^{-1}(\gamma) \cap \Sigma^o\ . 
        \end{align}
        By construction, every element of $\Gamma^{-1}(\gamma)$ is either contained in $R^\gamma_\pm$ or in $T^\gamma$, i.e.
        \begin{align} \label{eq:Ggdecomposition}
            \Gamma^{-1}(\gamma) = R_-^\gamma \sqcup  T^\gamma \sqcup R_+^\gamma\ . 
        \end{align}
        The freedom to choose three characters and two boundary states in the data $\varpi$ that specifies the interface stems from this decomposition of $\Gamma^{-1}(\gamma)$ into three, and its overlap with $\partial \Sigma$ into two, disjoint sets.
        Concretely, the boundary states $|a_\pm\rangle$ describe the boundary 
        conditions of the seed theory that are imposed along the components $R_\pm^\gamma$ of $\partial \Sigma$. As for the characters, note that the monodromy of the covering map $\Gamma$ along the curve $\gamma$ preserves the decomposition \eqref{eq:Ggdecomposition}, i.e. 
        \begin{align}
            M_\Gamma[\gamma] \in \text{Aut}[S \cap R_-^\gamma] \times \text{Aut}[S \cap T^\gamma] 
            \times \text{Aut}[S \cap R_+^\gamma] \subseteq \text{Aut}[S]
        \end{align}
        where $S := \Gamma^{-1}(\gamma(0))$. In other words, the monodromy 
        factorises into a product 
        \begin{align} 
        M_\Gamma[\gamma] = M_\Gamma^-[\gamma]M_\Gamma^P[\gamma]M_\Gamma^+[\gamma]\ . 
        \end{align} 
        The characters $\chi_\pm$ and $\chi_P$ enter correlations of $\mathcal{I}_{\varpi}[\gamma]$ with $M_\Gamma^\pm[\gamma]$ and $M_\Gamma^P[\gamma]$ as their arguments\footnote{As explained in the paragraph ``local gauge choice'' of Subsection \ref{sec:branched_cover}, the three factors may be identified with elements of the permutation groups $S_{N_\pm-P}$ and $S_P$, respectively, to evaluate the characters without ambiguity i.e.~independent of the choice of identification. See in particular eq.~\eqref{eq:M_Gamma_R}.~}.
        More explicitly, correlation functions 
        \begin{align}\label{eq:prod_I_corr}
            \langle \prod\limits_{i=1}^n \mathcal{I}_{\varpi_i}[\gamma_i] \rangle_{\text{Sym}^{N_-|N_+}(\mathcal{M})}^{\text{unnormalised}} 
            \quad \text{with} \quad 
            \varpi_i := (P^i,|a_-^i\rangle,|a_+^i\rangle;\chi_-^i,\chi_+^i, \chi_P^i)
        \end{align}
        of products of interfaces
        are defined through eq.~\eqref{eq:Interfaces_unnormalised_corr} with
        \begin{equation}\label{eq:interface_corr_C}
            \hspace{-5 pt }C\left[\prod\limits_{i=1}^n \mathcal{I}_{\varpi_i}[\gamma_i]\right] :=
            \Big\{ \; \parbox{20em}{$N_\pm$ sheeted interface coverings of $X = X_+ \cup X_-$ with 
            degree of transmissivity $P^i$ along $\gamma_i$. } \; \Big\} 
        \end{equation}
        and the characters $\chi_\pm^i$ and $\chi_P^i$ enter through the lift 
         \begin{align} \label{eq:interface_operators}
            \left[\prod\limits_{i=1}^n 
            \mathcal{I}_{\varpi_i}[\gamma_i]
            \right]^{\Gamma} := \alpha[\Gamma]  \left( \prod\limits_{i=1}^n \chi_-^{i}(M_\Gamma^-[\gamma_i]) \chi_P^{i}(M_\Gamma^P[\gamma_i]) \chi_+^{i}(M_\Gamma^+[\gamma_i]) \right) \mathds{1},
        \end{align}
        where $\mathds{1}$ is the identity operator of $\mathcal{M}$ on $\Sigma$ and the coefficient 
        $\alpha[\Gamma]$ is given by 
        \begin{align}  \label{eq:interface_alpha}
        \alpha[\Gamma]= \frac{\text{deg}[\Gamma_-]!\text{deg}[\Gamma_+]!}{\left|\text{Deck}[\Gamma]\right|}.
        \end{align}  
        The only data that do not manifestly appear on the r.h.s.~of eq.~\eqref{eq:interface_operators} and \eqref{eq:interface_corr_C} are
        the boundary states $|a^i_\pm\rangle$. 
        These enter in the evaluation of the seed theory expectation value of the r.h.s.~of eq.~\eqref{eq:interface_operators}.  
        As mentioned below eq.~\eqref{eq:Ggdecomposition}, these boundary states 
        describe the boundary conditions we impose along the component $R^{\gamma_i}_\pm$ of the boundary 
        $\partial \Sigma$. 
        The prescription of how to evaluate the vacuum expectation value \eqref{eq:prod_I_corr} of an arbitrary product of interfaces on an arbitrary Riemann surface $X$ fully defines the interface $\mathcal{I}_{\varpi}$, just as eqs.~\eqref{eq:C[1]} and \eqref{eq:1^Gamma} for arbitrary vacuum partition functions already fully define the bulk theory. Details on the relation of this definition to the one given in 
        Reference \cite{Harris:2025wak} are discussed in Appendix \ref{app:comments_on_normalisation}.

        \paragraph{Vacuum expectation value.} As a first example of an interface correlation function let us 
        consider the simplest setup, namely, a single interface $\mathcal{I}_\varpi$ with $\varpi = (P,|a_\pm \rangle; 
        \chi_\pm, \chi_P)$ inserted along the equator $\gamma$ of the sphere. The interface splits $X=\mathbb{S}^2$ 
        into two disks $X_{\pm}=\mathbb{D}_\pm$. The set $C[\mathcal{I}_{\varpi}]$ consists of a single map 
        $\Gamma$, which covers $\mathbb{S}^2$ by a disjoint union of $P$ spheres and $N_\pm -P$ disks, that are 
        mapped to the upper/lower hemisphere. Every deck transformations is a product of one of the $P!$ permutations 
        of spheres and two permutations of disks chosen from the $(N_\pm-P)!$ possible options. Thus we conclude that 
        \begin{align}\label{eq:D-D+_main_text}
            Z^{\mathbb{D}_-,\mathbb{D}_+}_{N_\pm,\varpi} 
            :=~ \langle \mathcal{I}_\varpi \rangle_{\text{Sym}^{N_-|N_+}(\mathcal{M})}^{\text{unnormalised}} 
            =~ \frac{\chi_-(id)\chi_P(id)\chi_+(id)
            ~ g_-^{N_--P}g_+^{N_+-P}}{(N_- - P)!(N_+ - P)!P!},
        \end{align}
        where $g_\pm := \langle 0 | a_\pm \rangle$ are the $g$-factors of the two boundary conditions $a_\pm$ of the 
        seed theory that are imposed along $R^\gamma_\pm$. Note that the latter consists of $N_\pm-P$ disconnected 
        components. The expectation value of the seed theory is evaluated in a normalisation in which the sphere 
        partition function of the seed theory $\mathcal{M}$ is set to 1. The same result can alternatively be 
        obtained from the construction in \cite{Harris:2025wak}, see eq.~\eqref{eq:spherevacuum}. 
        
        \paragraph{Torus partition function -- statement of the result.} Let us now discuss a much richer 
        example of an interface correlator, namely a pair of parallel interfaces inserted on a 
        torus $X = \mathbb{T}^2 = \mathbb{C} / (\mathbb{Z} + \tau \mathbb{Z})$. The interfaces are 
        located along two curves $\gamma_L: s \mapsto \tau s$ and $\gamma_R: s \mapsto \tau s + 1/2$.
        These cut the torus into two annuli $X_\pm$. The interface data $\varpi^L$ and $\varpi^R$ can 
        involve two different indices of transmissivity $P_L$ and $P_R$ and two different pairs of 
        boundary conditions $a^L_\pm$ and $a^R_\pm$. 
        To avoid notational clutter, we shall assume the characters 
        $\chi^{L/R}$ to be trivial, though it would not be difficult to consider the more general 
        case. According to the general definition, the interface correlation function takes the form 
        \begin{align}\label{eq:torus_definition}
            Z_{N_\pm,P_{L/R}}^{\mathbb{T}^2} & = \langle \mathcal{I}^{(P_L)}[\gamma_L] 
            \mathcal{I}^{(P_R)}[\gamma_R] \rangle_{\text{Sym}^{N-|N_+}(\mathcal{M})}^\text{unnormalised} = \sum\limits_{\Gamma \in C[\mathcal{I}^L \mathcal{I}^R]} 
            \frac{1}{\left|\text{Deck}[\Gamma]\right|} Z_\mathcal{M}^\Sigma ,
        \end{align}
        where $Z_\mathcal{M}^\Sigma$ is the vacuum partition function of $\mathcal{M}$ on $\Sigma$ with boundary conditions $|a_\pm^{L/R}\rangle$ imposed on $\partial \Sigma$. 
        It was found in \cite{Harris:2025wak} that these interface partition functions take a particularly useful shape if we pass to the 
        following grand-canonical partition function 
        \begin{align}\label{eq:gc_torus}
            Z^{\mathbb{T}^2}_{\mu_{L/R},\nu_{L/R,\pm}} := \sum\limits_{N_\pm = 0}^\infty \sum\limits_{P_{L/R} = 0}^{\text{min}(N_-,N_+)} Z^{\mathbb{T}^2}_{N_\pm,P_{L/R}} \prod\limits_{A\in\{L,R\}} \mu_A^{P_A}\nu_{A,-}^{N_--P_A}\nu_{A,+}^{N_+-P_A} 
        \end{align}
        where the dependence of the partition function on the integers $N_\pm$ and $P_{L/R}$ is traded 
        for a dependence on the fugacities $\mu_{L/R}$ and $\nu_{{L/R},\pm}$. The formula put forward in 
        \cite{Harris:2025wak} reads\footnote{We chose a slightly modified but equivalent way of writing 
        down the result of \cite{Harris:2025wak} here.} 
        \begin{align}\label{eq:torus_gc_result}
            Z^{\mathbb{T}^2}_{\mu_{L/R},\nu_{L/R,\pm}} = \exp\left( \sum\limits_{w,\ell=1}^\infty \sum\limits_{k=1}^w \frac{\mu^{w\ell}}{w\ell}Z(\tfrac{\ell \tau + k}{w}) + \sum\limits_{\substack{\omega,\ell \\ A,B}} \frac{1}{\ell} \left(\frac{\mu^\omega\nu_{AB} }{\mu_A\mu_B} \right)^\ell Z^{AB}_{o}(\tfrac{\ell \tau }{\omega})\right) 
        \end{align}
        where the sum in the second terms runs over $A,B \in\{L,R\}$,  $\omega \in 2\mathbb{Z}_{> 0}-1+\delta_A^B$ and $\ell \ge 1$. The variables $\mu$ and $\nu_{AB}$ are given by 
        $\mu:=\mu_L \mu_R$ and   
        \begin{align}
            \nu_{LL}:=\nu_{L,-}\nu_{L,+}, \quad \nu_{LR}:=\nu_{L,-}\nu_{R,-}, \quad \nu_{RL}:=\nu_{R,+}\nu_{L,+} \quad \text{and} \quad \nu_{RR}:=\nu_{R,-}\nu_{R,+}.
        \end{align}
        The dependence of the grand-canonical partition function on the seed theory and the choice of boundary conditions $a^{L/R}_\pm$ enters through the overlaps 
        \begin{align}
            Z_{a,b}(t)  = \langle a |\, \hat x^{\frac12 (L_0 + \bar L_0 - \frac{c_\mathcal{M}}{12})}
            \, |b\rangle =  \hat Z_{a,b}(\hat \tst) && \hat t:= -\frac{1}{t} && \hat x := e^{2\pi i \hat t},
        \end{align}
        between a pair of boundary states $|a\rangle$ and $|b\rangle$ of the seed theory. Through modular
        transformation, these overlaps are related to  annulus partition functions $\hat Z$, as usual. There 
        are four different annulus partition functions that appear in eq.\ \eqref{eq:torus_gc_result} which 
        we denoted by 
         \begin{align} \label{eq:seedZo}
            Z^{LL}_o = Z_{a^L_-,(a^L_+)^*}, \quad
            Z^{LR}_o = Z_{a^L_-,a^R_-}, 
            \quad
            Z^{RL}_o = Z_{a^L_+,a^R_+}
            \quad
            \text{and}
            \quad
            Z^{RR}_o = Z_{(a^R_+)^*,a^R_-}.
        \end{align}
        This concludes our description of the grand-canonical partition function \eqref{eq:torus_gc_result}
        for two interfaces $\mathcal{I}[\gamma_L]$ and $\mathcal{I}[\gamma_R]$ on the torus $\mathbb{T}^2$, restricted to the case of trivial characters.  
        
        \paragraph{Torus partition function -- derivation.}
        Let us now derive the formula \eqref{eq:torus_gc_result} from the general prescription in
        eq.~\eqref{eq:torus_definition}. The first ingredient there is the sum over interface 
        covering maps $\Gamma \in C[\mathcal{I}^L\mathcal{I}^R]$. 
        The relevant covering surfaces $\Sigma$ 
        can be decomposed as 
        \begin{align} 
        \Sigma = \Sigma_c \sqcup \Sigma^{LL}_o\sqcup\Sigma^{LR}_o\sqcup\Sigma^{RL}_o\sqcup\Sigma^{RR}_o
        \end{align}
        where the restriction of the covering map $\Gamma$ to $\Sigma_c$ is one of the coverings 
        of the torus by tori discussed in Sec.~\ref{sec:symm_orb} and each connected component of 
        the surfaces $\Sigma^{AB}_o$ is a cylinder
        \begin{align}
            \Sigma_{\omega,\ell} := [0,\tfrac{\omega}{2}] \times [0,\ell|\tau|]  / (x,0) \sim (x,\ell|\tau|).
        \end{align}
        The restriction of $\Gamma^{AB}_o:\Sigma^{AB}_o \rightarrow \mathbb{T}^2$ to the 
        connected component $\Sigma_{\omega,\ell}$ is given by 
        \begin{align}
            \Gamma^{AB}_{\omega,\ell} : \Sigma_{\omega,\ell} \rightarrow \mathbb{T}^2, \quad (x,y) 
            \mapsto \left[\left(x + \tfrac{1}{2} \delta_A^R, y \tfrac{\tau}{|\tau|}\right)\right],
        \end{align}
        where the value of $\omega$ is restricted to $\omega \in 2\mathbb{Z}+1-\delta_A^B$.
        Furthermore, $\text{Deck}[\Gamma^{AB}_{\omega,\ell}] \cong \mathbb{Z}_\ell$ consists of translations 
        $(x,y) \mapsto (x,y+k|\tau|)$. Let us denote by $C^k_{w,\ell}$ the number of copies of the surface 
        $\Sigma^{k}_{\omega,\ell}$ that appear within $\Sigma_c$ and similarly denote by $O^{AB}_{\omega,\ell}$
        the number of components of type $\Sigma_{\omega,\ell}$ that appear in $\Sigma^{AB}_o$. Then the 
        number of Deck transformations is given by 
        \begin{align}
            |\text{Deck}[\Gamma]| = \prod\limits_{w,\ell,k} C_{w,\ell}^k! (w\ell)^{C_{w,\ell}^k}
            \prod\limits_{\substack{\omega,\ell \\ A,B}} O^{AB}_{\omega,\ell}!\ell^{O^{AB}_{\omega,\ell}}\ . 
        \end{align}
        Inserting this formula into the general definition \eqref{eq:torus_definition} of the interface 
        partition functions on the torus we obtain  
        \begin{align}
            Z^{\mathbb{T}^2}_{N_\pm,P_{L/R}} =
            \hspace{-0.7 cm} \sum\limits_{\text{allowed } C,O^{AB}} 
            \left( \prod\limits_{w,\ell,k} \frac{\left(\tfrac{1}{w\ell}Z(\tfrac{\ell \tau + k}{w})\right)^{C_{w,\ell}^k}}{C_{w,\ell}^k !}
            \prod\limits_{\substack{\omega,\ell \\ A,B}} \frac{\left(\tfrac{1}{\ell}Z^{AB}_o(\tfrac{\ell \tau }{\omega})\right)^{O_{\omega,\ell}^{A,B}}}{O^{AB}_{\omega,\ell}!}\right)
        \end{align}
        where the allowed coefficients $C$ and $O^{A,B}$ are exactly those satisfying the constraints 
        \begin{align}
            N_- - P_L &=  \sum\limits_{\omega,\ell,B} O^{LB}_{\omega,\ell} \ell,  
            \quad\,\,
            N_+ - P_L = \sum\limits_{\omega,\ell,A} O^{AL}_{\omega,\ell} \ell ,
            \quad\,\,
            N_- - P_R = \sum\limits_{\omega,\ell,A} O^{AR}_{\omega,\ell} \ell ,  \\
            N_+ - P_R &= \sum\limits_{\omega,\ell,B} O^{RB}_{\omega,\ell} \ell,
            \quad
            \text{and} 
            \quad 
            P_A = \sum\limits_{w,\ell} \sum\limits_{k} w\ell C^k_{w,\ell} + \sum\limits_{\omega,\ell}\sum\limits_{A,B} (\omega-\delta_I^A-\delta_J^A)\ell O^{I,J}_{\omega,\ell} .
        \end{align}
        For the grand-canonical partition function \eqref{eq:gc_torus}, these constraints are lifted and 
        by performing the unconstrained sum over $C_{w,\ell}^k$ and $O^{AB}_{\omega,\ell}$ one immediately 
        obtains eq.~\eqref{eq:torus_gc_result}.

        \paragraph{Spectrum of interface changing operators.} 
        In this paragraph, we rewrite eq.~\eqref{eq:torus_gc_result} in a form that allows us to directly read off the spectrum of interface changing operators.
        To do so, we start by spelling out the seed partition functions in terms of the bulk and boundary spectra $I$ and $I_o^{AB}$ as well as the multiplicities $d_{h,\bar h}$ and $d_\Delta$ of the corresponding representations in the Hilbert space, i.e.
        \begin{align}
            Z(\tau,\bar \tau) = \sum_{(h,\bar{h}) \in I} d_{h,\bar{h}}~ q^{h-\tfrac{c_\mathcal{M}}{24}} ~\bar{q}^{\bar{h}-\tfrac{c_\mathcal{M}}{24}}
            \qquad \text{and} \qquad
            Z^{AB}_o(\tau) = \sum_{\Delta \in I^{AB}_o} d^{AB}_\Delta~ q^{\Delta-\tfrac{c_\mathcal{M}}{24}} .
        \end{align}
        Setting $\bar \tau = \tau$ for the bulk partition function and using the identities 
        \begin{align}
            \exp\left(\alpha \sum\limits_{\ell=1}^\infty \frac{x^\ell}{\ell}\right) = (1-x)^{-\alpha} && \text{and} && \frac{1}{w}\sum_{k=1}^w e^{2\pi i\frac{k}{w}(x-\bar{x})} = \begin{cases}
                1 & \text{if } x-\bar x \equiv 0 \text{ mod } w \\
                0 & \text{else}
            \end{cases}
        \end{align}
        then allows us to rewrite eq.~\eqref{eq:torus_gc_result} as
        \begin{align}\label{eq:Z_manifest_spectrum}
            Z^{\mathbb{T}^2}_{\mu_{L/R},\nu_{L/R,\pm}} \hspace{-0.1 cm} = \hspace{-0.1 cm}
            \prod_{w>0} \hspace{-0.4 cm} \prod_{\substack{(h,\bar{h})\in I \\ h-\bar{h}\equiv0 \text{ mod }w}}
            \hspace{-0.4 cm}
            \Bigg(1-\mu^w q^{\frac{1}{w}(h+\bar h - \frac{c_\mathcal{M}}{12})}\Bigg)^{-d_{h,\bar{h}}} 
            \hspace{-0.8 cm}
            \prod_{A,B}\prod_{\omega}\prod_{\Delta \in I^{AB}_o}
            \Bigg(1-\frac{\mu^\omega\nu_{AB} }{\mu_A\mu_B}q^{\frac{1}{\omega}(\Delta- \frac{c_\mathcal{M}}{24})}\Bigg)^{-d^{AB}_\Delta}\hspace{-0.7 cm},
        \end{align}
        where $\omega$ runs over $2\mathbb{Z}_{> 0}-1+\delta_A^B$. 
        The scaling dimensions of interface changing operators can be read off from eq.~\eqref{eq:Z_manifest_spectrum} by extracting the fixed $N_\pm$ partition functions and adding $\frac{c_\mathcal{M}}{24}(N_-+N_+)$ to the exponents of $q$. 
        From the grand-canonical perspective, this amounts to shifting the exponent of $q$ for the factors of the product over $w$ by $\frac{c_\mathcal{M}}{24}2w$ and likewise, shifting the factors of the $\omega$ product by $\frac{c_\mathcal{M}}{24} \omega$. 
        We conclude that the spectrum of local operators that can be inserted between our interfaces consists of products of single particle fields falling into the five families $\mathcal{O}_w$ and $\mathcal{O}_\omega^{AB}$ with $A,B \in \{L,R\}$.
        The former can simply be interpreted as bulk operators inserted on the transmissive part of the interfaces, while the latter are intrinsically attached to the interface.
        Quantitatively, the arguments given above tell us that for each $w>0$ and each seed theory bulk operator $\mathcal{O}$ of dimension $\bar h+h$, and spin $\bar h - h$ divisible by $w$, we obtain a single particle interface operator with dimension
        \begin{align}
            \Delta_w = \frac{\bar h+h}{w} + \frac{c_\mathcal{M}}{12}\left(w - \frac{1}{w}\right),
        \end{align}
        while each weight $\Delta \in I_o^{AB}$ leads to a family of fields with dimension
        \begin{align}\label{eq:interface_dimensions}
            \Delta_\omega^{AB} = \frac{\Delta}{\omega} + \frac{c_\mathcal{M}}{24}\left(\omega + \frac{1}{\omega }\right).
        \end{align}
        In the special case where the boundary states that enter the construction of the interfaces are chosen 
        to coincide, i.e. $a_+ = a_-$, the seed theory boundary spectra that are captured by the partition functions \eqref{eq:seedZo} contain an identity operator, i.e.~$0 
        \in I_o^{AB}$, and thus the spectrum of the symmetric product orbifold contains \emph{interface twist fields} $\sigma_\omega^{AB}$ whose conformal weight is given by 
        eq.~\eqref{eq:interface_dimensions} with $\Delta =0$. Studying correlation functions of these interface twist fields is the main focus of our work and hence we shall mostly consider interfaces with $|a_+\rangle = |a_-\rangle =:|a\rangle$ (and $\chi_\pm =  \chi_P = 1$). The associated interface operators will be denoted by $\mathcal{I}_{|a\rangle}^{(P)}$.
        
        \paragraph{Interface changing correlators.}
        We are finally ready to address the main topic of this work, namely correlation functions including twist
        field insertions along the interface (and in the bulk). To keep things simple we return to the case
        in which an interface operator $\mathcal{I}[\gamma]$ is inserted along a single closed curve $\gamma$ so
        that we are strictly in the setup we considered previously in Section \ref{sec:interface_coverings}. The extension to multiple
        interface insertions is straightforward. For simplicity we also restrict to the case of trivial
        characters $\chi$ and set $|a_+\rangle = |a_-\rangle =: |a\rangle$, see our discussion at the end of 
        the previous paragraph. This leaves the degree of transmissivity $P$ as the main parameter of the
        interface, besides the choice of the boundary state $|a\rangle$. 
        
        As we explained above, some interface twist fields can only be inserted at points at which the degree of
        transmissivity changes. To be more precise, the interface twists fields $\sigma^{LL}_\omega$ and
        $\sigma^{RR}_\omega$ can only be inserted at points along the curve $\gamma$ at which the degree of
        transmissivity $P$ jumps by one unit to $P \pm 1$, see also Figure \ref{fig:IntTwistRamif}. Put
        differently, if we want to insert these fields we have to break up the curve $\gamma$ into smaller
        segments and insert interface operators with different degree of transmissivity along these segments.
        Since this is notationally a bit cumbersome we shall keep using the interface operator
        $\mathcal{I}^{(P)}_{|a\rangle} [\gamma]$ where $P$ denotes the degree of transmissivity of 
        $\mathcal{I}[\gamma]$ at the base point $\gamma(0) = \gamma(1)$ of the curve. We shall sometimes 
        refer to this base point as the \emph{reference point} and denote it by $t^*$ in the rest of this paper.
        It is then understood that this degree jumps whenever $\gamma$ passes through the insertion point 
        of an interface changing twist field $\sigma^{LL}_\omega(t)$ or $\sigma^{RR}_\omega(t)$. 
        
        With this comment on notation in mind let us now spell out the family of observables $\mathcal{O}$ 
        whose expectation values we want to compute. These are given by 
        \begin{align}\label{interface_changing_correlator}
            \mathcal{O} = \mathcal{I}^{(P)}_{|a\rangle}[\gamma]\prod_{i=1}^{n_c}
            \sigma_{w_{i}}(x_{i}) \prod_{i=1}^{n_o}\sigma^{A_iB_i}_{\omega_{i}}(t_{i})
        \end{align}
        where we assume that the reference point $t^* = \gamma(0)$ is not the insertion point of a local operator. 
    
        As in our previous discussions, the expectation values of the observables 
        $\mathcal{O}$ we consider now can be evaluated by summing certain expectation values on the seed 
        theory over some set of covering maps $\Gamma \in C[\mathcal{O}]$. The insertion of the interface 
        changing operators $\sigma_\omega^{AB}(t_i)$ into the expectation value enforces that covering maps 
        that contribute to the correlator have a type $\omega^{AB}$ interface ramification point that maps 
        to the branch point $t_i \in X$. Thus,
        \begin{equation}\label{eq:interface_twist_field_C}
            \hspace{-5 pt }C\left[\mathcal{O}\right] =\left\{ \; \parbox{28em}{$N_\pm$ sheeted interface coverings of $X$ with deg.~of transmissivity~$P$ at $\gamma(0)$ and bulk branchpts.~$x_i$ lifting to ramification pts.~with ramification index $w_i$ as well as interface branchpts.~$t_i$ lifting to interface ramification pts.~with ramification index $\omega^{A_iB_i}_i$.
            } \; \right\}.
        \end{equation}
        Like the bulk vacuum twist fields, the interface vacuum twist fields lift to the identity operator 
        along the covering map, up to some coefficient as in eq.\ \eqref{eq:interface_operators}. Since
        we assumed that all three characters $\chi_\pm$ and $\chi_P$ are trivial, the factors that depend
        on the monodromies of the covering map is also trivial and hence we are left with $\alpha[\Gamma]$.
        In Sec.~\ref{sec:PRR}, we determine $C\left[\mathcal{O}\right]$ for any correlator with $X = \mathbb{S}^2$ explicitly. The explicit evaluation of the correlators from these data is then addressed in  
        Sec.~\ref{sec:Lunin_Mathur}. 

\section{Generalised Pakman-Rastelli-Razamat diagrammatics}\label{sec:PRR}
    In this section, we introduce diagrammatic rules that provide a systematic classification of the covering maps that contribute to any specific interface correlator.
    In the first subsection,  Sec.~\ref{sec:PRR_diagrams}, we review the analogous diagrams for bulk correlators which were introduced by Pakman, Rastelli and Razamat in \cite{Pakman:2009ab}.
    The main subsection, \ref{sec:Interface_PRR}, introduces the interface generalisation. 
    Sec.~\ref{sec:symmetry_factors} describes how to determine the symmetry group of an interface covering map, i.e.~the group $\textrm{Deck}[\Gamma]$ of its deck transformations, directly from the corresponding diagram. 
    Subsection \ref{sec:GaugeFixing} provides an alternative perspective on the same problem and
    determines the symmetry factors by counting different ways to label faces of diagrams, an approach that more directly connects to the bulk discussion of Reference \cite{Pakman:2009ab}. 
    The final subsection, \ref{sec:CanonicalLargeN}, is devoted to the large-$N$ limit of interface correlators.

    \subsection{Pakman-Rastelli-Razamat diagrams}\label{sec:PRR_diagrams}
        This section reviews the Pakman-Rastelli-Razamat (PRR) diagrammatic rules. 
        We first discuss how to construct the associated PRR diagram given a covering map.
        To do so, we take a slight detour of first constructing auxiliary diagrams with redundant extra vertices and then reducing these to their PRR form.
        While this makes the discussion of the case without interfaces slightly less efficient than the original description in Reference \cite{Pakman:2009ab}, the approach will pay off in Section \ref{sec:Interface_PRR} since it is well aligned with the 
        interface generalisation.    
        
        After we have explained how to extract diagrams form covering maps, the second paragraph of the section formulates constructive rules that allow us to generate the set of PRR diagrams without knowledge of the covering maps.
        Finally, we showcase how to apply the rules in simple examples. 
        Appendix \ref{app:proof_rules} sketches a proof of the validity of the rules.
        
        \begin{figure}
            \centering
            \begin{tikzpicture}
                \node [style=none] (0) at (-6, 0.5) {};
		\node [style=none] (1) at (6, 0.5) {};
		\node [style=none] (6) at (-4.5, 0.5) {};
		\node [style=none] (7) at (-6, -0.25) {};
		\node [style=none] (10) at (-4.5, 4.25) {};
		\node [style=none] (12) at (-3.5, 4.25) {};
		\node [style=none] (13) at (-4, 4.75) {};
		\node [style=none] (16) at (-2.5, -0.25) {};
		\node [style=none] (17) at (-3.5, 0.5) {};
		\node [style=none] (20) at (-0.75, 0.5) {};
		\node [style=none] (22) at (-0.75, 5) {};
		\node [style=none] (24) at (0.25, 5) {};
		\node [style=none] (25) at (-0.25, 5.5) {};
		\node [style=none] (28) at (6, -0.25) {};
		\node [style=none] (29) at (0.25, 0.5) {};
		\node [style=none] (32) at (1.25, 0.5) {};
		\node [style=none] (34) at (1.25, 2.75) {};
		\node [style=none] (36) at (2.25, 2.75) {};
		\node [style=none] (37) at (1.75, 3.25) {};
		\node [style=none] (41) at (2.25, 0.5) {};
		\node [style=none] (44) at (-0.75, 0.5) {};
		\node [style=none] (45) at (-1.5, -0.25) {};
		\node [style=none] (47) at (-2.5, -2.5) {};
		\node [style=none] (48) at (-1.5, -2.5) {};
		\node [style=none] (49) at (-2, -3) {};
		\node [style=none] (52) at (-2.5, -0.25) {};
		\node [style=White Dot] (56) at (-4, 4.25) {$w_+$};
		\node [style=White Dot] (57) at (-0.25, 5) {$w_+$};
		\node [style=White Dot] (58) at (1.75, 2.75) {$w_+$};
		\node [style=White Dot] (59) at (-2, -2.5) {$w_-$};
        
		\draw [style=SingleU] (0.center) to (6.center);
		\draw [style=SingleU, bend left=45] (10.center) to (13.center);
		\draw [style=SingleU, bend left=45, looseness=1.25] (13.center) to (12.center);
		\draw [style=SingleU] (12.center) to (17.center);
		\draw [style=SingleD] (7.center) to (16.center);
		\draw [style=SingleU, bend left=45] (22.center) to (25.center);
		\draw [style=SingleU, bend left=45, looseness=1.25] (25.center) to (24.center);
		\draw [style=SingleU] (24.center) to (29.center);
		\draw [style=SingleU, bend left=45] (34.center) to (37.center);
		\draw [style=SingleU, bend left=45, looseness=1.25] (37.center) to (36.center);
		\draw [style=SingleU] (36.center) to (41.center);
		\draw [style=SingleU] (6.center) to (10.center);
		\draw [style=SingleU] (20.center) to (22.center);
		\draw [style=SingleU] (32.center) to (34.center);
		\draw [style=SingleU] (29.center) to (32.center);
		\draw [style=SingleU] (41.center) to (1.center);
		\draw [style=SingleD] (52.center) to (47.center);
		\draw [style=SingleD, in=-180, out=-90] (47.center) to (49.center);
		\draw [style=SingleD, in=-90, out=0, looseness=1.25] (49.center) to (48.center);
		\draw [style=SingleD] (48.center) to (45.center);
		\draw [style=SingleD] (45.center) to (28.center);
		\draw [style=SingleU] (17.center) to (44.center);
            \end{tikzpicture}
            \caption{
                Lines on $\mathbb{S}^2$ whose preimage w.r.t.~a covering map $\Gamma$ produce $\mathcal{D}_\varepsilon[\Gamma]$. 
                Branch points in the upper and lower hemisphere are denoted by $w_+$ and $w_-$.
                Orientations are represented by half-arrows extending into the disk on whose boundary they are drawn.
            }
            \label{fig:illustration_Gamma_to_D}
        \end{figure}
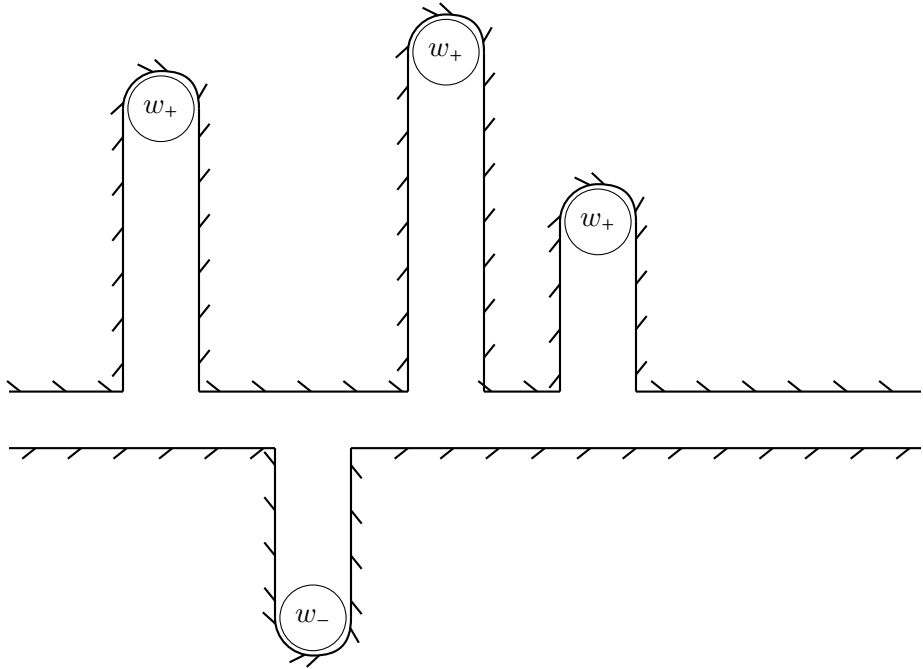
        
        \paragraph{Diagrams from covering maps.}
        To construct the PRR diagram $\mathcal{D}[\Gamma]$ of a covering map 
        $\Gamma: \Sigma \rightarrow \mathbb{S}^2$, we perform the following 
        three steps.
        \begin{enumerate}
            \item Draw lines with the same orientation, at distance $\varepsilon$
            above and below the equator, deformed as in Fig.~\ref{fig:illustration_Gamma_to_D} such that all branch points 
            of $\Gamma$ are in the strip between them.
            \item The preimage $\mathcal{D}_\varepsilon[\Gamma]$ of this configuration with respect to~$\Gamma$ is a network of oriented lines 
            on $\Sigma$. Take the limit  $\mathcal{D}[\Gamma] := \lim\limits_{\varepsilon \rightarrow 0}\mathcal{D}_\varepsilon[\Gamma]$.
            \item Each face of $\mathcal{D}[\Gamma]$ is either the preimage of the upper or the lower hemisphere with respect to $\Gamma$. 
            Label the faces with ``$+$'' and ``$-$'' respectively.
        \end{enumerate}
        If $\Gamma$ has degree $N$ then $\mathcal{D}[\Gamma]$ has $2N$ faces, 
        all of them with a consistently oriented boundary and exactly half
        labelled by $+$.
        The diagram has four classes of vertices, two of which are shown in
        Fig.~\ref{fig:PRR_vertices}. The other two may be obtained simply by 
        flipping all signs $\pm$ in the faces.  In the $\varepsilon \rightarrow 0$
        limit, the preimages of the lines on $\mathbb{S}^2$, reaching out to
        branch points from the equator of the sphere, degenerate into double 
        lines which end on the branch points. We shall refer to these double 
        lines as misaligned since the arrows along them point in opposite
        directions. Each misaligned double line lifts to $w$ copies on $\Sigma$, 
        leading to the vertex shown in Fig.~\ref{fig:bulkPRR_vertices_bulk_rammification}. 
        
        Note that the faces that touch at a misaligned double line possess 
        the same sign.
        Furthermore, each region where such a fold leaves the equator becomes a
        trivalent vertex of the type shown in Fig.~\ref{fig:bulkPRR_vertices_bulk_splitting}.
        We refer to the latter type of vertices as redundant vertices for reasons
        that will become clear in a moment. Let us first note, however, that these
        trivalent vertices involve a single misaligned double line and two aligned double lines
        for which the arrows on both sides of the double line point in the same 
        direction. Faces joined by aligned double lines carry opposite signs. 
        
\begin{figure}
            \begin{subfigure}{.5\textwidth}
                \centering
                \begin{tikzpicture}[scale=0.7]
                    \node [style=none] (1) at (2, -2.5) {};
                    \node [style=none] (5) at (-2, -2.5) {};
                    \node [style=none] (6) at (0, 3) {};
                    \node [style=none] (8) at (-3, 1) {};
                    \node [style=none] (14) at (3, 1) {};
                    \node [style=White Dot] (15) at (0, 0) {$w_+$};
                    \node [style=none] (16) at (2, -0.5) {+};
                    \node [style=none] (17) at (1.25, 1.75) {+};
                    \node [style=none] (18) at (-1.25, 1.75) {+};
                    \node [style=none] (19) at (-2, -0.5) {+};
                    \node [style=none] (20) at (0, -1.75) {$\dots$};
                    \node [style=none] (21) at (-1.5, 4.5) {};
                    \node [style=none] (22) at (1.5, 4.5) {};
                    \node [style=none] (23) at (-3, 3) {};
                    \node [style=none] (24) at (-5, 1) {};
                    \node [style=none] (25) at (-4, -2.5) {};
                    \node [style=none] (26) at (-2, -4.5) {};
                    \node [style=none] (27) at (2, -4.5) {};
                    \node [style=none] (28) at (4, -2.5) {};
                    \node [style=none] (29) at (5, 1) {};
                    \node [style=none] (30) at (3, 3) {};
                    \node [style=none] (31) at (0, 4.5) {$-$};
                    \node [style=none] (32) at (-4, 2) {$-$};
                    \node [style=none] (33) at (-3, -3.5) {$-$};
                    \node [style=none] (34) at (3, -3.5) {$-$};
                    \node [style=none] (35) at (4, 2) {$-$};
            
                    \draw [style=MisalignedU] (15) to (14.center);
                    \draw [style=MisalignedU] (15) to (6.center);
                    \draw [style=MisalignedU] (15) to (8.center);
                    \draw [style=MisalignedU] (15) to (5.center);
                    \draw [style=MisalignedU] (15) to (1.center);
                    \draw [style=Aligned] (8.center) to (24.center);
                    \draw [style=Aligned] (23.center) to (8.center);
                    \draw [style=Aligned] (6.center) to (21.center);
                    \draw [style=Aligned] (22.center) to (6.center);
                    \draw [style=Aligned] (14.center) to (30.center);
                    \draw [style=Aligned] (29.center) to (14.center);
                    \draw [style=Aligned] (1.center) to (28.center);
                    \draw [style=Aligned] (27.center) to (1.center);
                    \draw [style=Aligned] (5.center) to (26.center);
                    \draw [style=Aligned] (25.center) to (5.center);
                \end{tikzpicture}
                \caption{All edges connecting to $w$ vertices end on \\ the redundant trivalent vertices.}
                \label{fig:PRR_dressed_rammification_vertex}
            \end{subfigure}%
            \begin{subfigure}{.5\textwidth}
                \centering     
                \begin{tikzpicture}[scale=0.7]
        		\node [style=none] (16) at (2.5, -0.5) {+};
        		\node [style=none] (17) at (1.5, 2.5) {+};
        		\node [style=none] (18) at (-1.5, 2.5) {+};
        		\node [style=none] (19) at (-2.5, -0.5) {+};
        		\node [style=none] (20) at (0, -3) {$\dots$};
        		\node [style=none] (21) at (-1.5, 4.5) {};
        		\node [style=none] (22) at (1.5, 4.5) {};
        		\node [style=none] (23) at (-3, 3) {};
        		\node [style=none] (24) at (-5, 1) {};
        		\node [style=none] (25) at (-4, -2.5) {};
        		\node [style=none] (26) at (-2, -4.5) {};
        		\node [style=none] (27) at (2, -4.5) {};
        		\node [style=none] (28) at (4, -2.5) {};
        		\node [style=none] (29) at (5, 1) {};
        		\node [style=none] (30) at (3, 3) {};
        		\node [style=none] (31) at (0, 3) {$-$};
        		\node [style=none] (32) at (-2.5, 1.5) {$-$};
        		\node [style=none] (33) at (-2, -2) {$-$};
        		\node [style=none] (34) at (2, -2) {$-$};
        		\node [style=none] (35) at (2.5, 1.5) {$-$};
        		\node [style=White Dot] (36) at (0, 0) {$w$};
        
        		\draw [style=Aligned] (36) to (24.center);
        		\draw [style=Aligned] (23.center) to (36);
        		\draw [style=Aligned] (36) to (21.center);
        		\draw [style=Aligned] (22.center) to (36);
        		\draw [style=Aligned] (36) to (30.center);
        		\draw [style=Aligned] (29.center) to (36);
        		\draw [style=Aligned] (36) to (28.center);
        		\draw [style=Aligned] (27.center) to (36);
        		\draw [style=Aligned] (36) to (26.center);
        		\draw [style=Aligned] (25.center) to (36);
                \end{tikzpicture}
                \caption{Reduced $2w$-valent vertex obtained by contracting lines with misaligned orientation.}
                \label{fig:PRR_fused_vertex}
            \end{subfigure}
            \caption{Reduction to standard PRR diagrams.}
            \label{fig:PRR_reduction}
        \end{figure}
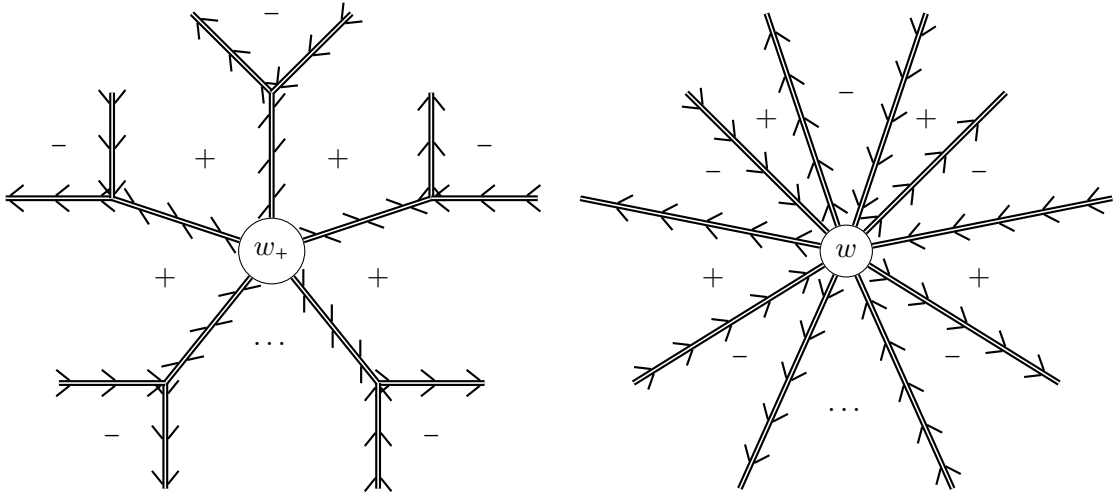

        As illustrated in Fig.~\ref{fig:PRR_dressed_rammification_vertex}, all
        edges emanating from a vertex that corresponds to a ramification point 
        end on redundant vertices. We can therefore simplify the diagrams without
        losing any information by moving the trivalent vertices like a zip 
        toward the ramification vertex until we have entirely eliminated the 
        misaligned double line. For each misaligned double line that was 
        entering the 
        ramification vertex before, this reduction produces two aligned double lines 
        that enter the vertex, with the sign of the faces flipping each time we
        cross an aligned double line. This fusion of ramification and redundant 
        vertices, leads to diagrams that only have one type of vertices, namely
        those shown in Fig.~\ref{fig:PRR_fused_vertex}. 
        The reduced graphs are the standard PRR diagrams.

        Let us close this subsection with further comments on this relation to the diagrams of Pakman, Rastelli and Razamat. 
        The original PRR diagrams use a single 
        type of double lines composed of one dashed and one solid line with no displayed orientation. 
        But in checking consistency of PRR diagrams it becomes 
        important that one goes through the dashed loops in clockwise 
        order and through solid loops in anti-clockwise order. In this 
        sense, the lines of PRR diagrams inherit an orientation from the
        orientation of $\Sigma$. We have decided here to make this orientation 
        manifest in our graphics. The PRR faces that are enclosed by solid loops
        correspond to the faces we label with $+$, while those surrounded by dashed 
        loops are the ones we label with $-$. The main difference in our approach 
        is that, initially, we have three types of double lines. Our aligned double lines correspond to the PRR double lines. But in addition we have two more
        double lines, namely the misaligned double lines that run between two 
        faces with label $+$, and the misaligned double lines that run between two 
        faces that carry a $-$ label. Following PRR graphics, one can also
        picture these as double lines with two solid and two dashed lines, 
        respectively. Our redundant vertex would then be represented as a 
        vertex in which the double line with e.g. two solid lines splits into 
        two double lines of PRR type, i.e.~with one dashed and one solid line.
        As long as there is no interface located at the equator, the misaligned double lines can be removed by the reduction procedure described in the previous paragraph.
        This leaves us with diagrams that only contain the aligned double lines used by Pakman, Rastelli and Razamat.

        \begin{figure}
            \begin{subfigure}{.5\textwidth}
                \centering
                \begin{tikzpicture}
                    \node [style=none] (1) at (2, -2.5) {};
                    \node [style=none] (5) at (-2, -2.5) {};
                    \node [style=none] (6) at (0, 3) {};
                    \node [style=none] (8) at (-3, 1) {};
                    \node [style=none] (14) at (3, 1) {};
                    \node [style=White Dot] (15) at (0, 0) {$w_+$};
                    \node [style=none] (16) at (2, -0.5) {$+$};
                    \node [style=none] (17) at (1.25, 1.75) {$+$};
                    \node [style=none] (18) at (-1.25, 1.75) {$+$};
                    \node [style=none] (19) at (-2, -0.5) {$+$};
                    \node [style=none] (20) at (0, -1.75) {$\dots$};
                    
                    \draw [style=MisalignedU] (15) to (14.center);
                    \draw [style=MisalignedU] (15) to (6.center);
                    \draw [style=MisalignedU] (15) to (8.center);
                    \draw [style=MisalignedU] (15) to (5.center);
                    \draw [style=MisalignedU] (15) to (1.center);
                \end{tikzpicture}
                \caption{A vertex dressing a ramification point \\ with index $w$ touches $w$ faces of $\mathcal{D}[\Gamma]$.}
                \label{fig:bulkPRR_vertices_bulk_rammification}
            \end{subfigure}%
            \begin{subfigure}{.5\textwidth}
                \centering     
                \begin{tikzpicture}
                    \node [style=none] (6) at (0, 3) {};
                    \node [style=none] (8) at (-3, -2) {};
                    \node [style=none] (14) at (3, -2) {};
                    \node [style=none] (17) at (1.5, 0.5) {$+$};
                    \node [style=none] (18) at (-1.5, 0.5) {$+$};
                    \node [style=none] (19) at (0, -1.5) {$-$};
                    \node [style=none] (20) at (0, 0) {};
                    
                    \draw [style=MisalignedU] (6.center) to (20.center);
                    \draw [style=Aligned] (20.center) to (14.center);
                    \draw [style=Aligned] (8.center) to (20.center);
                \end{tikzpicture}
                \caption{Redundant vertex created when the contour defining $\mathcal{D}[\Gamma]$ folds away from the equator. }
                \label{fig:bulkPRR_vertices_bulk_splitting}
            \end{subfigure}
            \caption{This figure shows half of the vertices of the PRR diagrams. Two more vertices can be generated by turning around all orientations and exchanging $+ \leftrightarrow -$.}\label{fig:PRR_vertices}
        \end{figure}
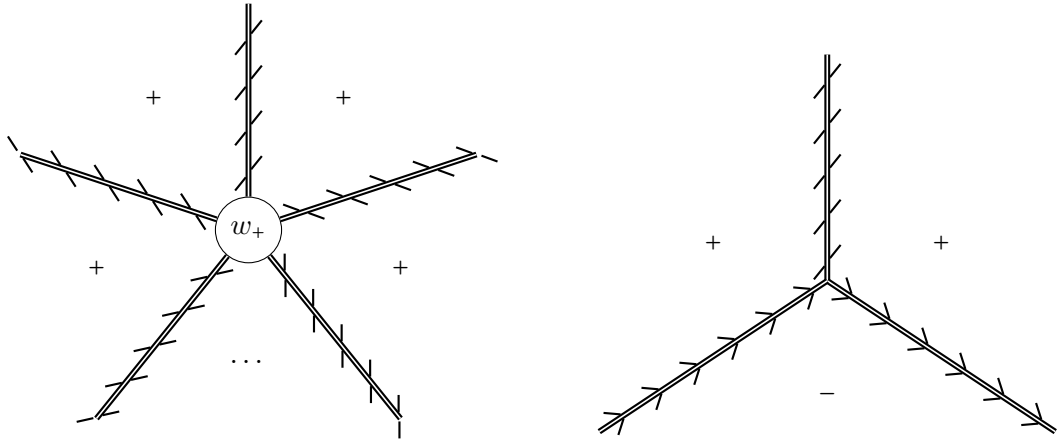

        \paragraph{Covering maps from diagrams.}
        The PRR diagrams for covering maps $\Gamma$ of degree $N$ can be constructed by using the vertices in Fig.~\ref{fig:PRR_fused_vertex}, as well as those with 
        flipped signs of the faces, choosing an order of the branch points and imposing the following three rules.
        \begin{enumerate}
            \item The boundary of a face contains each ramification point at most once. 
            \item Traversing the boundary of a face along its orientation, one encounters the ramification points  in the order of the corresponding branch points.
            \item Each face is simply connected and the number of faces equals the degree $N$ of the covering map.
        \end{enumerate}
        Clearly, the diagrams that classify equivalence classes of covering maps introduced in the previous paragraph satisfy the three rules above. In Appendix \ref{app:proof_rules}, we show that vice versa every diagram constructed in this way corresponds to an equivalence class of covering maps.
        
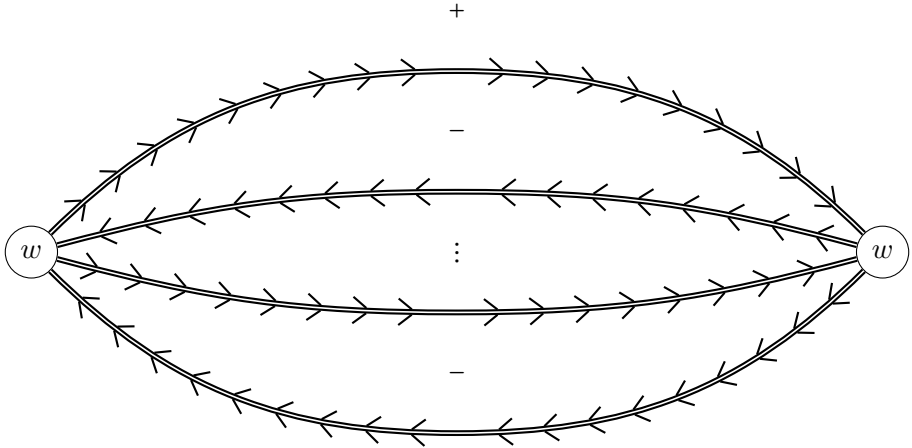
\begin{figure}
            \centering
            \begin{tikzpicture}[scale=0.8]
    		\node [style=none] (8) at (0, 3) {};
    		\node [style=none] (15) at (0, 1) {};
    		\node [style=none] (17) at (0, -3) {};
    		\node [style=none] (18) at (0, -1) {};
    		\node [style=White Dot] (21) at (7, 0) {$w$};
    		\node [style=White Dot] (22) at (-7, 0) {$w$};
    		\node [style=none] (23) at (0, 0) {$\vdots$};
    		\node [style=none] (24) at (0, 2) {$-$};
    		\node [style=none] (25) at (0, -2) {$-$};
    		\node [style=none] (26) at (0, 4) {$+$};
    
    		\draw [style=Aligned, in=-180, out=45] (22) to (8.center);
    		\draw [style=Aligned, in=180, out=-15] (22) to (18.center);
    		\draw [style=Aligned, in=15, out=-180] (15.center) to (22);
    		\draw [style=Aligned, in=-45, out=180] (17.center) to (22);
    		\draw [style=Aligned, in=135, out=0] (8.center) to (21);
    		\draw [style=Aligned, in=0, out=165] (21) to (15.center);
    		\draw [style=Aligned, in=195, out=0] (18.center) to (21);
    		\draw [style=Aligned, in=0, out=-135] (21) to (17.center);
            \end{tikzpicture}
            \caption{Diagram corresponding to the covering map that computes the two-point function of twist $w$ fields.}
            \label{fig:bulk_two_point}
        \end{figure}

        \paragraph{Example: Two-point correlator.}
        As a simple example, Figure \ref{fig:bulk_two_point} shows the covering map $\Gamma$ which computes the two-point function $\langle \sigma_w \sigma_w \rangle$ of two twist $w$ operators. 
        This map is a covering of the sphere by the sphere with two ramification points, both of them having ramification index $w$. 
        Thus, the diagram is a graph embedded in the sphere with two vertices that are connected by double lines of alternating orientation. 
        The graph has $2w$ faces, half of which with a ``$-$'' and half with a ``$+$'' label.

    \subsection{Interface PRR diagrams}\label{sec:Interface_PRR}

    \begin{figure}
        \centering
        \begin{tikzpicture}
    	\node [style=none] (0) at (-6, 0.5) {};
		\node [style=none] (1) at (6, 0.5) {};
		\node [style=none] (6) at (-4.5, 0.5) {};
		\node [style=none] (7) at (-6, -0.5) {};
		\node [style=none] (10) at (-4.5, 4.25) {};
		\node [style=none] (12) at (-3.5, 4.25) {};
		\node [style=none] (13) at (-4, 4.75) {};
		\node [style=none] (16) at (-2.5, -0.5) {};
		\node [style=none] (17) at (-3.5, 0.5) {};
		\node [style=none] (20) at (-0.75, 0.5) {};
		\node [style=none] (22) at (-0.75, 5) {};
		\node [style=none] (24) at (0.25, 5) {};
		\node [style=none] (25) at (-0.25, 5.5) {};
		\node [style=none] (28) at (6, -0.5) {};
		\node [style=none] (29) at (0.25, 0.5) {};
		\node [style=none] (32) at (1.25, 0.5) {};
		\node [style=none] (34) at (1.25, 2.75) {};
		\node [style=none] (36) at (2.25, 2.75) {};
		\node [style=none] (37) at (1.75, 3.25) {};
		\node [style=none] (41) at (2.25, 0.5) {};
		\node [style=none] (44) at (-0.75, 0.5) {};
		\node [style=none] (45) at (-1.5, -0.5) {};
		\node [style=none] (47) at (-2.5, -2.5) {};
		\node [style=none] (48) at (-1.5, -2.5) {};
		\node [style=none] (49) at (-2, -3) {};
		\node [style=none] (52) at (-2.5, -0.5) {};
		\node [style=White Dot] (56) at (-4, 4.25) {$w_+$};
		\node [style=White Dot] (57) at (-0.25, 5) {$w_+$};
		\node [style=White Dot] (58) at (1.75, 2.75) {$w_+$};
		\node [style=White Dot] (59) at (-2, -2.5) {$w_-$};
		\node [style=none] (60) at (6, 0) {};
		\node [style=none] (61) at (-6, 0) {};
		\node [style=none] (62) at (4, 0) {$\omega^{AB}$};
		\node [style=none] (63) at (4.5, 0) {};
		\node [style=none] (64) at (3.5, 0) {};
		\draw [style=SingleU] (0.center) to (6.center);
		\draw [style=SingleU, bend left=45] (10.center) to (13.center);
		\draw [style=SingleU, bend left=45, looseness=1.25] (13.center) to (12.center);
		\draw [style=SingleU] (12.center) to (17.center);
		\draw [style=SingleD] (7.center) to (16.center);
		\draw [style=SingleU, bend left=45] (22.center) to (25.center);
		\draw [style=SingleU, bend left=45, looseness=1.25] (25.center) to (24.center);
		\draw [style=SingleU] (24.center) to (29.center);
		\draw [style=SingleU, bend left=45] (34.center) to (37.center);
		\draw [style=SingleU, bend left=45, looseness=1.25] (37.center) to (36.center);
		\draw [style=SingleU] (36.center) to (41.center);
		\draw [style=SingleU] (6.center) to (10.center);
		\draw [style=SingleU] (20.center) to (22.center);
		\draw [style=SingleU] (32.center) to (34.center);
		\draw [style=SingleU] (29.center) to (32.center);
		\draw [style=SingleU] (41.center) to (1.center);
		\draw [style=SingleD] (52.center) to (47.center);
		\draw [style=SingleD, in=-180, out=-90] (47.center) to (49.center);
		\draw [style=SingleD, in=-90, out=0, looseness=1.25] (49.center) to (48.center);
		\draw [style=SingleD] (48.center) to (45.center);
		\draw [style=SingleD] (45.center) to (28.center);
		\draw [style=SingleU] (17.center) to (44.center);
		\draw [style=Dashed] (61.center) to (64.center);
		\draw [style=Dashed] (63.center) to (60.center);
        \end{tikzpicture}
        \caption{When the interface is inserted along the equator of the sphere, the contours defining $\mathcal{D}_\varepsilon[\Gamma]$ are identical to those shown in Fig.~\ref{fig:illustration_Gamma_to_D}.
        }
        \label{fig:interface_Gamma_to_D}
    \end{figure}
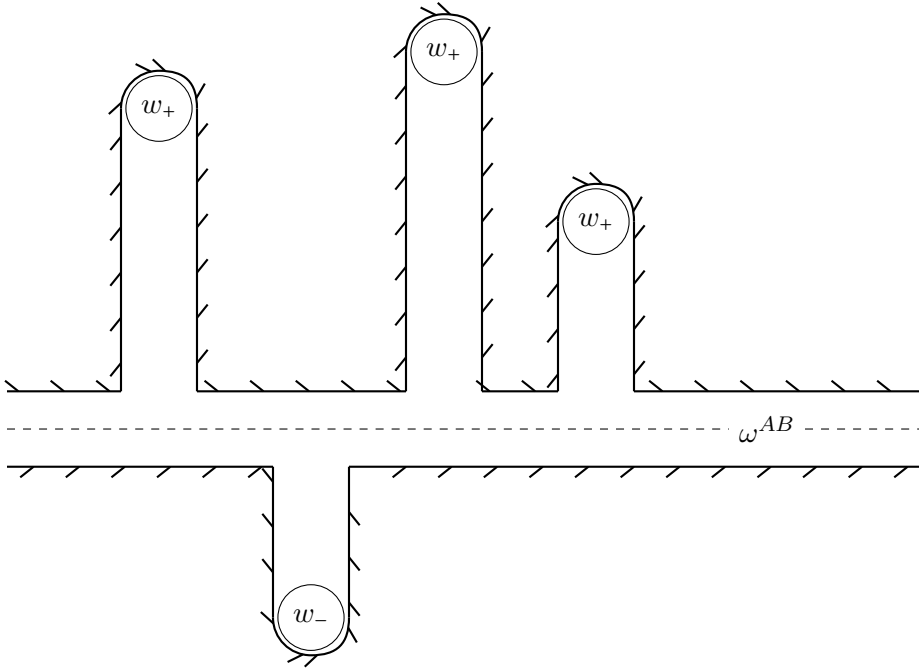

        Following the same strategy as in Section \ref{sec:PRR_diagrams}, we first construct diagrams from interface covering maps. 
        We then present a set of rules that allow us to systematically construct all possible diagrams given a set of prescribed ramification points.
        Finally, we demonstrate the rules in several examples.
        A proof for the validity of the rules is sketched in App.~\ref{app:proof_rules}.
        
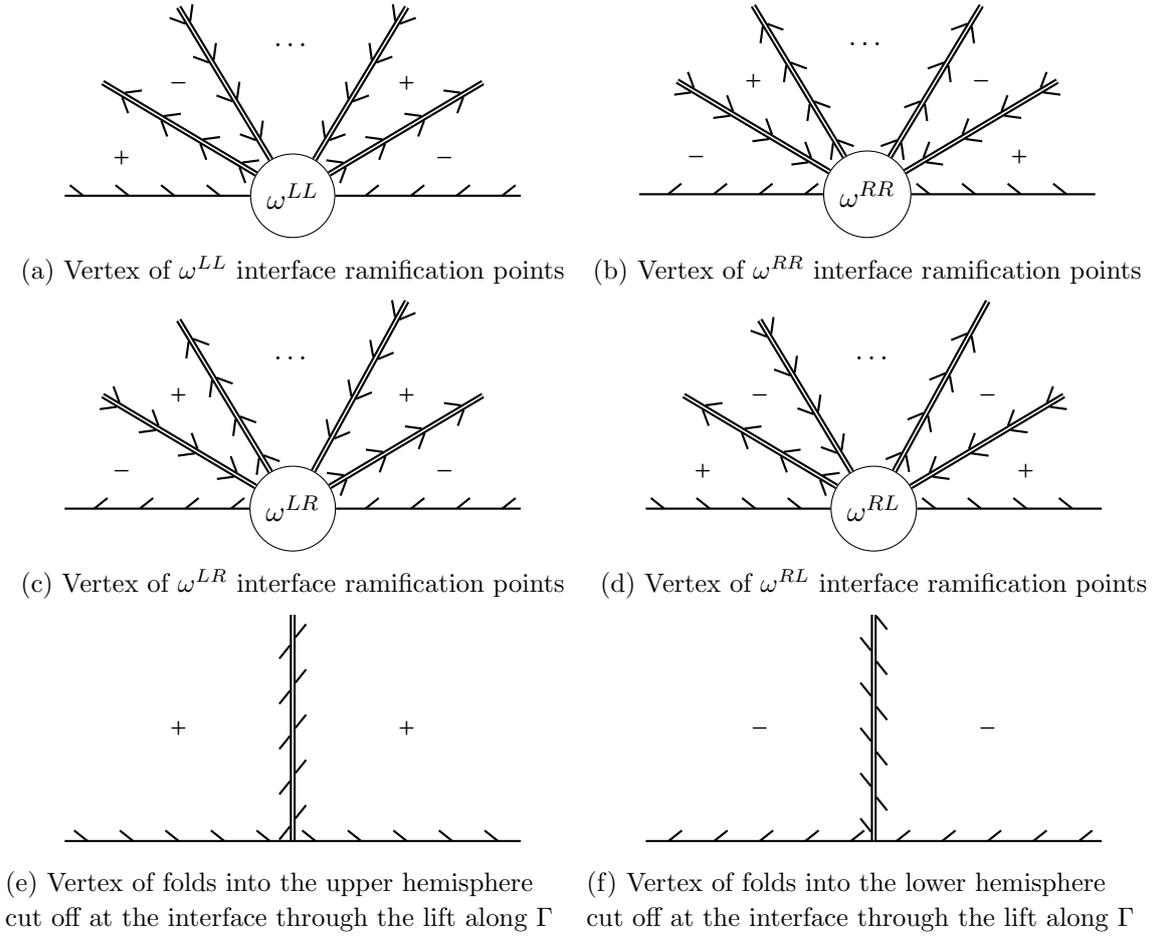
\begin{figure}
            \begin{subfigure}{.5\textwidth}
                \centering
                \begin{tikzpicture}
                    \node [style=none] (8) at (-3, 0) {};
                    \node [style=none] (14) at (3, 0) {};
                    \node [style=White Dot] (15) at (0, 0) {$\omega^{LL}$};
                    \node [style=none] (16) at (-2.5, 1.5) {};
                    \node [style=none] (17) at (-1.5, 2.5) {};
                    \node [style=none] (18) at (1.5, 2.5) {};
                    \node [style=none] (19) at (2.5, 1.5) {};
                    \node [style=none] (20) at (0, 2) {$\dots$};
                    \node [style=none] (21) at (-2.25, 0.5) {$+$};
                    \node [style=none] (22) at (2, 0.5) {};
                    \node [style=none] (23) at (2, 0.5) {$-$};
                    \node [style=none] (24) at (1.5, 1.5) {};
                    \node [style=none] (25) at (1.5, 1.5) {$+$};
                    \node [style=none] (26) at (-1.5, 1.5) {};
                    \node [style=none] (27) at (-1.5, 1.5) {$-$};
                    
                    \draw [style=SingleU] (8.center) to (15);
                    \draw [style=SingleD] (14.center) to (15);
                    \draw [style=Aligned] (15) to (16.center);
                    \draw [style=Aligned] (17.center) to (15);
                    \draw [style=Aligned] (15) to (19.center);
                    \draw [style=Aligned] (18.center) to (15);
                \end{tikzpicture}
                \caption{Vertex of $\omega^{LL}$ interface ramification points}
            \end{subfigure}%
            \begin{subfigure}{.5\textwidth}
                \centering     
                \begin{tikzpicture}
                    \node [style=none] (8) at (-3, 0) {};
                    \node [style=none] (14) at (3, 0) {};
                    \node [style=White Dot] (15) at (0, 0) {$\omega^{RR}$};
                    \node [style=none] (16) at (1.5, 2.5) {};
                    \node [style=none] (17) at (-2.5, 1.5) {};
                    \node [style=none] (18) at (2.5, 1.5) {};
                    \node [style=none] (19) at (-1.5, 2.5) {};
                    \node [style=none] (20) at (0, 2) {$\dots$};
                    \node [style=none] (21) at (-2.25, 0.5) {$-$};
                    \node [style=none] (22) at (2, 0.5) {};
                    \node [style=none] (23) at (2, 0.5) {$+$};
                    \node [style=none] (24) at (1.5, 1.5) {};
                    \node [style=none] (25) at (1.5, 1.5) {$-$};
                    \node [style=none] (26) at (-1.5, 1.5) {};
                    \node [style=none] (27) at (-1.5, 1.5) {$+$};
                
                    \draw [style=Aligned] (15) to (16.center);
                    \draw [style=Aligned] (17.center) to (15);
                    \draw [style=Aligned] (15) to (19.center);
                    \draw [style=Aligned] (18.center) to (15);
                    \draw [style=SingleU] (15) to (14.center);
                    \draw [style=SingleD] (15) to (8.center);
                \end{tikzpicture}
                \caption{Vertex of $\omega^{RR}$ interface ramification points}
            \end{subfigure}
            \begin{subfigure}{.5\textwidth}
                \centering     
                \begin{tikzpicture}
                    \node [style=none] (8) at (-3, 0) {};
                    \node [style=none] (14) at (3, 0) {};
                    \node [style=White Dot] (15) at (0, 0) {$\omega^{LR}$};
                    \node [style=none] (16) at (-2.5, 1.5) {};
                    \node [style=none] (17) at (-1.5, 2.5) {};
                    \node [style=none] (18) at (2.5, 1.5) {};
                    \node [style=none] (19) at (1.5, 2.75) {};
                    \node [style=none] (20) at (0, 2) {$\dots$};
                    \node [style=none] (21) at (-2.25, 0.5) {$-$};
                    \node [style=none] (22) at (2, 0.5) {};
                    \node [style=none] (23) at (2, 0.5) {$-$};
                    \node [style=none] (24) at (1.5, 1.5) {};
                    \node [style=none] (25) at (1.5, 1.5) {$+$};
                    \node [style=none] (26) at (-1.5, 1.5) {};
                    \node [style=none] (27) at (-1.5, 1.5) {$+$};
                
                    \draw [style=Aligned] (16.center) to (15);
                    \draw [style=Aligned] (15) to (17.center);
                    \draw [style=Aligned] (19.center) to (15);
                    \draw [style=Aligned] (15) to (18.center);
                    \draw [style=SingleD] (15) to (8.center);
                    \draw [style=SingleD] (14.center) to (15);
                \end{tikzpicture}
                \caption{Vertex of $\omega^{LR}$ interface ramification points}
            \end{subfigure}
            \begin{subfigure}{.5\textwidth}
                \centering     
                \begin{tikzpicture}
                    \node [style=none] (8) at (-3, 0) {};
                    \node [style=none] (14) at (3, 0) {};
                    \node [style=White Dot] (15) at (0, 0) {$\omega^{RL}$};
                    \node [style=none] (16) at (-2.5, 1.5) {};
                    \node [style=none] (17) at (-1.5, 2.5) {};
                    \node [style=none] (18) at (2.5, 1.5) {};
                    \node [style=none] (19) at (1.5, 2.75) {};
                    \node [style=none] (20) at (0, 2) {$\dots$};
                    \node [style=none] (21) at (-2.25, 0.5) {$+$};
                    \node [style=none] (22) at (2, 0.5) {};
                    \node [style=none] (23) at (2, 0.5) {$+$};
                    \node [style=none] (24) at (1.5, 1.5) {};
                    \node [style=none] (25) at (1.5, 1.5) {$-$};
                    \node [style=none] (26) at (-1.5, 1.5) {};
                    \node [style=none] (27) at (-1.5, 1.5) {$-$};
                
                    \draw [style=Aligned] (15) to (16.center);
                    \draw [style=Aligned] (17.center) to (15);
                    \draw [style=Aligned] (15) to (19.center);
                    \draw [style=Aligned] (18.center) to (15);
                    \draw [style=SingleU] (8.center) to (15);
                    \draw [style=SingleU] (15) to (14.center);
                \end{tikzpicture}
                \caption{Vertex of $\omega^{RL}$ interface ramification points}
            \end{subfigure}
            \begin{subfigure}{.5\textwidth}
                \centering     
                \begin{tikzpicture}
                    \node [style=none] (6) at (0, 3) {};
                    \node [style=none] (8) at (-3, 0) {};
                    \node [style=none] (14) at (3, 0) {};
                    \node [style=none] (17) at (1.5, 1.5) {$+$};
                    \node [style=none] (18) at (-1.5, 1.5) {$+$};
                    \node [style=none] (20) at (0, 0) {};
                
                    \draw [style=MisalignedU] (6.center) to (20.center);
                    \draw [style=SingleU] (8.center) to (20.center);
                    \draw [style=SingleU] (20.center) to (14.center);
                \end{tikzpicture}
                \caption{Vertex of folds into the upper hemisphere \\ cut off at the interface through the lift along $\Gamma$}
                \label{fig:misaligned_split_vertex+}
            \end{subfigure}
            \begin{subfigure}{.5\textwidth}
                \centering     
                \begin{tikzpicture}
                    \node [style=none] (6) at (0, 3) {};
                    \node [style=none] (8) at (-3, 0) {};
                    \node [style=none] (14) at (3, 0) {};
                    \node [style=none] (17) at (1.5, 1.5) {$-$};
                    \node [style=none] (18) at (-1.5, 1.5) {$-$};
                    \node [style=none] (20) at (0, 0) {};
                
                    \draw [style=MisalignedD] (20.center) to (6.center);
                    \draw [style=SingleD] (20.center) to (8.center);
                    \draw [style=SingleD] (14.center) to (20.center);
                \end{tikzpicture}
                \caption{Vertex of folds into the lower hemisphere \\ cut off at the interface through the lift along $\Gamma$}
                \label{fig:misaligned_split_vertex-}
            \end{subfigure}
            \caption{Additional vertices for interface covering maps.}
            \label{fig:IPRR_vertices}
        \end{figure}

        \paragraph{Diagrams from interface covering maps.}
        The following three steps construct a diagram\footnote{In mathematical terms the object that we construct is a combinatorial map i.e.~a graph with additional data describing the cyclic configuration of edges at each vertex.} $\mathcal{D}[\Gamma]$ on a Riemann surface $\Sigma$ given an interface covering map $\Gamma: \Sigma \rightarrow \mathbb{S}^2$.
        \begin{enumerate}
            \item Draw lines with the same orientation, at distance $\varepsilon$ above and below the interface, deformed as in Fig.~\ref{fig:interface_Gamma_to_D} such that all branch points of $\Gamma$ are in the strip between them.
            \item The preimage $\mathcal{D}_\varepsilon[\Gamma]$ of this configuration with respect to $\Gamma$ is a network of oriented lines on $\Sigma$. Take the limit  $\mathcal{D}[\Gamma] := \lim\limits_{\varepsilon \rightarrow 0}\mathcal{D}_\varepsilon[\Gamma]$.
            \item Each face of $\mathcal{D}[\Gamma]$ is either the preimage of the upper or the lower hemisphere with respect to $\Gamma$. 
            Label the faces with ``$+$'' and ``$-$'' respectively.
        \end{enumerate}
        In the interface case, the vertices of Fig.~\ref{fig:PRR_vertices} are augmented by the vertices of Fig.~\ref{fig:IPRR_vertices}. The first four vertices originate from boundary ramification points of the interface covering map. They involve two oriented (single) lines and $\omega-1$ aligned double lines. Since all the double lines are aligned, the label on the faces alternates. 
        There are four different assignments of signs in the two faces that involve the single line. These assignments distinguish the four vertices corresponding to the four classes of boundary ramification points. 
        Additionally, 
        there are two vertices in which a misaligned double line splits into a pair of single 
        lines. 
        The two variants of these vertices are distinguished by the sign of the two faces touching the misaligned line, which can be either 
        both $+$ or both $-$. 
        
        The trivalent vertices that we encountered in the absence of the interface were redundant since 
        there was a face with the opposite sign that we could use to split the misaligned double 
        line into a pair of aligned ones. Now, however, the interface can cause the surface to rip open. 
        When this happens, there is no face with opposite sign on the other side. Therefore, these 
        vertices cannot be removed. 
        
        As in the case without interface, the two halves of each double 
        line originating in bulk ramification points have opposite orientation, i.e.~these double 
        lines are misaligned. Again, we can remove all `redundant' trivalent vertices where 
        misaligned double lines split into a pair of aligned double lines by moving them on top of
        a bulk ramification point. In this process we can remove some of the misaligned double 
        lines. But if there is a non-trivial interface, it is no longer possible to remove all 
        the misaligned double lines through this process. Hence, the interface diagrams do involve 
        three different types of double lines in general, the aligned ones that also appear in 
        the bulk PRR diagrams and the two types of misaligned ones that run between faces 
        labelled by $+$ and by $-$. 
        
        The bulk vertices involve both aligned and misaligned 
        double lines. Their associated ramification index $w$ fixes the overall number of these double lines but not how they are distributed among the two types. More 
        precisely, $2w$ is the number of aligned double lines plus twice the number of misaligned
        ones.

        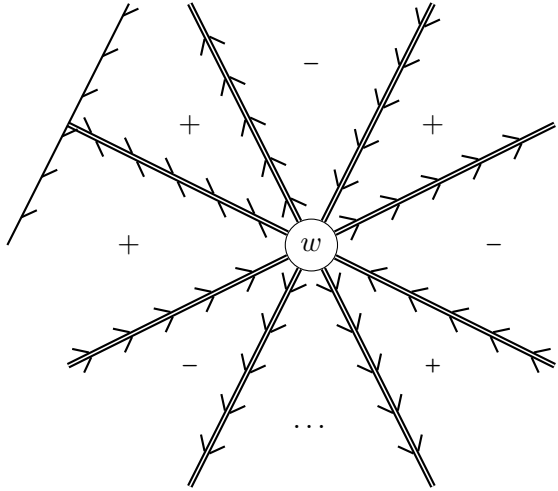
\begin{figure}
            \centering     
            \begin{tikzpicture}[scale=0.8]
                \node [style=none] (16) at (3, 0) {$-$};
                \node [style=none] (17) at (2, 2) {+};
                \node [style=none] (18) at (-2, 2) {+};
                \node [style=none] (19) at (-3, 0) {+};
                \node [style=none] (20) at (0, -3) {$\dots$};
                \node [style=none] (21) at (-2, 4) {};
                \node [style=none] (22) at (2, 4) {};
                \node [style=none] (25) at (-4, -2) {};
                \node [style=none] (26) at (-2, -4) {};
                \node [style=none] (27) at (2, -4) {};
                \node [style=none] (28) at (4, -2) {};
                \node [style=none] (29) at (4, 2) {};
                \node [style=none] (31) at (0, 3) {$-$};
                \node [style=none] (33) at (-2, -2) {$-$};
                \node [style=none] (34) at (2, -2) {$+$};
                \node [style=White Dot] (36) at (0, 0) {$w$};
                \node [style=none] (37) at (-4, 2) {};
                \node [style=none] (38) at (-5, 0) {};
                \node [style=none] (39) at (-3, 4) {};
                \draw [style=Aligned] (36) to (21.center);
                \draw [style=Aligned] (22.center) to (36);
                \draw [style=MisalignedU] (37.center) to (36);
                \draw [style=SingleU] (39.center) to (37.center);
                \draw [style=SingleU] (37.center) to (38.center);
                \draw [style=Aligned] (36) to (29.center);
                \draw [style=Aligned] (28.center) to (36);
                \draw [style=Aligned] (36) to (27.center);
                \draw [style=Aligned] (25.center) to (36);
                \draw [style=Aligned] (36) to (26.center);
            \end{tikzpicture}
            \caption{Bulk vertex in a reduced diagram obtained by removing redundant vertices. While this picture shows only a single misaligned line emanating from the bulk vertex to the boundary of $\Sigma$, there could in general be up to $w$ such lines attached to the same vertex.}
            \label{fig:reduced_bulk_vertex_i}
        \end{figure}
        
        \paragraph{Covering maps from diagrams.}
    
        The interface PRR diagrams of interface covering maps $\Gamma$ of degrees $N_\pm$ equipped with a consistent choice of order for the branch points 
        are the diagrams built out of the vertices in Figures \ref{fig:PRR_vertices} and 
        \ref{fig:IPRR_vertices}, satisfying the three rules given below.
        In this context, ``consistent choice of order'' means that the restriction of the ordering to interface branch points coincides with the order induced by the orientation of the interface. 
        \begin{enumerate}
            \item The boundary of a face contains each ramification point at most once. 
            \item Traversing the boundary of a face along its orientation, one encounters 
            the ramification points in the order of the corresponding branch points.
            \item Each face is simply connected and the number of ``$\pm$''-faces 
            is $N_\pm$.
        \end{enumerate}
     
        \noindent Clearly, the diagrams that classify equivalence classes of covering maps 
        introduced in the previous paragraph satisfy the three rules above. In Appendix 
        \ref{app:proof_rules} we show that, conversely, all interface PRR diagrams actually 
        correspond to interface covering maps. 
        
        \begin{figure}
            \begin{subfigure}{.5\textwidth}
                \centering
                 \begin{tikzpicture}
        		\node [style=none] (24) at (0, 3) {};
        		\node [style=none] (25) at (0, -3) {};
        		\node [style=White Dot] (26) at (0, 0) {$2^+$};
        		\node [style=none] (29) at (-1.25, 1.25) {$+$};
        		\node [style=none] (30) at (1.25, 1.25) {$+$};
        		\node [style=none] (31) at (0, -1.5) {$-$};
        		\node [style=White Dot] (32) at (3, 0) {$2^{RR}$};
        		\node [style=White Dot] (33) at (-3, 0) {$2^{LL}$};
        
        		\draw [style=Aligned] (33) to (26);
        		\draw [style=Aligned] (26) to (32);
        		\draw [style=SingleD, bend left=45] (32) to (25.center);
        		\draw [style=SingleD, bend left=45] (25.center) to (33);
        		\draw [style=MisalignedU] (26) to (24.center);
        		\draw [style=SingleU, bend right=45] (32) to (24.center);
        		\draw [style=SingleU, bend right=45] (24.center) to (33);
                \end{tikzpicture}

                \caption{The case $\text{Im}(z) > 0$.}
            \end{subfigure}%
            \begin{subfigure}{.5\textwidth}
                \centering     
                \begin{tikzpicture}
        		\node [style=none] (24) at (0, 3) {};
        		\node [style=none] (25) at (0, -3) {};
        		\node [style=White Dot] (26) at (0, 0) {$2^-$};
        		\node [style=none] (29) at (1.25, -1.5) {$-$};
        		\node [style=none] (30) at (0, 1.5) {$+$};
        		\node [style=none] (31) at (-1.25, -1.5) {$-$};
        		\node [style=White Dot] (32) at (3, 0) {$2^{RR}$};
        		\node [style=White Dot] (33) at (-3, 0) {$2^{LL}$};
        
        		\draw [style=Aligned] (33) to (26);
        		\draw [style=Aligned] (26) to (32);
        		\draw [style=SingleD, bend left=45] (32) to (25.center);
        		\draw [style=SingleD, bend left=45] (25.center) to (33);
        		\draw [style=SingleU, bend right=45] (32) to (24.center);
        		\draw [style=SingleU, bend right=45] (24.center) to (33);
        		\draw [style=MisalignedD] (26) to (25.center);
                \end{tikzpicture}
                \caption{The case $\text{Im}(z) < 0$.}
            \end{subfigure}
            \caption{Diagram of $\langle \sigma_2^{RR} \sigma_2(z) \sigma_2^{LL}\rangle$ with order $2^{LL} < 2 < 2^{RR}$.}
            \label{fig:LL2RR}
        \end{figure}
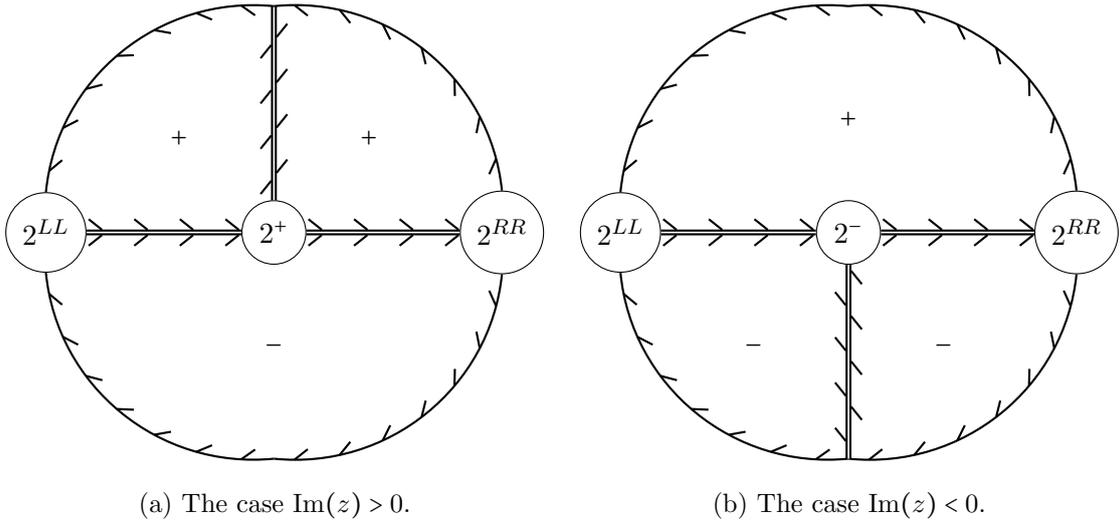

        \begin{figure}
            \begin{subfigure}{.5\textwidth}
                \centering
                \begin{tikzpicture}
        		\node [style=White Dot] (27) at (-3, 0) {$2^{LL}$};
        		\node [style=White Dot] (28) at (3, 0) {$2^{RR}$};
        		\node [style=none] (29) at (-2.25, 2) {};
        		\node [style=none] (30) at (2.25, 2) {};
        		\node [style=White Dot] (31) at (0, 1.5) {$2^+$};
        		\node [style=none] (32) at (0, 2.25) {$+$};
        		\node [style=none] (33) at (0, 0.5) {$+$};
        		\node [style=none] (34) at (0, -1.5) {$-$};

        		\draw [style=Aligned, bend right=15] (27) to (28);
        		\draw [style=SingleU, in=90, out=-135, looseness=0.75] (29.center) to (27);
        		\draw [style=SingleD, bend left=90, looseness=1.50] (28) to (27);
        		\draw [style=MisalignedU, bend left=15, looseness=0.75] (30.center) to (31);
        		\draw [style=MisalignedU, bend right=345, looseness=0.75] (31) to (29.center);
        		\draw [style=SingleU, in=-45, out=90, looseness=0.75] (28) to (30.center);
        		\draw [style=SingleU, bend right=45] (30.center) to (29.center);
                \end{tikzpicture}
                \vspace{-1 cm}
                \caption{The case $\text{Im}(z) > 0$.}
            \end{subfigure}
            \begin{subfigure}{.5\textwidth}
                \centering     
                \raisebox{0.8 cm}{
                \begin{tikzpicture}
                    \node [style=White Dot] (27) at (-3, 0) {$2^{LL}$};
                    \node [style=White Dot] (28) at (3, 0) {$2^{RR}$};
                    \node [style=none] (29) at (-2.25, -2) {};
                    \node [style=none] (30) at (2.25, -2) {};
                    \node [style=White Dot] (31) at (0, -1.5) {$2^-$};
                    \node [style=none] (32) at (0, 1) {$+$};
                    \node [style=none] (33) at (0, -0.25) {$-$};
                    \node [style=none] (34) at (0, -2.25) {$-$};
            
                    \draw [style=Aligned, bend left=15] (27) to (28);
                    \draw [style=SingleD, in=-90, out=135, looseness=0.75] (29.center) to (27);
                    \draw [style=SingleU, bend left=270, looseness=1.50] (28) to (27);
                    \draw [style=MisalignedD, bend right=15, looseness=0.75] (30.center) to (31);
                    \draw [style=MisalignedD, bend right=15, looseness=0.75] (31) to (29.center);
                    \draw [style=SingleD, in=45, out=-90, looseness=0.75] (28) to (30.center);
                    \draw [style=SingleD, bend left=45] (30.center) to (29.center);
                \end{tikzpicture}
                }
                \vspace{-1 cm}
                \caption{The case $\text{Im}(z) < 0$.}
            \end{subfigure}
            \caption{Diagram of $\langle \sigma_2^{RR} \sigma_2(z) \sigma_2^{LL}\rangle$ with order $2^{LL} < 2^{RR} < 2 $.}
            \label{fig:LLRR2}
        \end{figure}
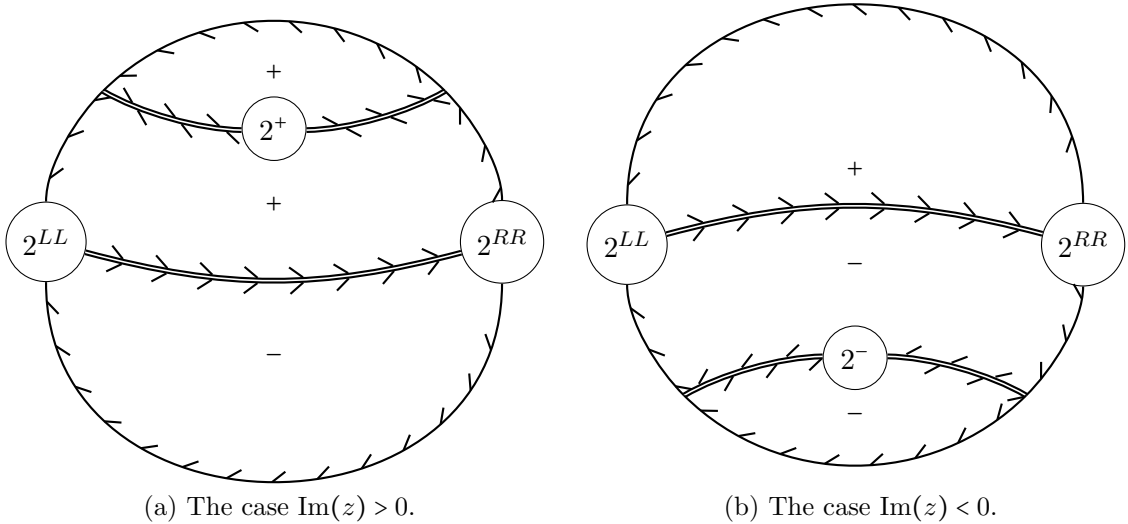
       
        \paragraph{Example 1.}
        As a first example, let us consider the diagrammatics for the interface covering 
        maps $\Gamma$ that can contribute to the correlation function 
        \begin{align}\label{eq:corr_RR2LL}
            \langle \sigma^{RR}_2~ \sigma_2(z) \sigma^{LL}_2\rangle.
        \end{align}
        The only connected covering space contributing to this correlator is a disk. If $z$ 
        is in the upper hemisphere, the corresponding interface covering map has $N_- = 1$ 
        and $N_+ = 2$. The assignment is reversed if $z$ is in the lower hemisphere, 
        i.e.~$N_- = 2$ and $N_+ = 1$ in that case. The two corresponding diagrams are shown in Figure 
        \ref{fig:LL2RR}. The correlator is a good example to highlight the relevance of the order that is assigned to the 
        bulk branch points through the contour in Figure \ref{fig:interface_Gamma_to_D}. 
        The diagrams shown in Figure \ref{fig:LL2RR} correspond to the 
        choice of ordering $2^{LL} < 2 < 2^{RR}$. The other option, i.e.~$2^{LL} < 2^{RR} < 2$, 
        is shown in Figure \ref{fig:LLRR2}. Let us stress that the diagrams arising from 
        different choices of ordering do not correspond to different types of covering 
        maps that contribute to the correlator \eqref{eq:corr_RR2LL}. Instead they are 
        different graphical representations of the same covering map. Finally, let us also 
        draw the attention of the reader to the subtle feature that performing a reflection 
        of the diagrams along the vertical axis that goes through the vertex corresponding 
        to the bulk field does \emph{not} result in valid diagrams. The vertices of valid 
        diagrams are precisely those listed in Figure \ref{fig:IPRR_vertices}. In particular, 
        a curve moving around a valid $\omega^{LL}$ vertex anticlockwise in a small semi-circle 
        has to start on the boundary of a ``$-$''-face and end on the boundary of a ``$+$''-face.
        The analogous statement with ``$+$'' and ``$-$'' exchanged holds for the $\omega^{RR}$ 
        vertices.

        \paragraph{Example 2.}
        Let us now discuss a less trivial example, where multiple diagrams contribute, namely the 
        four-point correlation function 
        \begin{align}\label{eq:RLRLRRLL}
            \langle \sigma_5^{RL}\sigma_5^{RL}\sigma_2^{RR}\sigma_2^{LL}\rangle
        \end{align}
        of interface operators. Whereas the ordering of bulk operators along the interface is an 
        arbitrary (but necessary) choice that does not have any direct physical meaning, the ordering 
        of interface operators is unambiguous (i.e.~different orderings correspond to physically 
        different correlation functions). With the ordering fixed in eq.~\eqref{eq:RLRLRRLL}, there 
        is a unique disk diagram corresponding to an interface covering map that contributes to the 
        correlator. This diagram is illustrated in Figure \ref{fig:RLRLRRLLDisk}. Furthermore, there 
        are two distinct cylinder diagrams that contribute to the connected part of the correlator.
        These are shown in Figure \ref{fig:RLRLRRLLCylinder}.
        
        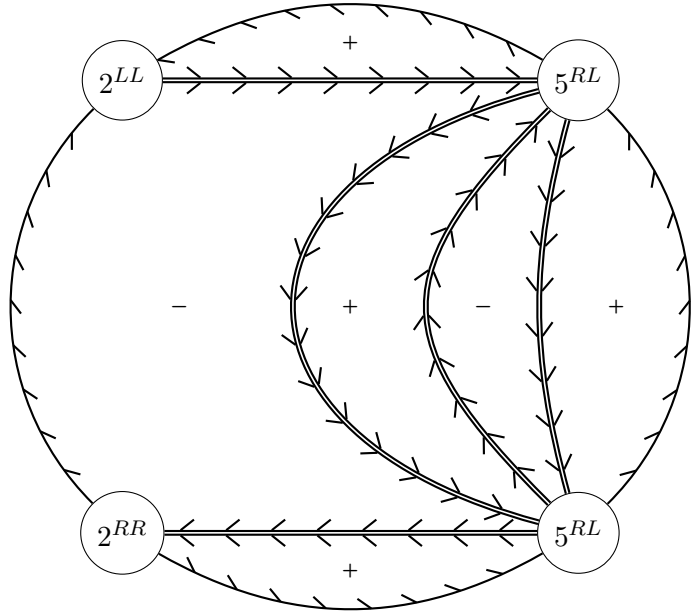
\begin{figure}
            \centering
            \begin{tikzpicture}
    		\node [style=White Dot] (0) at (-3, 3) {$2^{LL}$};
    		\node [style=White Dot] (1) at (-3, -3) {$2^{RR}$};
    		\node [style=White Dot] (2) at (3, -3) {$5^{RL}$};
    		\node [style=White Dot] (3) at (3, 3) {$5^{RL}$};
    		\node [style=none] (5) at (-2.25, 0) {$-$};
    		\node [style=none] (6) at (0, 0) {$+$};
    		\node [style=none] (7) at (1.75, 0) {$-$};
    		\node [style=none] (8) at (3.5, 0) {$+$};
    		\node [style=none] (9) at (0, 3.5) {$+$};
    		\node [style=none] (10) at (0, -3.5) {$+$};
    
    		\draw [style=SingleU, bend left=330] (3) to (0);
    		\draw [style=SingleD, bend left=45] (1) to (0);
    		\draw [style=SingleU, bend right] (1) to (2);
    		\draw [style=SingleU, bend right=45] (2) to (3);
    		\draw [style=Aligned] (2) to (1);
    		\draw [style=Aligned] (0) to (3);
    		\draw [style=Aligned, bend right=75, looseness=2.00] (3) to (2);
    		\draw [style=Aligned, bend left=45, looseness=1.50] (2) to (3);
    		\draw [style=Aligned, bend right=15] (3) to (2);
            \end{tikzpicture}
            \caption{Disk contribution to the correlator \eqref{eq:RLRLRRLL} at $N_- = 2$ and $N_+=4$.}
            \label{fig:RLRLRRLLDisk}
        \end{figure}
        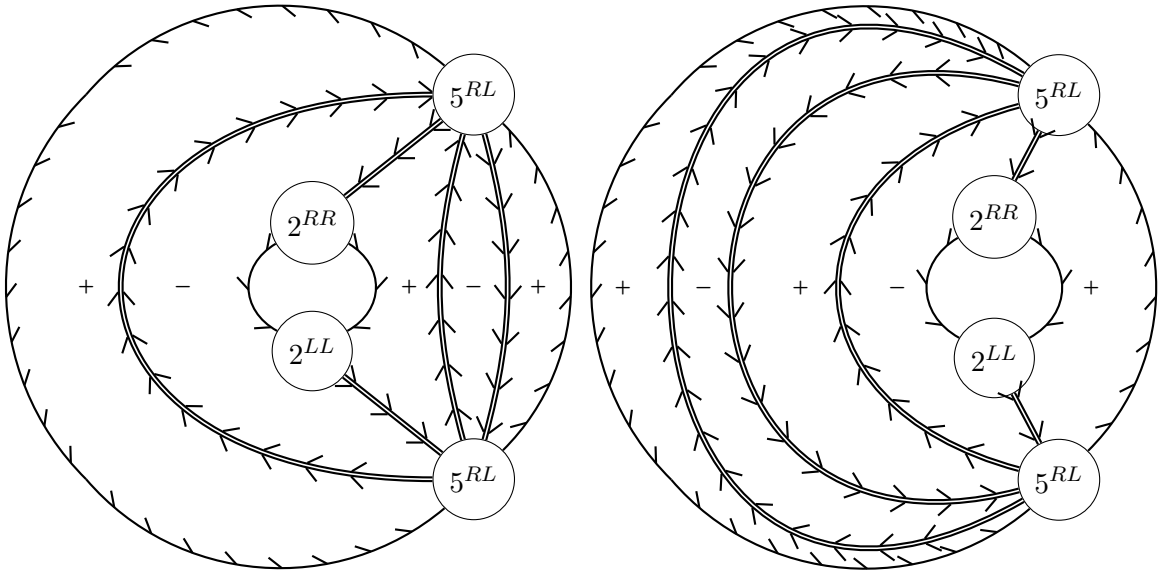
\begin{figure}
            \begin{subfigure}{.5\textwidth}
                \centering
                \raisebox{2.3 cm}{
                \begin{tikzpicture}[scale=0.85]
        		\node [style=White Dot] (0) at (0.5, -1) {$2^{LL}$};
        		\node [style=White Dot] (1) at (0.5, 1) {$2^{RR}$};
        		\node [style=White Dot] (2) at (3, 3) {$5^{RL}$};
        		\node [style=White Dot] (3) at (3, -3) {$5^{RL}$};
        		\node [style=none] (6) at (2, 0) {$+$};
        		\node [style=none] (7) at (3, 0) {$-$};
        		\node [style=none] (8) at (4, 0) {$+$};
        		\node [style=none] (9) at (-3, 3) {};
        		\node [style=none] (10) at (-3, -3) {};
        		\node [style=none] (13) at (-1.5, 0) {$-$};
        		\node [style=none] (14) at (-3, 0) {$+$};

        		\draw [style=SingleU, bend right=45] (10.center) to (3);
        		\draw [style=SingleU, bend right=45] (3) to (2);
        		\draw [style=SingleU, bend left=315] (2) to (9.center);
        		\draw [style=SingleU, bend right=45] (9.center) to (10.center);
        		\draw [style=Aligned, bend left=15] (2) to (3);
        		\draw [style=Aligned, bend left=15] (3) to (2);
        		\draw [style=Aligned] (2) to (1);
        		\draw [style=Aligned] (0) to (3);
        		\draw [style=Aligned, bend left=90, looseness=2.75] (3) to (2);
        		\draw [style=SingleU, bend left=60, looseness=1.25] (1) to (0);
        		\draw [style=SingleD, bend right=60, looseness=1.25] (1) to (0);
                \end{tikzpicture}
                }
                \vspace{-3 cm}
                \caption{Diagram with separated boundaries.}
            \end{subfigure}
            \begin{subfigure}{.5\textwidth}
                \centering     
                \begin{tikzpicture}[scale=0.85]
                    \node [style=White Dot] (0) at (2, -1.1) {$2^{LL}$};
                    \node [style=White Dot] (1) at (2, 1.1) {$2^{RR}$};
                    \node [style=White Dot] (2) at (3, 3) {$5^{RL}$};
                    \node [style=White Dot] (3) at (3, -3) {$5^{RL}$};
                    \node [style=none] (6) at (-1, 0) {$+$};
                    \node [style=none] (7) at (0.5, 0) {$-$};
                    \node [style=none] (8) at (3.5, 0) {$+$};
                    \node [style=none] (9) at (-3, 3) {};
                    \node [style=none] (10) at (-3, -3) {};
                    \node [style=none] (13) at (-2.5, 0) {$-$};
                    \node [style=none] (14) at (-3.75, 0) {$+$};
                    
                    \draw [style=SingleU, bend right=45] (3) to (2);
                    \draw [style=Aligned, bend left=75, looseness=1.75] (3) to (2);
                    \draw [style=Aligned, bend right=105, looseness=2.50] (2) to (3);
                    \draw [style=Aligned, bend right=240, looseness=3.25] (3) to (2);
                    \draw [style=Aligned] (0) to (3);
                    \draw [style=Aligned] (2) to (1);
                    \draw [style=SingleU, bend right=45] (2) to (9.center);
                    \draw [style=SingleU, bend right=45] (9.center) to (10.center);
                    \draw [style=SingleU, bend right=45] (10.center) to (3);
                    \draw [style=SingleU, bend left=60, looseness=1.25] (1) to (0);
                    \draw [style=SingleD, bend right=60, looseness=1.25] (1) to (0);
                \end{tikzpicture}
                \vspace{-3 cm}
                \caption{Diagram with adjacent boundaries.}
            \end{subfigure}
            \caption{Cylinder contributions to the correlator \eqref{eq:RLRLRRLL} at $N_- = 2$ and $N_+=3$.}
            \label{fig:RLRLRRLLCylinder}
        \end{figure}
        
        \begin{figure}
            \centering
            \begin{tikzpicture}
    		\node [style=White Dot] (0) at (0, -6) {$7^{RL}$};
    		\node [style=White Dot] (1) at (0, 3) {$3^+$};
    		\node [style=White Dot] (2) at (0, 0) {$2^-$};
    		\node [style=White Dot] (3) at (0, -3) {$2^+$};
    		\node [style=none] (4) at (6, 0) {};
    		\node [style=none] (5) at (-6, 0) {};
    		\node [style=none] (6) at (0, 6) {};
    		\node [style=none] (7) at (-2, 4) {$+$};
    		\node [style=none] (8) at (-1.5, 0.5) {$-$};
    		\node [style=none] (9) at (1.5, 0.5) {$+$};
    		\node [style=none] (10) at (-1, -2) {$+$};
    		\node [style=none] (11) at (1, -2) {$-$};
    		\node [style=none] (12) at (-0.25, -4.5) {$-$};
    		\node [style=none] (13) at (0.25, -4.5) {$+$};
    		\node [style=none] (14) at (2, 4) {$+$};
    
    		\draw [style=SingleU, bend right=45] (0) to (4.center);
    		\draw [style=SingleU, bend right=45] (5.center) to (0);
    		\draw [style=SingleU, bend left=315] (6.center) to (5.center);
    		\draw [style=SingleU, bend right=45] (4.center) to (6.center);
    		\draw [style=MisalignedU] (1) to (6.center);
    		\draw [style=Aligned, bend left=75, looseness=1.50] (0) to (1);
    		\draw [style=Aligned] (1) to (2);
    		\draw [style=Aligned, bend right=60, looseness=1.25] (2) to (0);
    		\draw [style=Aligned, bend left=60, looseness=1.25] (2) to (0);
    		\draw [style=Aligned] (3) to (2);
    		\draw [style=Aligned, bend left=45] (0) to (3);
    		\draw [style=Aligned, bend right=45] (0) to (3);
    		\draw [style=Aligned] (3) to (0);
    		\draw [style=MisalignedU] (1) to (4.center);
            \end{tikzpicture}
            \caption{Diagram for one of the disk covering maps contributing to \eqref{eq:complicated_four_point}.}
            \label{fig:placeholder}
        \end{figure}
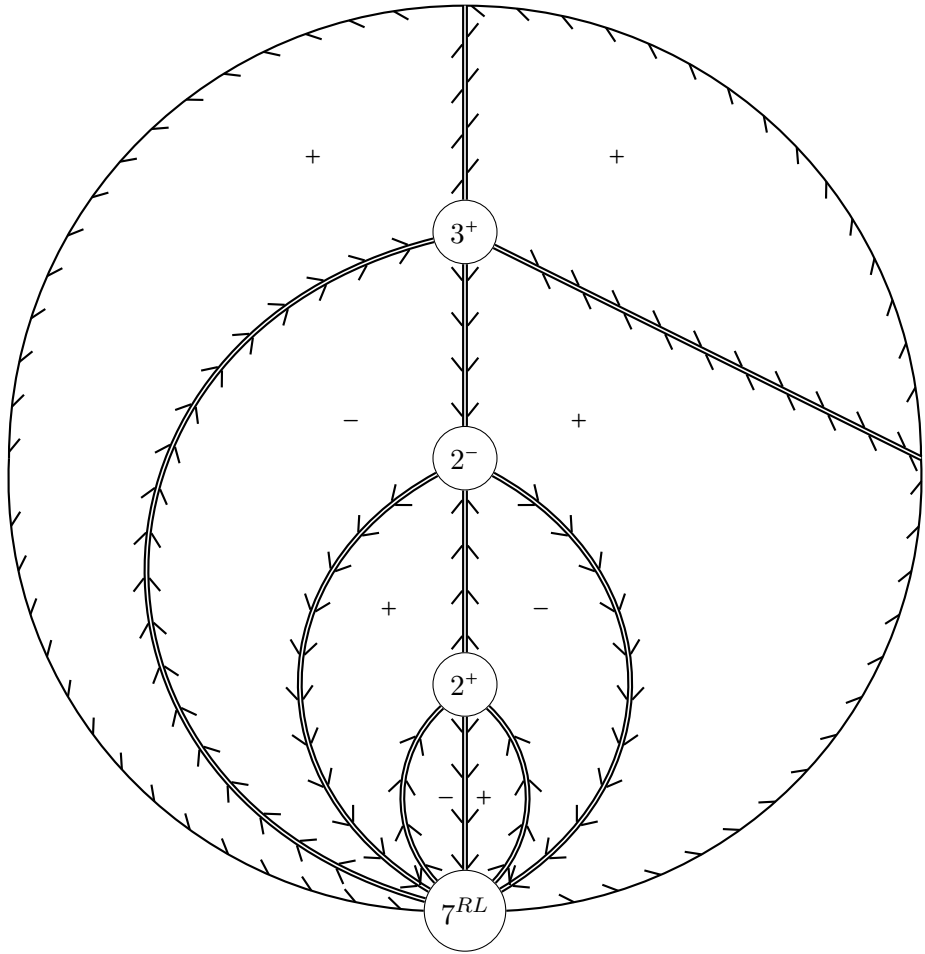

        \paragraph{Example 3.}
        As the complexity of correlators increases, the number of diagrams that contributes for each covering 
        surface also grows. Figure \ref{fig:placeholder} shows only one of many diagrams that contribute at disk 
        level to the four-point function 
        \begin{align}\label{eq:complicated_four_point}
            \langle\sigma_7^{RL}\sigma_2(z_2^+)\sigma_3(z_3^+)\sigma_2(z_2^-)\rangle.
        \end{align}    
        We hope that these three example suffice to illustrate the diagrammatic rules that we have introduced in 
        order to describe interface covering maps. 
        
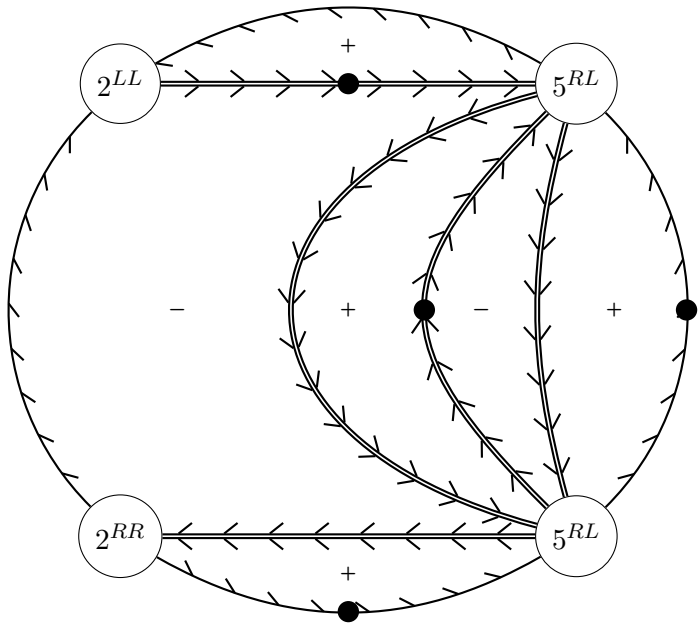
\begin{figure}[ht]
            \centering
            \begin{tikzpicture}
        		\node [style=White Dot] (0) at (-3, 3) {$2^{LL}$};
        		\node [style=White Dot] (1) at (-3, -3) {$2^{RR}$};
        		\node [style=White Dot] (2) at (3, -3) {$5^{RL}$};
        		\node [style=White Dot] (3) at (3, 3) {$5^{RL}$};
        		\node [style=none] (5) at (-2.25, 0) {$-$};
        		\node [style=none] (6) at (0, 0) {$+$};
        		\node [style=none] (7) at (1.75, 0) {$-$};
        		\node [style=none] (8) at (3.5, 0) {$+$};
        		\node [style=none] (9) at (0, 3.5) {$+$};
        		\node [style=none] (10) at (0, -3.5) {$+$};
        
        		\draw [style=SingleU, bend left=330] (3) to (0);
        		\draw [style=SingleD, bend left=45] (1) to (0);
        		\draw [style=SingleU, bend right] (1) to (2);
        		\draw [style=SingleU, bend right=45] (2) to (3);
        		\draw [style=Aligned] (2) to (1);
        		\draw [style=Aligned] (0) to (3);
        		\draw [style=Aligned, bend right=75, looseness=2.00] (3) to (2);
        		\draw [style=Aligned, bend left=45, looseness=1.50] (2) to (3);
        		\draw [style=Aligned, bend right=15] (3) to (2);
    
                \fill (0,3) circle (4pt);
                \fill (0,-4) circle (4pt);
                \fill (1,0) circle (4pt);
                \fill (4.45,0) circle (4pt);
            \end{tikzpicture}
            \caption{Dressed version of Fig.~\ref{fig:RLRLRRLLDisk}. Preimages $\Gamma^{-1}(t^*)$ of the reference point $t^*$ under the covering map are depicted by black dots.}
            \label{fig:RLRLRRLLDiskDressed}
        \end{figure}

        \paragraph{Diagrams with reference point.}
        By construction, we can read off $N_\pm$ from a diagram by counting the faces 
        labelled by $+$ and $-$. We now explain how one can read off the degree 
        of transmissivity $P$. 
        Recall that, when moving 
        along the interface, $P$ changes by one unit whenever we cross an interface branch point of type $LL$ 
        or $RR$, see Section \ref{sec:symmetric_orbifold_boundaries_interfaces}. 
        Thus, the integer $P$ is not a global property of the covering map, but a local property associated to a reference point $t^*$ on the interface, as also discussed below 
        eq.~\eqref{interface_changing_correlator}.
        Given a choice of $t^*$, we decorate our diagrams
        with the elements of the preimage $\Gamma^{-1}(t^*)$ under the covering map. 
        The degree of transmissivity $P$ at $t^*$ is then given by the number of these preimages occurring in the interior of the covering space, i.e.~on double lines. 
        The relevant data encoded in the choice of $t^*$ is its position in the ordering of interface and bulk branch points.
        Using this data, the reference points have to be distributed on the diagrams 
        according to the following three rules
        \begin{enumerate}
            \item The boundary of each face includes exactly one reference point.
            \item Moving along the boundary of a face along its orientation, one 
            encounters not only the ramification points but also the reference 
            point in the correct order.
            \item Reference points cannot occur on misaligned double lines.
        \end{enumerate}
        As an example, Fig.~\ref{fig:RLRLRRLLDiskDressed} shows a $t^*$ dressed 
        version of the diagram shown in Fig.~\ref{fig:RLRLRRLLDisk}. The decoration 
        shown here amounts to the ordering $t^*<5^{RL}<5^{RL}<2^{RR}<2^{LL}$ for 
        which we read off that $P=2$ since two of the black dots are placed on 
        double lines. The reader is invited to draw the diagram that is associated 
        with the order $5^{RL}<5^{RL}<2^{RR}< t^* <2^{LL}$ and to verify that this 
        returns $P=1$. 
     
    \subsection{Symmetry group of interface PRR diagrams}\label{sec:symmetry_factors}
        In this short subsection, we discuss how to determine the symmetry group of an (interface) covering map from the corresponding diagram. 
        To do so, we first define the automorphism group of (interface) PRR diagrams.
        From the definition, it will directly be clear that the automorphism group of a diagram is isomorphic to the group of deck transformations of the covering map.
        We end the section by applying this isomorphism in two examples.
        
        \paragraph{Definition of the symmetry group.}
        As already mentioned above, an interface PRR diagram is not a just graph but comes with some 
        additional structure. 
        To begin with it is a fat 
        graph -- the edges and faces that meet at 
        any vertex come with a cyclic order. 
        In addition, the vertices of the diagram are labelled by the associated branch points of the covering map. 
        For covering maps that arise from single-cycle twist fields, this labelling is injective, i.e.~no two vertices possess the same 
        label. 
        However, in case of multi-cycle twist fields, vertices that carry the same label can arise. 
        
        Now that 
        we understand the relevant structure of the interface PRR diagram, we can define its
        symmetry group: It is the group of all permutations of (vertices), edges and faces of the diagram that respect the labelling of vertices, the cyclic ordering at each vertex and of course the structure of the underlying graph.  
        In terms of an explicit embedding representing a PRR diagram in $\Sigma$, the symmetries of the diagram are exactly the permutations of (vertices), edges and faces that descend from automorphisms of $\Sigma$ which preserve the embedded labelled graph. 
        Let us also note that the symmetry group of an interface PRR diagram is much smaller than the symmetry group of an associated graph which one obtains from the diagram by dropping the additional structure. 
        \paragraph{Identification with deck transformations.}
        We now argue that the symmetry group of an interface PRR diagram is isomorphic to the 
        group of deck transformations of the corresponding covering maps.
        Given a diagram, we pick a representative $\Gamma$ of the equivalence class of covering maps that the diagram corresponds to. 
        Clearly, each deck transformation gives a symmetry of the associated interface PRR diagram.
        Vice versa, if $\phi$ is an automorphism of $\Sigma$ that descends to an symmetry of the diagram, then $\Gamma$ and $\Gamma \circ \phi$ are two holomorphic maps that by construction coincide on the vertices and edges of the embedded interface PRR diagram. 
        But since both $\Gamma$ and $\Gamma \circ \phi$ are holomorphic (and the faces of diagrams are simply connected), this means that they also need to coincide on the faces, which implies that $\phi$ is a deck transformation of 
        $\Gamma$.
        
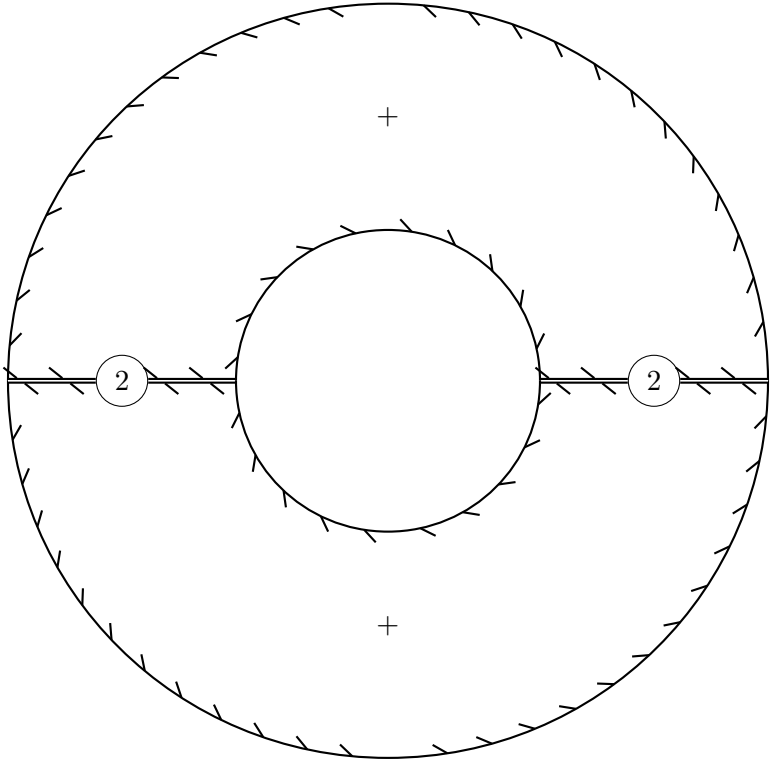
\begin{figure}
            \centering
            \begin{tikzpicture}
    		\node [style=none] (4) at (5, 0) {};
    		\node [style=none] (5) at (-5, 0) {};
    		\node [style=none] (6) at (0, 5) {};
    		\node [style=none] (11) at (0, -5) {};
    		\node [style=none] (12) at (0, -2) {};
    		\node [style=none] (13) at (-2, 0) {};
    		\node [style=none] (14) at (0, 2) {};
    		\node [style=none] (15) at (2, 0) {};
    		\node [style=White Dot] (16) at (-3.5, 0) {2};
    		\node [style=White Dot] (17) at (3.5, 0) {2};
    		\node [style=none] (19) at (0, -3.25) {+};
    		\node [style=none] (20) at (0, 3.5) {+};
    
    		\draw [style=SingleU, bend left=315] (6.center) to (5.center);
    		\draw [style=SingleU, bend right=45] (4.center) to (6.center);
    		\draw [style=SingleU, bend right=45] (5.center) to (11.center);
    		\draw [style=SingleU, bend right=45] (11.center) to (4.center);
    		\draw [style=MisalignedU] (16) to (13.center);
    		\draw [style=MisalignedU] (5.center) to (16);
    		\draw [style=MisalignedU] (15.center) to (17);
    		\draw [style=MisalignedU] (17) to (4.center);
    		\draw [style=SingleU, bend left=45] (12.center) to (13.center);
    		\draw [style=SingleU, bend left=45] (13.center) to (14.center);
    		\draw [style=SingleU, bend left=45] (14.center) to (15.center);
    		\draw [style=SingleU, bend left=45] (15.center) to (12.center);
            \end{tikzpicture}
            \caption{Cylinder contribution to $\langle \sigma_2 \sigma_2 \rangle$ at $N_+ = 2$, $N_- = 0$.}
            \label{fig:cylinder}
        \end{figure}

        \paragraph{Examples.}
        The only diagram with a non trivial symmetry group discussed so far in this section is the one shown in Fig.~\ref{fig:bulk_two_point}.
        The symmetry group of this diagram is the $\mathbb{Z}_w$ sub-group of the rotations of the sphere that fix the two ramification points and cyclically permute the edges of the graph. 
        On higher genus covering spaces the symmetry groups can be more general transformations than rotations. 
        For instance, the cylinder diagram contributing to the two-point function $\langle \sigma_2 \sigma_2 \rangle$ of bulk twist 2 operators shown in Fig.~\ref{fig:cylinder} has a $\mathbb{Z}_2$ symmetry group that is generated by a modular $S$-transformation of the cylinder which flips the two boundaries but leaves the ramification points fixed. 

    \subsection{Gauge fixing of interface PRR diagrams}
    \label{sec:GaugeFixing}
        In equation \eqref{eq:conjugacy_class}, we identified the coefficient 
        $\alpha[\Gamma]$ that weighs the contribution of the covering map $\Gamma$ to 
        a connected sphere correlator, see eq.\ \eqref{eq:twist_field_Gamma}, with the 
        size of a particular orbit under an $S_N$ action. This subsection provides an 
        analogous interpretation for the interface coefficient $\alpha[\Gamma]$, see eqs.~\eqref{eq:interface_operators} and \eqref{eq:interface_alpha}, as the 
        size of an orbit of some $S_{N_{-}} \times S_{N_{+}}$ action. 
        It is straight forward to obtain such an interpretation from a generalisation 
        of the monodromy based arguments that led to eq.~ \eqref{eq:conjugacy_class}. 
        However, now that we have developed the PRR diagrammatics of interface 
        covering maps we can do better and provide a more transparent description 
        in terms of graphs.

        \paragraph{Gauge fixed PRR diagrams.} Consider an interface covering map $\Gamma$ 
        with $\text{deg}[\Gamma_\pm] = N_\pm$. As in the case of bulk covering maps, we 
        refer to a bijective assignment of integers in $\underline{N}_{\pm}:=\{1,\dots,
        N_{\pm}\}$ to sheets of the covering as a gauge choice. In terms of interface 
        PRR diagrams, such a gauge choice assigns integers to the $\pm$-faces. We call 
        a PRR diagram whose faces are labelled in such a way \textit{gauge fixed}.
        
        \paragraph{Gauge fixed ramification indices.} 
        Gauge fixed PRR diagrams associate ramification points of $\Gamma$ with cyclic 
        permutations of $\underline{N}_{-} \sqcup \underline{N}_+$. These permutations 
        are determined by circumnavigating the associated vertices of the diagram 
        anti-clockwise and reading off the encountered $\underline{N}_{\pm}$ labels of 
        the traversed faces. We refer to these permutations as \emph{gauge fixed 
        ramification indices}. To give a more detailed description it is useful introduce 
        different variables $j$ and $a$ to denote elements of $\underline{N}_{-}$ and 
        $\underline{N}_{+}$, respectively.
        
        With this convention, every ramification point of $\Gamma$ with ramification 
        index $w$ whose branch point is in the upper half plane, is associated with 
        a permutation $\pmb{g}=(a_1 j_1 a_2 a_3 \dots a_w)$ of length $|\pmb{g}|$. 
        The length is constrained by $w \leq |\pmb{g}| \leq 2w$. There must be 
        exactly $w$  $a$-labels. In addition one may have up to $w$ $j$-labels which, 
        however, must always be separated by $a$-labels in the permutation. The labels 
        of interior ramification points whose branch point is in the lower half plane 
        are defined analogously with $a$ and $j$ labels interchanged. The labels 
        corresponding to branch points on the interface are also defined in the 
        same manner. In this case, however, the associated permutation necessarily 
        alternates between $a$- and $j$-labels and has length $|\pmb{g}|=2w$.

        Finally, every type $\omega^{AB}$ boundary ramification point is associated 
        with a sequence $\pmb{\gamma}^{AB}$ of length $\omega$, whose entries 
        alternate between $a$- and $j$-labels. We choose sequences instead of 
        permutations to describe gauge fixed boundary ramification since the 
        boundary case involves more data than just the cyclic ordering: Boundary 
        ramification points also distinguish a particular first and last element 
        of the associated sequence of $\underline{N}_\pm$ labelled faces.
        At these end-point the sequence stops since the boundary of the covering 
        space is reached.
        
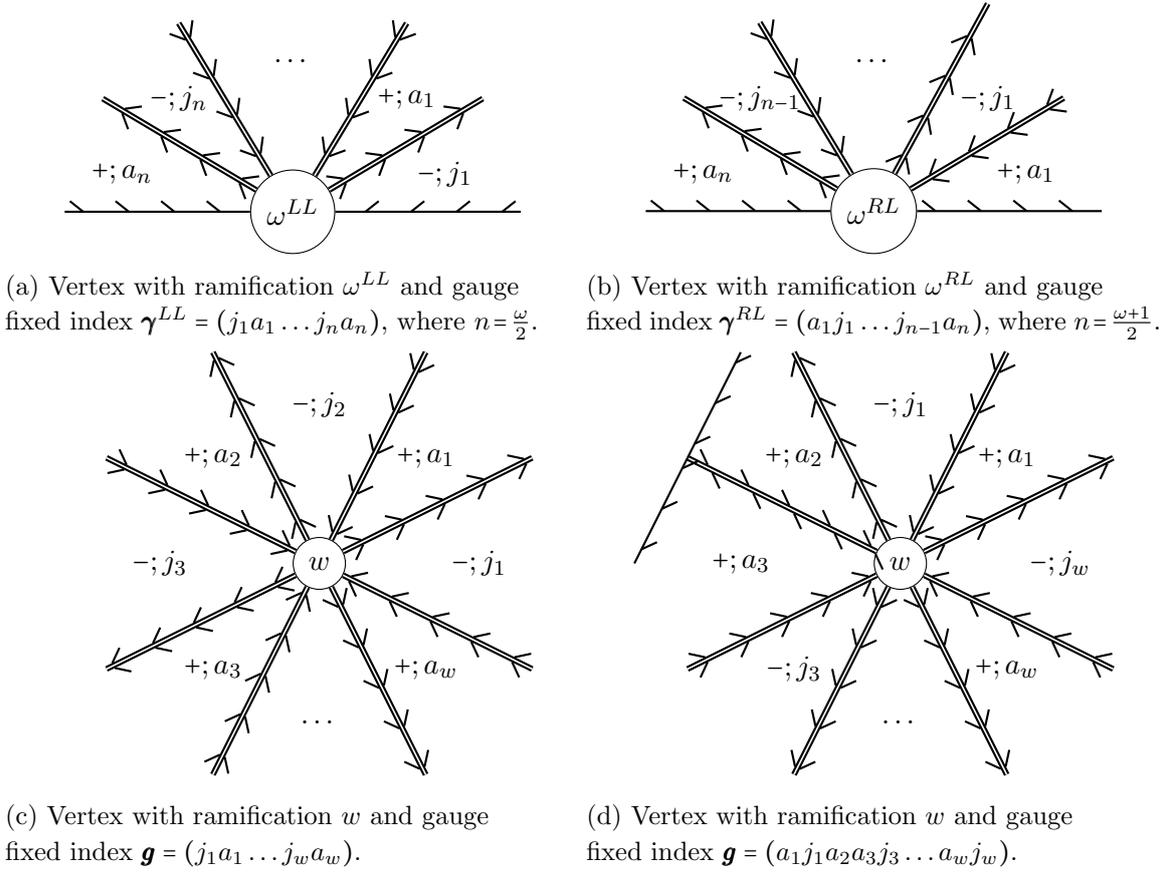
\begin{figure}
            \begin{subfigure}{.5\textwidth}
                \centering
                \begin{tikzpicture}
                    \node [style=none] (8) at (-3, 0) {};
                    \node [style=none] (14) at (3, 0) {};
                    \node [style=White Dot] (15) at (0, 0) {$\omega^{LL}$};
                    \node [style=none] (16) at (-2.5, 1.5) {};
                    \node [style=none] (17) at (-1.5, 2.5) {};
                    \node [style=none] (18) at (1.5, 2.5) {};
                    \node [style=none] (19) at (2.5, 1.5) {};
                    \node [style=none] (20) at (0, 2) {$\dots$};
                    \node [style=none] (21) at (-2.25, 0.5) {$+;a_n$};
                    \node [style=none] (22) at (2, 0.5) {};
                    \node [style=none] (23) at (2, 0.5) {$-;j_1$};
                    \node [style=none] (24) at (1.5, 1.5) {};
                    \node [style=none] (25) at (1.5, 1.5) {$+;a_1$};
                    \node [style=none] (26) at (-1.5, 1.5) {};
                    \node [style=none] (27) at (-1.5, 1.5) {$-;j_n$};
                    
                    \draw [style=SingleU] (8.center) to (15);
                    \draw [style=SingleD] (14.center) to (15);
                    \draw [style=Aligned] (15) to (16.center);
                    \draw [style=Aligned] (17.center) to (15);
                    \draw [style=Aligned] (15) to (19.center);
                    \draw [style=Aligned] (18.center) to (15);
                \end{tikzpicture}
                \caption{Vertex with ramification $\omega^{LL}$ 
                and gauge \\ fixed index $\pmb{\gamma}^{LL}=(j_1a_1\dots j_na_n)$, where $n \hspace{-0.5 mm}=\hspace{-0.5 mm} \tfrac{\omega}{2}$.}
            \end{subfigure}
            \begin{subfigure}{.5\textwidth}
                \centering     
                \begin{tikzpicture}
                    \node [style=none] (8) at (-3, 0) {};
                    \node [style=none] (14) at (3, 0) {};
                    \node [style=White Dot] (15) at (0, 0) {$\omega^{RL}$};
                    \node [style=none] (16) at (-2.5, 1.5) {};
                    \node [style=none] (17) at (-1.5, 2.5) {};
                    \node [style=none] (18) at (2.5, 1.5) {};
                    \node [style=none] (19) at (1.5, 2.75) {};
                    \node [style=none] (20) at (0, 2) {$\dots$};
                    \node [style=none] (21) at (-2.25, 0.5) {$+;a_n$};
                    \node [style=none] (22) at (2, 0.5) {};
                    \node [style=none] (23) at (2, 0.5) {$+;a_1$};
                    \node [style=none] (24) at (1.5, 1.5) {};
                    \node [style=none] (25) at (1.5, 1.5) {$-;j_1$};
                    \node [style=none] (26) at (-1.5, 1.5) {};
                    \node [style=none] (27) at (-1.5, 1.5) {$-;j_{n-1}$};
                
                    \draw [style=Aligned] (15) to (16.center);
                    \draw [style=Aligned] (17.center) to (15);
                    \draw [style=Aligned] (15) to (19.center);
                    \draw [style=Aligned] (18.center) to (15);
                    \draw [style=SingleU] (8.center) to (15);
                    \draw [style=SingleU] (15) to (14.center);
                \end{tikzpicture}
                \caption{Vertex with ramification  $\omega^{RL}$ and gauge \\ fixed index $\pmb{\gamma}^{RL}=(a_1j_1\dots j_{n-1}a_n)$, where $n \hspace{-0.5 mm} =  \hspace{-0.5 mm} \tfrac{\omega +1}{2}$.}
            \end{subfigure}
            \begin{subfigure}{.5\textwidth}
                \centering     
                \begin{tikzpicture}[scale=0.7]
                    \node [style=none] (16) at (3, 0) {$-;j_1$};
                    \node [style=none] (17) at (2, 2) {$+;a_1$};
                    \node [style=none] (18) at (-2, 2) {$+;a_2$};
                    \node [style=none] (19) at (-3, 0) {$-;j_3$};
                    \node [style=none] (20) at (0, -3) {$\dots$};
                    \node [style=none] (21) at (-2, 4) {};
                    \node [style=none] (22) at (2, 4) {};
                    \node [style=none] (25) at (-4, -2) {};
                    \node [style=none] (26) at (-2, -4) {};
                    \node [style=none] (27) at (2, -4) {};
                    \node [style=none] (28) at (4, -2) {};
                    \node [style=none] (29) at (4, 2) {};
                    \node [style=none] (31) at (0, 3) {$-;j_2$};
                    \node [style=none] (33) at (-2, -2) {$+;a_3$};
                    \node [style=none] (34) at (2, -2) {$+;a_w$};
                    \node [style=White Dot] (36) at (0, 0) {$w$};
                    \node [style=none] (37) at (-4, 2) {};
                    \node [style=none] (38) at (-5, 0) {};
                    \node [style=none] (39) at (-3, 4) {};
                    \draw [style=Aligned] (36) to (27.center);
                    \draw [style=Aligned] (28.center) to (36);
                    \draw [style=Aligned] (36) to (29.center);
                    \draw [style=Aligned] (22.center) to (36);
                    \draw [style=Aligned] (36) to (21.center);
                    \draw [style=Aligned] (37.center) to (36);
                    \draw [style=Aligned] (36) to (25.center);
                    \draw [style=Aligned] (26.center) to (36);
                \end{tikzpicture}
                \caption{Vertex with ramification $w$ and gauge \\ fixed index $\pmb{g}=(j_1a_1\dots j_wa_w)$.}
            \end{subfigure}
            \begin{subfigure}{.5\textwidth}
                \centering     
                \begin{tikzpicture}[scale=0.7]
                    \node [style=none] (16) at (3, 0) {$-;j_w$};
                    \node [style=none] (17) at (2, 2) {$+;a_1$};
                    \node [style=none] (18) at (-2, 2) {$+;a_2$};
                    \node [style=none] (19) at (-3, 0) {$+;a_3$};
                    \node [style=none] (20) at (0, -3) {$\dots$};
                    \node [style=none] (21) at (-2, 4) {};
                    \node [style=none] (22) at (2, 4) {};
                    \node [style=none] (25) at (-4, -2) {};
                    \node [style=none] (26) at (-2, -4) {};
                    \node [style=none] (27) at (2, -4) {};
                    \node [style=none] (28) at (4, -2) {};
                    \node [style=none] (29) at (4, 2) {};
                    \node [style=none] (31) at (0, 3) {$-;j_1$};
                    \node [style=none] (33) at (-2, -2) {$-;j_3$};
                    \node [style=none] (34) at (2, -2) {$+;a_w$};
                    \node [style=White Dot] (36) at (0, 0) {$w$};
                    \node [style=none] (37) at (-4, 2) {};
                    \node [style=none] (38) at (-5, 0) {};
                    \node [style=none] (39) at (-3, 4) {};
                    \draw [style=Aligned] (36) to (21.center);
                    \draw [style=Aligned] (22.center) to (36);
                    \draw [style=MisalignedU] (37.center) to (36);
                    \draw [style=SingleU] (39.center) to (37.center);
                    \draw [style=SingleU] (37.center) to (38.center);
                    \draw [style=Aligned] (36) to (29.center);
                    \draw [style=Aligned] (28.center) to (36);
                    \draw [style=Aligned] (36) to (27.center);
                    \draw [style=Aligned] (25.center) to (36);
                    \draw [style=Aligned] (36) to (26.center);
                \end{tikzpicture}
                \caption{Vertex with ramification $w$ and gauge \\ fixed index $\pmb{g}=(a_1j_1a_2a_3 j_3\dots a_w j_w)$.}
            \end{subfigure}
            \caption{Examples of vertices and corresponding sequences.}
            \label{fig:vertexandsequence}
        \end{figure}

        Examples of various types of gauge fixed ramification indices are illustrated in Fig.~\ref{fig:vertexandsequence}.

        \paragraph{Gauge fixed twist fields.} Gauge fixed ramification indices allow 
        us to define gauge-fixed twist fields labelled by such indices. These are non-local 
        operators whose correlators are computed by a restriction of the sum over covering 
        maps of local twist fields to gauge fixed PRR diagrams compatible with the gauge 
        fixed ramification indices. Clearly, the constraint can be lifted by taking a linear
        combination of all possible ways to fix the gauge. Thus, local twist fields can be 
        expressed as linear combinations of their gauge fixed counterpart. In fact, this is 
        how we initially defined the interface changing operators in \cite{Harris:2025wak}.
        
        \paragraph{Gauge fixed interfaces.} 
        The gauge fixed PRR diagram also allows us to associate a permutation to the interface itself. For a covering map with degree of transmissivity $P$, the permutation $\pmb{\iota}$ that describes the gauge fixed glueing condition of the interface at the reference point $t^*$ is 
        a product $\pmb{\iota} = (j_1a_1) (j_2a_2) \dots (j_P a_P)$ of $P$ commuting transpositions $(j_i a_i)$. Recall from the our discussion at the end of 
        Section \ref{sec:Interface_PRR} that the preimage $\Gamma^{-1}(t^*)$ can 
        be represented by placing black dots on the edges of the PRR diagram 
        according to some rules we spelled out before. For an interface covering 
        with the degree of transmissivity $P$ at $t^*$, there will be exactly $P$ 
        black dots on aligned double lines. Hence we can associate $P$ pairs $(j_i a_i)$ 
        to the choice of $t^*$. Note that the permutations that would describe the analogous 
        data for any other choices of the reference point is fixed by 
        the gauge fixed ramification indices along with the $\pmb{\iota}$ of 
        one specific choice $t^*$ of the reference point. 

        \paragraph{Symmetry factors from gauge fixed diagrams.} Now we are actually 
        ready to determine the weight factor $\alpha(\Gamma)$ we are after. Consider an arbitrary gauge fixed PRR diagram for $\Gamma$ and read off the corresponding gauge fixed interior ramification indices $\pmb{g}_{1}, \dots, \pmb{g}_{n_c}$ and gauge fixed boundary ramification indices $\pmb{\gamma}^{A_1B_1}_1, \dots,\pmb{\gamma}^{A_{n_o}B_{n_o}}_{n_o}$ as well as the permutation $\pmb{\iota}$ describing the glueing data at the reference 
        point $t^*$. Given all this data, the weight is given by 
        \begin{align}\label{eq:alpha_orbit}
            \alpha[\Gamma] = \frac{\text{deg}[\Gamma_-]!\text{deg}[\Gamma_+]!}{\left|\text{Deck}[\Gamma]\right|} = \left|[\pmb{g}_{1}, \dots, \pmb{g}_{n_c},\pmb{\iota},  \pmb{\gamma}^{A_1B_1}_1, \dots, \pmb{\gamma}^{A_{n_o}B_{n_o}}_{n_o}]\right|.
        \end{align}
        The number on the r.h.s.~is the size of the orbit 
        \begin{align} \label{eq:orbit}
            [\pmb{g}_{1}, \dots, \pmb{\iota},  \pmb{\gamma}^{A_1B_1}_1, \dots] := 
            \{(g\pmb{g}_{1}g^{-1}, \dots, g\pmb{\iota}g^{-1},  g.\pmb{\gamma}^{A_1B_1}_1, \dots)| g \in S_{N_-} \times S_{N_-}\}.
        \end{align}
        that is obtained form the gauged fixed ramification indices and 
        $\pmb{\iota}$ by simultaneous action with elements $g$ in the 
        product $S_{N_-} \times S_{N_+}$ of the permutation groups. On 
        the interior ramification indices $\pmb{g}$ and on $\pmb{\iota}$
        elements $g$ act through the adjoint action. Recall that $\pmb{g}_i$
        and $\pmb{\iota}$ are themselves elements of $S_{N_-} \times S_{N_+}$
        so that the adjoint action is well defined. The action of $g$ on the 
        gauged fixed boundary ramification indices $\pmb{\gamma}_i$ is through 
        the defining action of $S_{N_-} \times S_{N_+}$ on the entries of the 
        $\pmb{\gamma}^{A_iB_i}_i$ sequences. 
        
        Diagrammatically, this group action just corresponds to the relabelling 
        of faces in the gauge fixed PRR diagram. The action of a deck transformation on the covering space amounts to a permutation of faces of the diagram.
        Therefore, a gauge choice for the diagram allows us to identify 
        $\text{Deck}[\Gamma]$ with a subgroup of $S_{N_-} \times S_{N_+}$.
        Since deck transformations leave the structure of the covering map invariant, this subgroup of $S_{N_-} \times S_{N_+}$ is exactly the stabiliser of the orbit \eqref{eq:orbit}  
        under the permutation action. Hence, equation \eqref{eq:alpha_orbit} directly follows from the orbit-stabiliser theorem.

    \subsection{The large \texorpdfstring{$N$}{N} expansion}
    \label{sec:CanonicalLargeN}
        
        In this section, we discuss a particular limit in which the computation of our interface correlation functions simplifies and non-planar diagrams are suppressed. 
        Concretely, we study connected correlators on the sphere $X=\mathbb{S}^2$ with an interface $\mathcal{I}^{(P)}_{|a\rangle}$ inserted along the equator and the parameters $N_\pm$ and $P$ scaling as
        \begin{align}\label{eq:large_N_P}
            N_\pm \sim N, \quad P \sim N \quad \text{and} \quad N_\pm - P \sim N^{\tfrac{1}{2}}
        \end{align}
        with an expansion parameter $N$ that we will take to be large.
        Our main finding is that contributions of genus $g$ covering spaces with $b$ boundaries to connected correlators of $n_c$ bulk and $n_o$ interface twist fields scale, to leading order in $N$, as
        \begin{align}\label{eq:Large_N_expansion}
            N^{1-g-\tfrac{b}{2} - \tfrac{n_c}{2} - \tfrac{n_o}{4}}
        \end{align}
        if the two point-functions of the operators are normalised to $1$.
        Equation \eqref{eq:Large_N_expansion} is suggestive of an identification $N \sim g_s^{-2}$ of the parameter $N$ with the string coupling of a potential holographic dual.
        As in the bulk case, the true relationship between the parameters of the symmetric orbifold and the string coupling is however slightly more subtle.
        In Section \ref{sec:saddle_point}, we will show that the resemblance of eq.~\eqref{eq:Large_N_expansion} to a string genus expansion can be traced back to the fact that eq.~\eqref{eq:large_N_P} with $N\sim g_s^{-2}$ describes the expectation values of $N_\pm$ and $P$ in the truly holographic grand-canonical ensemble of interfaces to leading order in small~$g_s$.

        \paragraph{Connected sphere correlators.}
        Towards the end of Section \ref{sec:symm_orb}, we defined connected sphere correlators for bulk twist fields in symmetric product orbifolds.
        Connected correlation functions in the presence of the interface $\mathcal{I}_{|a\rangle}^{(P)}$ are defined analogously, i.e.
        \begin{align}\label{eq:conn_interface}
            \langle \mathcal{I}_{|a\rangle}^{(P)} \mathcal{O}\rangle^\text{conn,unnormalised}_{\text{Sym}^{N_-|N_+}(\mathcal{M})} := 
            \frac{1}{N_-!N_+!} \hspace{-0.2 cm} \sum_{\substack{n_\pm=0 \\ n_-+n_+ \neq 0}}^{N_\pm} \hspace{-0.2 cm} \sum_{p=0}^{\text{min}(n_\pm,P)} \hspace{-0.3 cm} \sum_{\Gamma \in C_{n_\pm,p}^{\text{conn}}[\mathcal{I}_{|a\rangle}^{(P)}\mathcal{O}]}
            \langle \mathcal{O}^{\Gamma}\rangle_\mathcal{M}, 
        \end{align} 
        where the third sum runs over the subset of connected interface coverings maps $\Gamma$. 
        By definition, these are interface covering maps
        \begin{align} \label{eq:decomposition}
            \Gamma: \overline{\mathbb{C}}^{\sqcup(P-p)}
            \sqcup \mathbb{D}_{-}^{\sqcup (N_--P-b_-)}
            \sqcup \mathbb{D}_{+}^{\sqcup (N_+-P-b_+)}
            \sqcup \Sigma \rightarrow \overline{\mathbb{C}},
        \end{align}
        with $b_\pm = n_\pm - p$ that satisfy three conditions: First of all, they restrict to the identity map on the $P-p$ spheres $\overline{\mathbb{C}}$. Furthermore they are also required to  restrict to the identity embedding of the upper and lower hemispheres on the $N_\pm - P - b_\pm$ disconnected discs $\mathbb{D}_\pm$. Finally, 
        we impose that $\Sigma$ should be connected. The parameters $n_\pm$ and $p$ correspond to the degrees and transmissivity of the restriction
        $\Gamma|_\Sigma$.  For $\Gamma \in C_{n_\pm,p}^{\text{conn}}[\mathcal{I}_{|a\rangle}^{(P)}\mathcal{O}]$, we have
        \begin{align}\label{eq:DeckDecomposition}
            \left|\text{Deck}[\Gamma]\right|=(N_--P-b_{-})!(N_+-P-b_{+})!(P-p)!\left|\text{Deck}[\Gamma|_\Sigma]\right|
        \end{align}
        which we can use to partially evaluate the correlator $\langle \mathcal{O}^\Gamma\rangle_\mathcal{M}$ on the r.h.s.~of eq.~\eqref{eq:conn_interface} as
        \begin{align}\label{eq:O_disc_conn}
            \langle \mathcal{O}^\Gamma\rangle_\mathcal{M} = \frac{
            N_-!N_+!~ 
            g^{N_-+N_+-2P - 
            (b_- + b_+)
            }}{(N_--P-b_{-})!(N_+-P-b_{+})!(P-p)!}  \, \frac{\langle\mathcal{O}^{\Gamma|_\Sigma}\rangle_\mathcal{M}}{  n_-! n_+!
            }.
        \end{align}
        The factors in front of the remaining expectation value $\langle \mathcal{O}^{\Gamma|_\Sigma}\rangle_\mathcal{M}$ account for the contribution of the trivial connected components of the covering space. 
        When inserted into eq.~\eqref{eq:conn_interface}, the formula \eqref{eq:O_disc_conn} leads to
        \begin{align}\label{eq:normalised_conn_corr}
            \langle \mathcal{I}_{|a\rangle}^{(P)} \mathcal{O}\rangle^\text{conn}_{\text{Sym}^{N_-|N_+}(\mathcal{M})} = 
            \hspace{-0.3 cm} \sum_{\substack{n_\pm=0 \\ n_-+n_+ \neq 0}}^{N_\pm} \hspace{-0.2 cm} \sum_{p=0}^{\text{min}(n_\pm,P)} \hspace{-0.3 cm} \sum_{\Gamma \in C_{n_\pm,p}^{\text{conn}}[\mathcal{I}_{|a\rangle}^{(P)}\mathcal{O}]}
            C_\Gamma(N_\pm,P)  \, \frac{g^{
            - (b_- + b_+)
            }}{ n_-! n_+!
            }
            \langle\mathcal{O}^{\Gamma|_\Sigma}\rangle_\mathcal{M}
        \end{align}
        for normalised connected correlators, where
        \begin{align}\label{eq:def_CGamma}
             C_\Gamma(N_\pm,P) := \frac{
            (N_--P)!(N_+-P)!P!}{(N_--P-b_{-})!(N_+-P-b_{+})!(P-p)!} 
        \end{align}
        
        \paragraph{Normalisation of twist fields.}
        Eq.~\eqref{eq:normalised_conn_corr} makes the $N_\pm$ and $P$ dependence and hence, with the scaling \eqref{eq:large_N_P}, the $N$ dependence of correlators fully manifest, since $\langle \mathcal{O}^{\Gamma|_\Sigma}\rangle_\mathcal{M}$ is independent of the parameter $N$.
        To have a well behaved large $N$ expansion, we would like to replace bulk and interface twist fields by normalised versions 
        \begin{align}
            \hat \sigma^{AB}_\omega = \mathcal{N}_\omega^{AB} \sigma^{AB}_\omega && \text{and} && \hat \sigma_w(x) = \mathcal{N}_w^s \sigma_w(x)
        \end{align}
        whose two-point functions are of order $1$ in the large $N$ limit. 
        The superscript $s$ we placed on the normalization constants $\mathcal{N}_w^s$ 
        is the sign of the imaginary part of the insertion point $x$. It can take one of the three values $-$, $0$ and $+$, depending on 
        whether $x$ is in the lower half-plane, on the real line or in the upper half-plane. 
        The correlators defining the normalizations 
        $\mathcal{N}_\omega^{AB}$ of the interface operators are
        \begin{align}
            \langle \sigma^{RR}_\omega(0) \sigma^{LL}_\omega(1) \rangle, && \langle \sigma^{LR}_\omega(0) \sigma^{LR}_\omega(1) \rangle && \text{and} && \langle \sigma^{RL}_\omega(0) \sigma^{RL}_\omega(1) \rangle,
        \end{align}
        each of which is computed by a single covering map. 
        Let us refer to the three covering maps as $\Gamma^{RR}_\omega = \Gamma^{LL}_\omega$,  $\Gamma^{LR}_\omega$ and  $\Gamma^{RL}_\omega$.
        With this notation, we have 
        \begin{align}\label{eq:N_C}
            \mathcal{N}_\omega^{AB} := C_{\Gamma^{AB}_\omega}(N_\pm,P)^{-\tfrac{1}{2}} \, 
        \end{align}
        with $C$ as defined in eq.\ \eqref{eq:def_CGamma}. For $A\neq B$, the index 
        $\omega$ is odd and our formula for $C_\Gamma$ evaluates to
        \begin{align}\label{eq:NAB}
            \mathcal{N}_\omega^{LR} = \sqrt{\frac{(P-\tfrac{\omega-1}{2})!}{
            (N_--P)P!}} && \text{and} && \mathcal{N}_\omega^{RL} = \sqrt{\frac{(P-\tfrac{\omega-1}{2})!}{
            (N_+-P)P!}},
        \end{align}
        independently of the choice of reference point defining $P$ and $p$. In the first case with $AB=LR$, for example, the degrees $n_\pm$ of 
        $\Gamma^{AB}$ are given by $n_- = \omega/2 +1/2$ 
        and $n_+=\omega/2-1/2$ while the transmissivity is $p = \omega/2 - 1/2$. Hence, the parameters 
        $b_\pm$ take the values $b_- = 1$ and $b_+ = 0$. 
        Insertion into eq.\ \eqref{eq:def_CGamma} gives 
        the first formula in eq.~\eqref{eq:NAB}. The second formula is obtained in the same way with the role of $\pm$ exchanged. If $A=B$,  
        on the other hand, the label $\omega$ must 
        be even and the resulting normalizations 
        \begin{align}\label{eq:NAA}
            \mathcal{N}_\omega^{AA} = 
            \begin{cases}
                 \sqrt{\frac{(P-\tfrac{\omega-2}{2})!}{(N_--P)(N_+-P)P!}} & \text{if } t^* \in ]0,1[ \\
                 \sqrt{\frac{(P-\tfrac{\omega}{2})!}{P!}} & \text{else}
            \end{cases}
        \end{align}
        explicitly depend on the choice of the reference point $t^*$. The derivation is similar to the previous one but now the degrees $n_\pm$ coincide, i.e. $n_- = n_+ = \omega/2$. The 
        two possible decorations of the corresponding interface PRR 
        diagrams with black dots lead to a degree of transmissivity 
        $p = n_\pm+1$ and $p=n_\pm$, respectively. This 
        gives the two possible values of the normalization constant. 
        
        For bulk fields $\hat \sigma_w$ the normalization is determined along the same lines. The resulting normalization constants are given 
        by 
        \begin{align}\label{eq:Nw}
            \mathcal{N}^\pm_w = \sqrt{\frac{w\,(N_\pm-w)!}{N_\pm!}} \quad \text{and} \quad
            \mathcal{N}^0_w = \sqrt{\frac{w\,(P-w)!}{P!} } \, .
        \end{align}
        
        \paragraph{Large $N$ expansion.}
        If we let $N_\pm$ and $P$ scale with a large parameter $N$ as in eq.~\eqref{eq:large_N_P}, then, to leading order in $N$, 
        \begin{align}
            \mathcal{N}^{AB}_\omega = N^{-\tfrac{\omega}{4}}, \quad \mathcal{N}_w^s = N^{-\tfrac{w}{2}} \quad \text{and} \quad C_\Gamma(N_\pm,P) \sim N^{
            \tfrac{n_-+n_+}{2} 
            },
        \end{align}
        independently of the choice of the reference point, as one can directly read off from eqs.~\eqref{eq:def_CGamma}, \eqref{eq:NAB}, \eqref{eq:NAA} and \eqref{eq:Nw}.
        Thus, we find that the contribution of elements $\Gamma \in C_{n_\pm,p}^{\text{conn}}$ to the correlator
        \begin{align}
            \langle \mathcal{I}_{|a\rangle}^{(P)} \prod\limits_{i=1}^{n_o} \hat{\sigma}_{\omega_i}^{A_i B_i}  \prod\limits_{i=1}^{n_c} \hat{\sigma}_{w_i} \rangle^\text{conn}_{\text{Sym}^{N_-|N_+}(\mathcal{M})} 
        \end{align}
        scales, to leading order in $N$, as
        \begin{align}\label{eq:N_RH}
            N^{-\sum\limits_{i=1}^{n_o} \frac{\omega_i}{4} - \sum\limits_{i=1}^{n_c} \frac{w_i}{2} + \tfrac{n_-+n_+}{2} 
            } = N^{1-g-\tfrac{b}{2} - \tfrac{n_c}{2} - \tfrac{n_o}{4}},
        \end{align}
        where $g$ and $b$ are the genus and number of boundaries of the non-trivial connected component $\Sigma$ of the covering space associated with $\Gamma$. 
        The equality between the l.h.s.~and r.h.s.~of eq.~\eqref{eq:N_RH} follows directly from the interface Riemann-Hurwitz formula \eqref{eq:interface_RH}.
        As advertised in the beginning of this section, higher genus and multi-boundary covering spaces are hence systematically suppressed in the large $N$ limit.

\section{Generalised Lunin-Mathur construction}\label{sec:Lunin_Mathur}
    
    In this short section, we present an interface version of the Lunin-Mathur construction  to 
    compute correlation functions of the seed theory on the covering surface $\Sigma$ or 
    rather partition functions which are relevant for pure twist field insertions in the 
    symmetric product orbifold. Its first half, i.e.~Subsection \ref{sec:bulk_LM}, is 
    devoted to a review of the bulk method. In the second half, i.e.~Subsection
    \ref{sec:interface_LM}, we introduce the interface generalisation and then illustrate 
    its usage in several examples. Appendix \ref{app:LM_computations} contains detailed
    computations supporting the main text.
    Throughout this section, whenever we write $c$ without any further specifications, we mean the central charge $c_\mathcal{M}$ of the seed theory. 
    
    \subsection{Review of the bulk method}\label{sec:bulk_LM}
    
    As reviewed in Section \ref{sec:symmetric_orbifold_boundaries_interfaces}, correlation
    functions in symmetric orbifolds are defined in terms of the data of associated covering
    maps. More specifically, correlators of (single cycle) twist fields on a Riemann surface 
    $X$ can, through the definitions \eqref{eq:twist_field_C} and \eqref{eq:twist_field_Gamma}, 
    be expressed as a sum
    \begin{align}\label{eq:correlator_to_evaluate}
        \langle \sigma_{w_1}(x_1) \dots \sigma_{w_n}(x_n) 
        \rangle_{\text{Sym}^N(\mathcal{M})}^{\text{unnormalised}}   = 
        \sum\limits_{\Gamma} Z[\Gamma] && \text{with} && Z[\Gamma] 
        := \frac{Z_1^\Sigma}{\text{Deck}[\Gamma]} 
    \end{align}
    where $Z_1^\Sigma$ denotes the partition function $Z^\Sigma_1$ of the seed theory on 
    the covering space $\Sigma$. As before, the  sum runs over coverings $\Gamma : \Sigma
    \rightarrow X$ of $X$ whose branch points are exactly $x_1, \dots, x_n$ with 
    indices $w_1, \dots w_n$. In this subsection, we review the Lunin-Mathur method
    \cite{Lunin:2000yv} for the evaluation of $Z_1^\Sigma$.  
        
    \paragraph{General setup.} 
    We assume that for each conformal equivalence class of Riemannian structures $ds^2_\Sigma$
    on the covering surface $\Sigma$, there is a representative $ds^2_{\hat\Sigma}$ for which 
    the value of the vacuum partition function of the seed theory is known. The problem whose
    solution we describe in the paragraphs below is how to express $Z_1^\Sigma$ in terms of 
    $Z_1^{\hat\Sigma}$. Two steps need to be performed to achieve this goal. First, we need to 
    compute the conformal factor $\phi$ in
         \begin{align} \label{eq:gtoghat}
                ds^2_\Sigma = e^\phi d s^2_{\hat \Sigma} \, ,
        \end{align}
    relating the metric $ds^2_{\Sigma} = \Gamma^*ds^2_X$ to the reference metric $ds^2_{\hat \Sigma}$. Once $\phi$ has been computed, we need to calculate the 
    value of the Liouville action on $\phi$ in a second step,     
        \begin{align}\label{eq:liouville}
            S_L[\phi] = \frac{c}{96 \pi} \int\limits_\Sigma d^2 t 
            \sqrt{g_{\hat\Sigma}} \left( \partial_\mu \phi \partial_\nu \phi 
            g^{\mu \nu}_{\hat\Sigma} + 2 R^{{\hat\Sigma}}\phi\right)\ . 
        \end{align}
    The partition function $Z^\Sigma_1$ that appears on the right hand side of 
    eq.~\eqref{eq:correlator_to_evaluate} is then given by
    \begin{align} \label{eq:Z1toZ1hat}
        Z^\Sigma_1 = e^{S_L[\phi]} Z^{\hat \Sigma}_{1}.  
    \end{align}
     It is thus reduced to the reference partition function $Z^{\hat \Sigma}_{1}$ which 
     we assumed to be known. 
        
    \paragraph{The sphere.}
    For concreteness, we focus on the case where both $X$ and $\Sigma$ are spheres.
    Following \cite{Lunin:2000yv}, we introduce a family of metrics $ds^2_{\delta}$ 
    on the Riemann sphere $\overline{\mathbb{C}}$ which are parametrised by some 
    $\delta>0$. On $\mathbb{C}$, these metrics are given by
    \begin{align}
        ds^2_{\delta} = 
        \begin{cases}
             dzd\bar{z} & \text{if } |\delta z| < 1 \\
             \frac{dzd\bar{z}}{|z\delta |^4} & \text{else}\ . 
       \end{cases}
    \end{align}
     They extend continuously to the Riemann sphere. Given a seed conformal field 
     theory with central charge $c$, its partition function on $\overline{\mathbb{C}}$ 
     equipped with $ds^2_\delta$ is
    \begin{align}
        Z^{\overline{\mathbb{C}}_\delta}_1 = Q \delta^{- \tfrac{c}{3}},
    \end{align}
    where $Q$ is an arbitrary normalisation constant \cite{Lunin:2000yv}. 
    We take $ds^2_{\delta}$ as the reference metric on $\hat \Sigma 
    = {\overline{\mathbb{C}}_\delta}$. 
    When dealing with the sphere $X$ that is covered by $\Sigma$ we shall use the 
    same metric, but denote the parameter by $\eta$ instead of $\delta$, i.e. 
    \begin{align}\label{eq:Xeta}
        X = \overline{\mathbb{C}}_{\eta} \quad\quad \text{for some small} 
        \quad\quad \eta>0.
    \end{align}
    We fix the normalization constant to $Q = \eta^{\frac{c}{3}}$ such that the sphere 
    partition function of the seed theory is trivial, $Z_1^X = 1$. Without loss of generality, we assume 
    that all twist fields of the correlator that we would like to compute are inserted 
    in the region $|\eta z| <1$. Since we are only interested in covering maps up to 
    the action of automorphisms of $\Sigma$, we can fix 
    \begin{align}\label{eq:Gammainf=inf}
        \Gamma(\infty) = \infty.
    \end{align}
    As explained above, we compute $Z_1^\Sigma$ by relating it to the partition function 
    $Z^{\hat \Sigma}_1$ with $\hat \Sigma ={\overline{\mathbb{C}}_\delta}$ for an 
    appropriately chosen $\delta$.
        
    \paragraph{Regulating the Liouville action.} Consider the subset $B_{\delta^{-1}} 
    \subseteq \mathbb{C}$ that consists of all points $z \in X$ with $|z \delta|<1$. 
    We denote the preimage of this ball with respect to $\Gamma$ by $\Sigma_0 :=
    \Gamma^{-1}(B_{\delta^{-1}})$. On the subset $\Sigma_0 \subseteq \Sigma$, the metric 
    is given by
    \begin{align}
        ds^2_{\Sigma}|_{\Sigma_0} = \Gamma^*ds^2_{X}|_{|z \delta|<1} = d\Gamma(t) 
        d\bar \Gamma(t) = \left|\tfrac{d\Gamma}{dt}\right|^2 dtd\bar t
    \end{align}
    and therefore
    \begin{align}
        \phi = \log\left(\tfrac{d\Gamma}{dt} \right) + \text{h.c.}\,,
    \end{align}
    which diverges if the derivative $\Gamma'$ vanishes and the metric of $\Sigma$ 
    degenerates. 
    Thus, at the ramification points $R[\Gamma]$ of $\Gamma$, it is 
    necessary to regulate the Liouville action. We explain 
    the details of this regularisation in App.~\ref{app:details_regularisation} following \cite{Lunin:2000yv}. 
    It involves cutting out small neighbourhoods of each ramification point from $\Sigma$. 
    The result of the detailed computation performed in that appendix is
    \begin{align}\label{eq:S_L_boundary_integral}
        Z_{1}^\Sigma = \eta^{-\frac{c}{3}(\text{Deg}[\Gamma]-1)} 
        \prod\limits_{r \in R[\Gamma]} w^{-\frac{c}{12}(w-1)} 
        |a_r|^{-\frac{c}{12}(1-\frac{1}{w})}\prod\limits_{p \in P[\Gamma]}
        |b_p|^{-\frac{c}{6}} \, ,
    \end{align}
    where $\eta$ parametrises the metric of $X$ (see eq.~\eqref{eq:Xeta}), 
    $R[\Gamma]$ is the set of ramification points of $\Gamma$, $P[\Gamma]$ is 
    the set of poles of $\Gamma$ in $\Sigma \setminus \{\infty\}$ and the 
    complex numbers $a_r$ and $b_p$ are defined via the local behaviour
    \begin{align}\label{eq_a_r_b_p}
        \Gamma(r+t) = \Gamma(r) + a_r t^w + O(t^{w+1}) \quad \text{and} \quad 
        \Gamma(p+t) = \frac{b_p}{t} + O(1)
    \end{align}
    of the covering map $\Gamma$ near the ramification points and poles. While 
    formula \eqref{eq:S_L_boundary_integral} was derived in a few examples in the 
    original work \cite{Lunin:2000yv} an explicit, fully detailed derivation was only recently published in \cite{Hikida:2020kil}. 
    While the 
    latter uses somewhat different methods, the derivation we spell out in Appendix 
    \ref{app:details_regularisation} follows exactly the route of \cite{Lunin:2000yv}. It 
    will also turn out to directly extend to the interface case. Before we go there, let us 
    illustrate the Lunin-Mathur construction through an example. 
        
    \paragraph{Example: Two-point functions.} As an application of
    eq.~\eqref{eq:S_L_boundary_integral}, let us compute the two-point functions of 
    two bulk fields $\sigma_w$ with identical index $w_1 = w = w_2$ in the symmetric 
    product orbifold. Without loss of generality, we can insert the first field at 
    $z_1 = 0$ and then denote the insertion point of the second twist field by $z_2 = 
    z$. In this setup, only a single class of coverings contributes. 
    We choose the representative
    \begin{align}\label{eq:no_interface_2_cover}
        \Gamma(t) = z \frac{t^w}{t^w-(t-1)^w}.
    \end{align}
    This covering map has two ramification points $R[\Gamma]=\{0,1\}$, around which
    it expands as 
    \begin{align}
        \Gamma(t) = 0 + z (-1)^{w+1} t^w + O(t^{w+1}) && \text{and} &&\Gamma(1+t) 
        = z + z t^w + O(t^{w+1}).
    \end{align}
    Moreover, $\Gamma$ has $w-1$ finite poles which are located at the zeroes of the 
    denominator, 
    \begin{align}
        t_k := \left(1-e^{2\pi i \tfrac{k}{w}}\right)^{-1} \  
    \end{align}
    where $k$ runs from $k=1$ to $k=w-1$, i.e.  $P[\Gamma]=\{t_1, \dots, t_{w-1}\}$.  
    The residues at these poles are easy to determine, 
    \begin{align}\label{eq:b_k_for_twopoint}
        b_k = \frac{z}{4 w \sin^2(\tfrac{k \pi}{w}) }.
    \end{align}
    Now we have assembles all the data we need in order to evaluate the formula 
    ~\eqref{eq:S_L_boundary_integral}. The result is given by 
    \begin{align}
        Z_{1}^\Sigma = |z|^{-2 \Delta^{(w)}} w^\frac{c}{3} \eta^{-\frac{c}{3}(w-1)}
    \end{align} 
    with
    \begin{align}\label{eq:Delta_N}
        \Delta^{(w)} = \frac{c}{12}\left(w-\frac{1}{w}\right) .
    \end{align}
    The last ingredient we need in order to evaluate the unnormalised correlator 
    \eqref{eq:correlator_to_evaluate} is the number of Deck transformations, which 
    is given by 
    \begin{align}
        |\text{Deck}[\Gamma]| = w (N-w)! \ . 
    \end{align}
    Hence, the the unnormalised two-point function is given by
    \begin{align}\label{eq:bulk_two_point}
        \langle \sigma_w(0) \sigma_w(z)\rangle_{\text{Sym}^N(\mathcal{M})}^{\text{unnormalised}} = \frac{ w^\frac{c}{3} \eta^{-\frac{c}{3}(w-1)} }{w(N-w)!} |z|^{-2 \Delta^{(w)}}.
    \end{align}
    We use the term ``unnormalised'' in the sense of definition \eqref{eq:norm_vs_unnorm}. 
    In particular, the normalised two-point function is 
    \begin{align}
        \langle \sigma_w(0) \sigma_w(z)\rangle_{\text{Sym}^N(\mathcal{M})} = N!\langle \sigma_w(0) \sigma_w(z)\rangle_{\text{Sym}^N(\mathcal{M})}^{\text{unnormalised}} 
    \end{align}
    and not just $|z|^{-2\Delta^{(w)}}$.
        
    \subsection{The interface generalisation}\label{sec:interface_LM}
    Let us now turn to the interface generalisation of the Lunin-Mathur construction.
    The interface generalisation introduces two modifications to the evaluation 
    of correlators which we detail in the next two paragraphs. The first 
    modification arises from the fact that we now need to evaluate the Liouville
    action on surfaces with boundaries which requires in particular to add an 
    an additional boundary term, see eq.~\eqref{eq:Liouville_bdry}. The second 
    modification is that bulk covering maps are replaced by interface covering 
    maps. This can be dealt with by applying the method of images. After generally discussing 
    disk correlators in the third paragraph of the section, we conclude by 
    computing several example correlation functions.
    
    \paragraph{Boundary Liouville action.}
    As in Section \ref{sec:bulk_LM}, we assume that there is a reference 
    geometry $\hat \Sigma$ such that the partition function $Z_1^{\hat \Sigma}$ 
    on $\hat \Sigma$ is known. As before, the partition function $Z_1^\Sigma$ on $\Sigma$ 
    with the induced metric \eqref{eq:gtoghat} is then given by eq.~\eqref{eq:Z1toZ1hat}. The difference to the bulk case is 
    that the Liouville action now also has a boundary contribution. Concretely,
    \begin{align}\label{eq:Liouville_bdry}
        S_L[\phi] = \frac{c}{96 \pi} \int\limits_\Sigma d^2 t \sqrt{g_{\hat\Sigma}} \left( \partial_\mu \phi \partial_\nu \phi g^{\mu \nu}_{\hat\Sigma} + 2 R^{{\hat\Sigma}}\phi\right) + \frac{c}{24 \pi} \int\limits_{\partial \Sigma} \sqrt{\gamma_{\hat \Sigma}} K^{\hat \Sigma} \phi,
    \end{align}
    where $\gamma$ is the induced metric on the boundary $\partial \Sigma$ of the 
    covering surface and $K^{\hat \Sigma}$ is the extrinsic curvature of the boundary.
    
    \paragraph{The method of images.}  
    In the case where $X$ is a disk, we can use the method of images in order to reduce the discussion of interface covering maps $\Gamma: \Sigma \rightarrow X$ to that of coverings $\Gamma'$ of the sphere by the Schottky\footnote{The Schottky double is a closed Riemann surface obtained by taking two copies $\Sigma$ and $\overline{\Sigma}$ of the Riemann surface $\Sigma$ and glueing them along the boundary.} double $\Sigma \# \overline{\Sigma}$.
    The restriction of the doubled map to $\Sigma$ coincides with $\Gamma$. 
    On $\overline{\Sigma}$, the covering $\Gamma'$ is given by $\Gamma'(\bar z) := \overline{\Gamma(z)}$.
    The doubled map $\Gamma'$ has two copies of each interior ramification point of $\Gamma$. 
    Additionally, each boundary ramification point with index $\omega$ leads to a single ramification point with index $w = \omega$ for $\Gamma'$.
    Below, we use the method of images for two purposes. 
    First, it allows us to construct interface covering maps as restrictions of bulk covering maps.   
    Second, it allows us to evaluate the bulk part of the Liouville action \eqref{eq:Liouville_bdry} by applying eq.~\eqref{eq:S_L_boundary_integral} to $\Gamma'$.
    
    \paragraph{The hemisphere.}
    In the case of coverings of the upper hemisphere in $\overline{\mathbb{C}}_\eta$ by another hemisphere, where we take the reference metric to again be of the type $ds_\delta^2$, the extrinsic curvature vanishes along the entire boundary of the covering space. 
    Thus, only the bulk part contributes to the Liouville action.
    In Appendix \ref{app:details_regularisation_disk}, we evaluate this bulk part using the method of images. 
    This leads to the interface generalisation 
    \begin{align}\label{eq:Z_1_Sigma_Interface_master}
        Z_1^\Sigma = \eta^{-\frac{c}{6}(N_-+N_+-1)}\left( \prod\limits_{r \in R_B[\Gamma]} \hspace{-0.2 cm} w^{w-1} a_r^{1-\frac{1}{w}}\hspace{-0.2 cm}\prod\limits_{p \in P_B[\Gamma]} \hspace{-0.2 cm} b_p^{2}\right)^{-\frac{c}{12}} \left( \prod\limits_{r \in R_I[\Gamma]} \hspace{-0.2 cm} \omega^{\omega-1} a_r^{1-\frac{1}{\omega}}\hspace{-0.2 cm}\prod\limits_{p \in P_I[\Gamma]} \hspace{-0.2 cm} b_p^{2}\right)^{-\frac{c}{24}}  
    \end{align}
    of eq.~\eqref{eq:S_L_boundary_integral} for interface coverings of the upper half plane by itself. 
    Here, $N_+$ and $N_-$ are the degrees of the covering map $\Gamma$ on the upper and lower half plane. 
    The sets $R_{B}[\Gamma]$ and $R_{I}[\Gamma]$ contain the bulk and interface ramification points of $\Gamma$. 
    Likewise, the sets $P_B[\Gamma]$ and $P_I[\Gamma]$ contain all complex and real poles, except for $\infty$ (which we however continue to w.l.o.g.~assume to be a pole).
    The coefficients $a_r$ and $b_p$ for interface ramification points and poles are defined as we defined the analogous bulk coefficients in eq.~\eqref{eq_a_r_b_p}.
    
    \paragraph{Example: Bulk one-point functions.}
    Let us consider the one-point function 
    \begin{align}
        \langle \mathcal{I}^{(P)}_{|a_\pm\rangle}
        \sigma_w(z) \rangle^{\text{unnormalised}}_{\text{Sym}^{N_-|N_+}(\mathcal{M})}
    \end{align}
    of a bulk twist field $\sigma_w(z)$ inserted on the upper half plane, with the interface $\mathcal{I}^{(P)}_{|a_\pm\rangle}$ inserted along the real line. 
    The correlation function can be non-trivial only if $N_+ -P \ge w$. 
    In that case, it evaluates to 
    \begin{align}
        \langle \mathcal{I}^{(P)}_{|a_\pm\rangle}
        \sigma_w(z) \rangle^{\text{unnormalised}}_{\text{Sym}^{N_-|N_+}(\mathcal{M})} = \frac{g_-^{N_--P}g_+^{N_+-P-w} Z^{(w)}_{|a_+\rangle}}{P! (N_--P)!(N_+-P-w)! w},
    \end{align}
    where $g_\pm$ are the $g$-functions of $|a_\pm\rangle$ and $Z^{(w)}_{|a_\pm\rangle}$ captures the contribution of a $w$-fold cover of the upper half plane by itself with a ramification point at $z$. 
    To compute $Z^{(w)}_{|a_\pm\rangle}$, we choose the explicit covering map 
     \begin{align}
        \Gamma(t) :&= z \frac{(it-1)^w}{(it-1)^w - (it +1)^w}+ \bar z \frac{(-it-1)^w}{(-it-1)^w - (-it +1)^w} .
    \end{align}
    The way $\Gamma$ is written above makes manifest two important properties, namely
    \begin{align}
        \Gamma(\bar t) = \overline{\Gamma(t)} \quad \text{and} \quad \Gamma(i) = z.
    \end{align}
     A more economical way of writing $\Gamma$ is
    \begin{align}
        \Gamma(t) =  \frac{z (it-1)^w-\bar z (it+1)^w}{(it-1)^w - (it +1)^w}.
    \end{align}
    The coefficient associated with the ramification point $i$ is
    \begin{align}
        a_i = \frac{z - \bar z}{(2i)^w}. 
    \end{align}
    The poles and residues are
    \begin{align}
        P[\Gamma] = \{\cot(\tfrac{k\pi}{w})\}_{k=1}^{w-1} && \text{and} && b_k = \frac{i(z- \bar z)}{2w \sin^2(\tfrac{k \pi}{w})} \,.
    \end{align}
    Plugging this into eq.~\eqref{eq:Z_1_Sigma_Interface_master} gives
    \begin{align}
        Z_{|a_+\rangle}^{(w)} = g_+ \eta^{- \tfrac{c}{6} (w-1)} |z-\bar z|^{-\Delta^{(w)}} w^{\tfrac{c}{6}}
    \end{align}
   and therefore, 
   \begin{align}
        \langle \mathcal{I}^{(P)}_{|a_\pm\rangle}
        \sigma_w(z) \rangle^{\text{unnormalised}}_{\text{Sym}^{N_-|N_+}(\mathcal{M})} = \frac{g_-^{N_--P}g_+^{N_+-P-(w-1)}  \eta^{- \tfrac{c}{6} (w-1)} w^{\tfrac{c}{6}}}{P! (N_--P)!(N_+-P-w)! w} |z-\bar z|^{-\Delta^{(w)}}.
    \end{align}
    
    \paragraph{Example: Interface two-point functions.}
    The evaluation of the two-point functions
    \begin{align}
        \langle \mathcal{I}^{(P)}_{|a\rangle} \sigma_\omega^{RR}(0) \sigma_\omega^{LL}(x) \rangle_{\text{Sym}^{N_-|N_+}(\mathcal{M})}^{\text{unnormalised}} && \text{and} && \langle \mathcal{I}^{(P)}_{|a\rangle}\sigma_\omega^{RL}(0) \sigma_\omega^{RL}(x) \rangle_{\text{Sym}^{N_-|N_+}(\mathcal{M})}^{\text{unnormalised}}
    \end{align}
    directly reduces to that of the bulk two-point function \eqref{eq:bulk_two_point} for even and odd $w=\omega$ respectively via the method of images. 
    We thus deduce that the two correlators above are given by\footnote{Assuming of course a sufficiently large number of sheets $N_-$ and $N_+$ on the upper/lower hemisphere as well as a sufficient number of reflecting and transmitting boundary conditions to ensure that the correlator is non vanishing.}
    \begin{align}
        \langle \mathcal{I}^{(P)}_{|a\rangle} \sigma_\omega^{RR}(0) \sigma_\omega^{LL}(x) \rangle_{\text{Sym}^{N_-|N_+}(\mathcal{M})}^{\text{unnormalised}} = \frac{ g^{N_-+N_+-2P+1}\omega^\frac{c}{6} \eta^{-\frac{c}{6}(\omega-1)} }{(N_-- P)! (N_+-P)! (P - \tfrac{\omega}{2})!} |x|^{-2 \Delta_{\omega}^{AB}},
    \end{align}
     where the reference point for the interface is $-1$ and, with the same reference point, 
    \begin{align}
        \langle \mathcal{I}^{(P)}_{|a\rangle} \sigma_\omega^{RL}(0) \sigma_\omega^{RL}(x) \rangle_{\text{Sym}^{N_-|N_+}(\mathcal{M})}^{\text{unnormalised}} = \frac{ g^{N_-+N_+-2P}\omega^\frac{c}{6} \eta^{-\frac{c}{6}(\omega-1)} }{(N_--P)! (N_+-P-1)! (P- \tfrac{\omega-1}{2})!} |x|^{-2 \Delta_{\omega}^{AB}}.
    \end{align}
    These results follow from applying \eqref{eq:Z_1_Sigma_Interface_master} to the covering map \eqref{eq:no_interface_2_cover} with real $z = x$.
    Here,
    \begin{align}
        \Delta_{\omega}^{AB} = \frac{c}{24}\left(\omega - \frac{1}{\omega}\right),
    \end{align}
    which agrees with the special case $\Delta = 0$ of eq.~\eqref{eq:interface_dimensions}. 
    For $\omega = 2$, the interface two-point function was essentially already computed in \cite{Martinec:2022ofs}, see Section 5.3 of that reference.
    
    \paragraph{Example: Interface four-point functions.}

    The disk contribution to general interface four-point functions 
    \begin{align}
        \langle \mathcal{I}^{(P)}_{|a\rangle}\sigma_{\omega_1}^{A_1,B_1}(x_1) \sigma_{\omega_2}^{A_2,B_2}(x_2) \sigma_{\omega_3}^{A_3,B_3}(x_3) \sigma_{\omega_4}^{A_4,B_4}(x_4) \rangle_{\text{Sym}^{N_-|N_+}(\mathcal{M})}^{\text{unnormalised}}
    \end{align}
    can be evaluated using section 2 of the notebook \texttt{CoveringMaps.nb} provided in the ancillary files of this publication.
    The corresponding covering maps, via the method of images, reduce to covering maps of the sphere in the absence of an interface, which satisfy the extra constraint $\Gamma(s) = \overline{\Gamma(\bar s)}$.
    The latter can be computed by first using the approach described in Appendix \ref{app:compute_cover} to compute generic covering maps and then restricting to the subset that satisfies the additional reality constraint.
    By applying $SL(2;\mathbb{R})$ transformations of the upper hemisphere, we can w.l.o.g.~ensure that $x_1 = 0$, $x_3 = 1$ and $x_4 = -1$ with corresponding ramification points $s_i = x_i$ for $i \neq 2$.
    The value of $s_2$ on the other hand is a non trivial function of the parameter $x_2$. 
    Sections 2.2 and 2.3 of the notebook automatically generate plots from the covering maps, which resemble the diagrams discussed in Section \ref{sec:PRR}.
    Using eq.~\eqref{eq:Z_1_Sigma_Interface_master} it is straight forward to evaluate correlation functions by simply determining the poles as well as the local expansions around ramification points of the explicitly computed covering maps. 
    For eq.~\eqref{eq:Z_1_Sigma_Interface_master} to be applicable, one has to, however, ensure that $\Gamma(\infty) = \infty$ i.e.~pick an arbitrary pole and conformally map it to $\infty$. 
    This operation is performed in Section 2.4.~of the notebook. 
    
    \paragraph{Example: Interface-Interface-Bulk three-point functions.} To study in\-ter\-face-in\-ter\-face-bulk three-point functions, we fix the SL$(2,\mathbb{R})$ symmetry of the disk by mapping the two interface operators to $0$ and $1$ and the real part of the bulk operator to $\tfrac{1}{2}$. 
    The remaining kinematical freedom consists then of a single positive real number $\chi$, the imaginary part of the position of the bulk twist field. 
    Hence, the correlators take the form
    \begin{align}
        \langle \mathcal{I}^{(P)}_{|a\rangle}\sigma_{\omega_1}^{A_1,B_1}(0) \sigma_{w}(\tfrac{1}{2} + i \chi) \sigma_{\omega_2}^{A_2,B_2}(1) \rangle_{\text{Sym}^{N_-|N_+}(\mathcal{M})}^{\text{unnormalised}}
    \end{align}
    and are computable by determining the bulk covering maps of the sphere with ramification points $0$ and $1$ of index $\omega_1$ and $\omega_2$ that are mapped to $0$ and $1$, as well as ramification points at $\tfrac{1}{2} \pm i s$ for some real number $s$, which map to $\tfrac{1}{2} \pm i \chi$ with ramification index $w$. 
    The maps that contribute non-trivially to the correlator are again those that satisfy $\Gamma(z) = \overline{\Gamma(\bar z)}$. 
    They are computed and visualised in Section 3 of the \texttt{CoveringMaps.nb} notebook.
    
    \paragraph{Example: Bulk two-point functions.}
    The last type of correlators whose evaluation reduces to the computation of sphere coverings with four ramification points are bulk two-point functions. 
    Again, the kinematical freedom reduces to a single real number, which we choose to be the imaginary part of the insertion points of the two bulk fields after using the SL$(2,\mathbb{R})$ symmetry to fix the real parts to $-1$ and $1$ and sending the intersection of the line through the two insertion points with Im$(z)=0$ to $\infty$. 
    The resulting correlator
    \begin{align}
        \langle \mathcal{I}^{(P)}_{|a\rangle}\sigma_{w_1}(-1 + i \chi) \sigma_{w_2}(1 + i \chi) \rangle_{\text{Sym}^{N_-|N_+}(\mathcal{M})}^{\text{unnormalised}}
    \end{align}
    can be computed from covering maps with ramification points at $-1 \pm i s$ and $1 \pm i s$ for some $s>0$, which are mapped to $-1 \pm i \chi$ and $1 \pm i \chi$ with ramification index $w_1$ and $w_2$. 
    Section 4 of the \texttt{CoveringMaps.nb} notebook computes and visualises the relevant $\Gamma(z) = \overline{\Gamma(\bar z)}$ maps.
    
\section{String theory and the interface grand-canonical ensemble}\label{sec:string_theory_&_GCE}

    So far, we have described correlation functions at fixed $N_\pm$ and $P$. 
    The purpose of this section is to discuss a grand-canonical ensemble in which the parameters $N_\pm$ and $P$ are averaged with a fixed chemical potential.
    As reviewed in Subsection \ref{sec:holographic_motivation}, studying such an ensemble is strongly motivated by holography.
    In fact, super string theory on AdS$_3 \times S_3 \times \mathbb{T}^4$ at minimal pure NS-NS flux $k=1$ is known to be precisely dual to a grand-canonical ensemble of symmetric product orbifolds whose construction we review in Subsection \ref{sec:bulkgce}. 
    The key result of this section is a novel grand-canonical ensemble of interfaces introduced in Subsection \ref{sec:interfacegce}. 
    Using this new construction, we show 
    that the grand-canonical ensemble of correlation functions studied in this paper structurally matches the genus expansion of string scattering amplitudes.
    We discuss the expectation values of $N_\pm$ and $P$ at the end of Subsection \ref{sec:saddle_point}, and show the leading-order agreement to the canonical ensemble result shown in Sec.~\ref{sec:CanonicalLargeN}.
    
    \subsection{Holographic motivation}\label{sec:holographic_motivation}
        Pure NSNS AdS$_3 \times S^3 \times \mathbb{T}^4$ can be obtained from $\mathbb{R}^6 \times \mathbb{T}^4$ as the near horizon limit of $Q_1$ fundamental strings and $Q_5$ NS5 branes, with four directions of the latter wrapping $\mathbb{T}^4$.
        String perturbation theory around this pure NSNS background is described in the hybrid formalism by a world sheet theory based on the PSU$(1,1|2)_k$ WZNW model.
        The level $k$ of the model can be directly identified with the number $Q_5$ of the brane construction.
        However, the naive expectation that the string coupling $g_s$ is fixed in terms of $Q_1$ is not realised. 
        Instead of being determined by the expectation values of fields in the target space, the string coupling is a continuous parameter of the world sheet theory\footnote{For more details consult e.g.~\cite{Giveon:1998ns,Kutasov:1999xu,Giveon:2001up,Kim:2015gak} on pure NSNS AdS$_3$ strings in the RNS formalism for $Q_5 \ge 2$ and \cite{Eberhardt:2020bgq,Eberhardt:2021jvj} on the special case $Q_5=1$ in the hybrid formalism.}.
        This parameter can be interpreted as a chemical potential controlling the averaging over different values of $Q_1$ in a grand-canonical ensemble.

        Since string perturbation theory therefore does not compute scattering amplitudes in a fixed $Q_1$ background, but in the grand-canonical ensemble, one should expect the space time CFT description of these amplitudes to analogously be given in terms of correlation functions in a grand-canonical ensemble.
        For minimal NSNS flux $k=1$, this expectation has been confirmed in great detail.
        Concretely, it was established \cite{Gaberdiel:2018rqv,Eberhardt:2019qcl,Eberhardt:2019ywk,Eberhardt:2020bgq, Hikida:2023jyc} that string theory on AdS$_3 \times S^3 \times \mathbb{T}^4$ at $k=1$ is dual to a grand-canonical ensemble of free symmetric orbifold theories $\text{Sym}^{Q_1}(\mathbb{T}^4)$. 
        Section 3 of Reference \cite{Aharony:2024fid} provides a detailed description of this grand-canonical ensemble, which we partially review in the next subsection, as well as a broader discussion of the string theoretic context that we have sketched above. 

        Reference \cite{Harris:2025wak} provided evidence strongly suggesting that AdS$_2$ branes in the $k=1$ background are similarly not dual to a single interface, but rather an ensemble of the interfaces $\mathcal{I}^{(P)}$. 
        We conjecture this to be true not only in the sense that $N_\pm$ need to be averaged over since the bulk theory is not a fixed $N$ symmetric product orbifold, but also in the sense that $P$ is averaged over all its possible values for each fixed $N_\pm$. 
        
        To qualitatively motivate this expectation, consider that, in string theory, the AdS$_2$ branes are realised by the insertion of a probe string on which the fundamental strings which create the AdS$_3 \times S^3 \times \mathbb{T}^4$ background can end \cite{Bachas:2001vj}.
        The fixed $P$ interface $\mathcal{I}^{(P)}$ corresponds to a setup where $P$ parallel fundamental strings go through the probe string and another $Q_1^- - P$ and $Q_1 ^+-P$ end on it extending into either of the two directions parallel to the other fundamental strings and orthogonal to the probe.
        As recalled above, the issue that the PSU$(1,1|2)_1$ world sheet theory at a first glance cannot account for the parameter $Q_1$, or for that matter $Q_1^\pm$, is resolved by understanding that it describes a grand-canonical ensemble.
        Since the AdS$_2$ boundary states of the world sheet theory do not have a parameter describing the value of $P$, it is natural to also expect this parameter to be controlled by a chemical potential, which is set by to the only available world sheet parameter -- the string coupling.
        While the result for the torus partition function obtained in Reference \cite{Harris:2025wak} is consistent with this expectation, it is not possible to directly infer the precise relation between the chemical potential controlling $P$ and the string coupling from it.
        The purpose of this section is to go beyond these prior results by fixing the interface chemical potentials in terms of the string coupling, and showing that the ensemble defined in this way produces correlators that structurally reproduce the string genus expansion. 
        
    \subsection{The bulk grand-canonical ensemble}\label{sec:bulkgce}

        Among the first attempts to connect symmetric product orbifolds to potentially dual string theories it was common to consider correlators of twist fields $\tilde{\sigma}_w$ with the normalisation
        \begin{align}
            \tilde{\sigma}_w = \sqrt{\tfrac{(N-w)! w}{N!}} \sigma_w
        \end{align}
        at fixed $N$.
        Here, $\sigma_w$ is the twist field with the canonical normalisation used in Sec.~\ref{sec:symm_orb}.
        At leading order in $N \gg 1$, the connected part of these correlation functions indeed shares some features of scattering amplitudes in string perturbation theory.
        As observed in  \cite{Lunin:2000yv,Pakman:2009zz}, for instance,
        the contribution of genus $g$ covering spaces to a connected correlation function of bulk twist fields scales as
        \begin{align}
            \left\langle \prod\limits_{i=1}^{n_c} \tilde\sigma_{w_i}\right\rangle_{\text{conn}}^{N,g} \sim N^{1 - g - \frac{n_c}{2}}
        \end{align}
        with $N$.
        Upon the identification $g_s^2 \sim N^{-1}$, this matches the scaling of genus $g$ contributions to string scattering amplitudes with $g_s$.
        However, this proposal breaks down as soon as subleading $N$ corrections or disconnected contributions are considered. 
        
        These issues can be fixed by working in the grand-canonical ensemble, as argued in Sec.~3.3 of \cite{Aharony:2024fid}.
        To arrive at this conclusion, consider the partition function
        \begin{align}\label{eq:def_ZNX}
            Z_N^X[J] := \left\langle \exp\left( \sum\limits_{w=1}^\infty \int d^2z \, g_s^w J_w(z) \sigma_w(z) \right) \right\rangle_{\text{Sym}^N(\mathcal{M})}^{\text{unnormalised}}
        \end{align}
        of $\text{Sym}^N(\mathcal{M})$ with sources $J_1, \dots, J_w$ on a Riemann surface $X$ of genus $G$ defined using the unnormalised symmetric orbifold correlators of eq.~\eqref{eq:norm_vs_unnorm}. 
        Here, $g_s$ is the parameter that turns out 
        to control the genus expansion of symmetric orbifold  correlators. It does not only multiply the sources in the definition \eqref{eq:def_ZNX} but, more importantly, also parametrises the chemical potential $\mu$ which controls $N$ in the grand-canonical ensemble.
        More concretely, the grand-canonical partition function on $X$ is defined as
        \begin{align}\label{eq:def_ZgsX}
            Z_{g_s}^X[J] := \sum\limits_{N=0}^\infty \mu^{N} Z_N^X[J] \quad\quad 
            \text{where} \quad\quad \mu := g_s^{2G-2}. 
        \end{align}
        The key feature of this ensemble is that grand-canonical correlation functions
        \begin{align}
            \left\langle \prod\limits_{i=1}^{n_c} \sigma_{w_i}(z_i) \right\rangle_{g_s}^X := 
            \left[\prod\limits_{i=1}^{n_c}\frac{\delta}{\delta J_{w_i}(z_i)} \right] 
            \left. \frac{Z_{g_s}^X[J]}{Z_{g_s}^X[0]} \right|_{J=0}
        \end{align}
        have a genus expansion that, to all orders in $g_s$ and for both connected and disconnected contributions, structurally matches the genus expansion of string perturbation theory with string coupling $g_s$. 
        This can be shown in three simple steps. 
        
        First, we expand the exponential in eq.~\eqref{eq:def_ZNX} and express the resulting correlation functions in terms of covering maps, which leads to
        \begin{align}\label{eq:expand_sum}
            Z_N^X[J] = \sum\limits_{n=0}^\infty \frac{1}{n!}\left( \sum\limits_{w=1}^\infty \int d^2z J_w(z) \right)^n \sum\limits_{\Gamma} Z_{g_s}[\Gamma] \, .
        \end{align}
        The sum runs over all coverings $\Gamma$ of $X$ with the appropriate\footnote{There is some slight abuse of notation in eq.~\eqref{eq:expand_sum}. In a more literal formulation, we should have expanded the $n^\text{th}$ power of sums over source terms, into a sum of product of sources, corresponding to twist fields whose order and insertion location determines the ramification points.} ramification points and 
        \begin{align}\label{def:ZgsGamma}
            Z_{g_s}[\Gamma] := \frac{g_s^{w[\Gamma]}}{\text{Deck}[\Gamma]} \langle \mathds{1} \rangle_\mathcal{M}^{\Sigma},
        \end{align}
        where $w[\Gamma]$ is the total ramification index of $\Gamma$ i.e.~the sum of the indices of all ramification points.
        Each covering space that is involved has 
        \begin{align}
            c_\Gamma = \sum\limits_{i=1}^k n_i
        \end{align}
        connected components falling into $k$ isomorphism classes of sizes $n_1, \dots n_k$.
        The restriction $\Gamma_i$ of $\Gamma$ to any element of the $i^\text{th}$ isomorphism class has some degree $N_i$ and is associated with the contribution $Z_{g_s}[\Gamma_i]$ to the correlation function.
        In terms of this decomposition,
        \begin{align}
            \text{Deck}[\Gamma] = \prod\limits_{i=1}^k S_{n_i} \ltimes \text{Deck}[\Gamma_i]^{n_i}
        \end{align}
        and thus $Z_{g_s}[\Gamma]$ can be expressed as
        \begin{align}
            Z_{g_s}[\Gamma] =\prod\limits_{i=1}^k \frac{1}{n_i!}(Z_{g_s}[\Gamma_i])^{n_i} .
        \end{align} 
        Therefore, 
        \begin{align}\label{eq:step1}
            Z_N^X[J] = \sum\limits_{n=0}^\infty \frac{1}{n!}\left( \sum\limits_{w=1}^\infty \int d^2z J_w(z) \right)^n \sum\limits_{\Gamma}  \prod\limits_{i=1}^k \frac{1}{n_i!}(Z_{g_s}[\Gamma_i])^{n_i}\, .
        \end{align}
        Note that,  since the overall degree of $\Gamma$ is $N$, $n_i$ and $N_i$ are constrained by
        \begin{align}
            N = \sum\limits_{i=1}^k n_i N_i .
        \end{align}
        
        Next, observe that since $N$ is no longer fixed in the grand-canonical ensemble, the sum over $n_i$ and $N_i$ becomes unconstrained for $Z_{g_s}^X$. 
        Upon inserting eq.~\eqref{eq:step1} into the definition of $Z_{g_s}$, one therefore finds that the grand-canonical partition function is an exponential
        \begin{align}\label{eq:step2}
            Z_{g_s}^X[J] = \exp\left( \sum\limits_{n=0}^\infty  \frac{1}{n!}\left( \sum\limits_{w=1}^\infty \int d^2z J_w(z) \right)^n \sum\limits_{\Gamma \text{ conn.}} \mu^{N} Z_{g_s}[\Gamma] \right)
        \end{align}
        of a sum over connected coverings with covering maps $\Gamma$ of arbitrary degree $N$.

        Finally, we infer from Definition \eqref{def:ZgsGamma} that 
        \begin{align}
            Z_{g_s}[\Gamma] = g_s^{w[\Gamma]} Z[\Gamma] && \text{with} && w[\Gamma]=\sum\limits_{i=1}^{n_c} w_i && \text{and} && Z[\Gamma] := Z_1[\Gamma].
        \end{align}
        Thus, the power of $g_s$ weighing the contribution of $\Gamma$ to $Z_{g_s}$ can be expressed using the Riemann-Hurwitz formula \eqref{eq:RH} as
        \begin{align}
            N(2G-2) + \sum\limits_{i=1}^{n_c} w_i = - 2 + 2g + n_c,
        \end{align}
        where $G$ is the genus of $X$. 
        Putting all this together we obtain the following formula for the grand-canonical partition function 
        \begin{align}
            Z_{g_s}^X[J] = \exp\left( \sum\limits_{n=0}^\infty  \frac{1}{n!}\left( \sum\limits_{w=1}^\infty \int d^2z J_w(z) \right)^n \sum\limits_{\Gamma \text{ conn.}} g_s^{-2 + 2g + n_c}Z[\Gamma]  \right) \, .
        \end{align}
        We hence conclude that the logarithm of the grand-canonical partition function is the sum over all connected coverings, or equivalently, connected PRR diagrams. 
        The individual contributions to this sum 
        are weighted with the parameter $g_s$ taken to the power of (minus) the Euler characteristic of the surface $\Sigma$ on which 
        the PRR diagram is drawn.
        In the equation, we have expressed the Euler characteristic in terms of number $n_c$ of bulk vertices of the diagram or punctures of the surface $\Sigma$ and the genus 
        $g$ of $\Sigma$. 
        Correlation functions in the grand-canonical ensemble therefore possess a topological expansion which structurally resembles the 't Hooft large $N$ expansion and, crucially, string perturbation theory. 
        This the result that we aimed to establish. 
        
    \subsection{The grand-canonical ensemble for interfaces}\label{sec:interfacegce}

        As we have shown in Section \ref{sec:CanonicalLargeN}, it is possible to fix the normalisation of the interface changing operators of our interfaces in such a way that the corresponding connected correlation functions, to leading order in a $1/N$ expansion, match the structure of string perturbation theory. 
        However, this approach leads to the same issues as the bulk fixed $N$ genus expansion, i.e.~the match is restricted to the leading large $N$ contribution. 
        In the three paragraphs of this section, we first introduce the interface grand-canonical ensemble, then show that our definition remedies this deficiency, leading to a fully consistent topological expansion, and finally reinterpret the results of Section \ref{sec:CanonicalLargeN} as a saddle-point approximation of the full ensemble.
        
        \paragraph{Definition of the interface grand-canonical partition function.}
        This paragraph constructs an interface analogue of the grand-canonical ensemble reviewed in Sec.~\ref{sec:bulkgce}.
        In Sec.~\ref{sec:interfaces}, we defined the interface $\mathcal{I}^{(P)}_{|a\rangle}$ as interpolating between $\text{Sym}^{N_-}(\mathcal{M})$ and $\text{Sym}^{N_+}(\mathcal{M})$ on Riemann surfaces $X_\pm$ of genus $G_\pm$ that are glued along their boundaries to form a closed Riemann surface $X$ of genus $G$.
        In order to study correlators with a single interface insertion, let us restrict to $X_\pm$ with boundaries
        \begin{align}
            \partial X_- \cong \partial X_+ \cong \mathbb{S}^1.
        \end{align}
        In this setup, we use the notation introduced in Section \ref{sec:interfaces} to define
        \begin{align}\label{eq:ZXpmNpmJ}
            Z^{X_-,X_+}_{N_\pm,P}[J] := \left\langle \mathcal{I}^{(P)}_{|a\rangle}\exp\left(\sumint g_s^w J_w \sigma_w + \sumint \sqrt{g_s^\omega} J_\omega^{AB} \sigma_\omega^{AB}\right)\right\rangle_{\text{Sym}^{N_-|N_+}(\mathcal{M})}^{\text{unnormalised}}
        \end{align}
        with
        \begin{align}
        \label{eq:sourcebulktwist}
            \sumint g_s^w J_w \sigma_w := \sum\limits_{w=1}^{\infty} \int\limits_{X} d^2z g_s^w J_w(z) \sigma_w(z)
        \end{align}
        and
        \begin{align}
        \label{eq:sourceinterfacetwist}
            \sumint \sqrt{g_s^\omega} J_\omega^{AB} \sigma_\omega^{AB} := \hspace{-0.5 cm}\sum\limits_{A,B \in \{L,R\}}\sum\limits_{\omega} 
            \,\, \int\limits_{\partial X_\pm} dx \sqrt{g_s^\omega} J_\omega^{AB}(x) \sigma_\omega^{AB}(x)
        \end{align}
        where $\omega$ runs over $2\mathbb{Z}_{> 0}-1+\delta_A^B$. Given this definition \eqref{eq:ZXpmNpmJ} of the interface partition functions we now introduce its grand-canonical 
        counterpart which we define as
        \begin{align}\label{eq:ZXpmJ}
            Z^{X_-,X_+}_{g_s}[J] := \sum\limits_{N_\pm=0}^\infty\sum\limits_{P=0}^{\text{min}(N_-,N_+)} \mu^{P} \nu_-^{N_- - P} \nu_+^{N_+ - P} Z_{N_\pm,P}^{X_-,X_+}[J].
        \end{align}
        It involves three fugacities $\mu, \nu_\pm$ which are associated to the integers $N_\pm$ that determine the symmetric product orbifolds on both sides of the interface and on the degree of transmissivity $P$ that characterizes the interface itself. We suggest to control these three fugacities through a single loop counting 
        parameter $g_s$ as 
        \begin{equation} \label{eq:munug}
        \mu:=g_s^{2G-2} \quad , \quad \nu_\pm:=g_s^{2G_\pm-1} \  
        \end{equation}
        where $G$ and $G_\pm$ denote the genus of the 
        surface $X = X_+ \cup_{\mathbb{S}^1} X_-$ and of 
        the two components $X_\pm$ that $X$ is glued from.

    \paragraph{The string genus expansion.}\label{sec:string_genus_gce}
        We now show that the $g_s$ expansion of correlation functions in the grand-canonical ensemble \eqref{eq:ZXpmJ} has the structure of a string genus expansion describing the scattering of open and closed strings in the presence of a  $D$-brane.  

        As in the bulk case, our first step is to expand the exponential in the fixed $N_\pm$ partition function \eqref{eq:ZXpmNpmJ} and to express the result in terms of covering maps. 
        This leads to
        \begin{align}\label{eq:XpmStep1}
            Z^{X_-,X_+}_{N_\pm,P}[J] =& 
            \sum\limits_{n_o,n_c = 0}^\infty \frac{\left(\sumint \hspace{-0.1 cm} J_w \right)^{n_c}
            \left(\sumint \hspace{-0.1 cm} J_\omega^{AB}\right)^{n_o} }{n_o! n_c!} 
            \sum\limits_{\Gamma} Z_{g_s}[\Gamma].
        \end{align}
        Here, the sum runs over all coverings $\Gamma$ of $X$ with the appropriate ramification points and
        \begin{align}
            Z_{g_s}[\Gamma] := \frac{g_s^{w[\Gamma]+\omega[\Gamma]}}{\text{Deck}[\Gamma]} \langle \mathds{1} \rangle_\mathcal{M}^{\Sigma},
        \end{align}
        where $w[\Gamma]$ is the sum of bulk ramification indices and $\omega[\Gamma]$ is half the sum of boundary ramification indices of $\Gamma$.
        
        As in the bulk case, the covering maps generically have several connected components, which fall into $k$ isomorphism classes $n_1, \dots, n_k$. 
        Let us denote by 
        $N_i^\pm$ the order of the covering maps $\Gamma_i$ in the $i^\text{th}$ isomorphism class restricted to the preimage of $X^\pm$. 
        Analogously, let $p_i$ be the degree of transmissivity of $\Gamma_i$ at a reference point $t_*$. 
        Then,
        \begin{align}
             \mu^{P} \nu_-^{N_- - P} \nu_+^{N_+ - P} Z_{g_s}[\Gamma] =
            \sum\limits_{i=1}^k \frac{1}{n_i!} \left(\mu^{p_i} \nu_-^{N_i^- - p_i}\nu_+^{N_i^+ - p_i}Z_{g_s}[\Gamma_i]\right)^{n_i}.
        \end{align}
        From the powers of $\mu$ and $\nu_\pm$ we can read off the constraints
        \begin{align}
            \sum\limits_{i=1}^k n_i (N_i^\pm-p_i) = N_\pm - P&&
            \text{and}
            &&
            \sum\limits_{i=1}^k n_i p_i= P   
        \end{align}
        on $n_i$, $N_i^\pm$ and $p_i$. The sum over $P$ and $N_\pm$ in the grand-canonical ensemble lifts these constraints such that the grand-canonical partition function becomes the exponential
        \begin{align}\label{eq:XpmStep2}
            Z_{g_s}^{X_-,X_+}[J] = \exp\left( \sum\limits_{n_o,n_c = 0}^\infty \frac{\left(\sumint \hspace{-0.1 cm} J_w \right)^{n_c}
            \left(\sumint \hspace{-0.1 cm} J_\omega^{AB}\right)^{n_o} }{n_o! n_c!} 
            \sum\limits_{\Gamma \text{ conn.}} \mu^{P}\nu_-^{N_--P}\nu_+^{N_+-P}Z_{g_s}[\Gamma] \right)
        \end{align}
        of a sum of connected covering maps with orders $N_\pm$ on the preimage of $X_\pm$ and the degree of transmissivity $P$ at $t^*$.

        Applied to this setup, the interface Riemann-Hurwitz formula \eqref{eq:interface_RH} gives
        \begin{align}
            N_-(2G_--1)+N_+(2G_+-1) + \sum\limits_{i=1}^{n_c} w_i + \frac{1}{2} \sum\limits_{i=1}^{n_o} \omega_i = -2 + 2g+b+ n_c + \frac{1}{2} n_o,
        \end{align}
        where $g$ is the genus of the covering space $b$ and is the number of its boundaries. With 
        \begin{align}
            Z_{g_s}[\Gamma] = g_s^{w[\Gamma]+\omega[\Gamma]} Z[\Gamma], && \text{where} && w[\Gamma]=\sum\limits_{i=1}^{n_c} w_i, && \omega[\Gamma]=\frac{1}{2}\sum\limits_{i=1}^{n_o} \omega_i 
        \end{align}
        and $Z[\Gamma] := Z_1[\Gamma]$ carries no more dependence on the coupling $g_s$. Using the  definitions \eqref{eq:munug} of the fugacities  in terms of $g_s$ we conclude that 
        \begin{align}
            Z_{g_s}^{X_-,X_+}[J] =  \exp\left( \sum\limits_{n_o,n_c = 0}^\infty \frac{\left(\sumint \hspace{-0.1 cm} J_w \right)^{n_c}
            \left(\sumint \hspace{-0.1 cm} J_w^{AB}\right)^{n_o} }{n_o! n_c!} 
            \sum\limits_{\Gamma \text{ conn.}} g_s^{-2+2g+b+n_c + \frac{n_o}{2}} Z[\Gamma] \right),
        \end{align}
        which is the result that we wanted to derive. Once again it contains a sum over connected covering maps, or interface PRR diagrams, in the exponent. Each contribution is weighted by a specific power of the parameter $g_s$ that can be read off from the diagram. It involves the number $n_c$ and $n_o$ of bulk and interface 
        operator insertions, respectively, as well as 
        the genus $g$ and number $b$ of boundary components for the surface $\Sigma$ the interface PRR diagram is drawn on. As we have anticipated, the exponent mimics precisely that of a string genus expansion for scattering of open strings in the presence of a $D$-brane. 
        
    \paragraph{Expectation value and saddle point approximation.}\label{sec:saddle_point}
        Before we conclude this section we would like to make contact with our analysis in Section  \ref{sec:CanonicalLargeN}. To this end we 
        compute the expectation values of $N_\pm$ 
        and $P$ in the grand-canonical ensemble 
        through saddle-point approximation of the interface partition function at small $g_s$.
        We will see that the leading contribution to 
        this saddle point approximation agrees with 
        the results of Section \ref{sec:CanonicalLargeN}.
         
        For the rest of this section, we focus on the sphere case $X=\mathbb{S}^2$, $X_{\pm}=\mathbb{D}_\pm$. Let us first consider the vacuum grand-canonical partition function
        \begin{align}
        \label{eq:sphereGPF}
            Z^{\mathbb{D}_\pm}_{g_s}[0] = \exp\left[\mu z + \nu_- g_- + \nu_+ g_+\right],
        \end{align}
        where $z$ is the sphere partition function of the seed theory, and $g_\pm = \langle 0 | a_\pm \rangle$ are the boundary $g$-functions of two boundary states $|a_\pm\rangle$ therein.
        The expectation value of $P$ (the number of seed spheres) and $N_\pm-P$ (the number of disks $\mathbb{D}_\pm$) can directly be evaluated as 
        \begin{align}
            \overline{P} = \mu \partial_\mu \log(Z^{\mathbb{D}_\pm}_{g_s}[0]) =  \mu z, &&  \overline{N_\pm-P} = \nu_\pm \partial_{\nu_\pm} \log(Z^{\mathbb{D}_\pm}_{g_s}[0]) =  \nu_\pm g_\pm.
        \end{align}
        After setting $\mu = \nu_\pm^2 = g^{-2}_s$ and $z=1$, we find that $ \overline{N_\pm-P}$ is proportional to $(\overline{P})^{1/2}$ with a constant of proportionality controlled by the boundary conditions $a_\pm$. Hence, the scaling behaviour for the averages agrees with the one we used for the symbols $N_\pm, P$ and $N\pm-P$ in eq.\ \eqref{eq:large_N_P} with $N \sim g_s^{-2}$. Let us also note that the ratio 
        \begin{align}
        \label{eq:ratioNpm}
            \frac{\overline{N_+-P}}{\overline{N_--P}} = 
            \frac{\langle 0 | a_+ \rangle}{\langle 0 | a_- \rangle}.
        \end{align}
        is controlled by the ratio of boundary entropies in the seed theory. 
        To leading order in $g_s$, the normalised correlation functions 
        \begin{align}
            \left\langle \prod\limits_{i=1}^{n_o} \sigma_{\omega_i}^{A_iB_i}(t_i) \prod\limits_{i=1}^{n_c} \sigma_{w_i}(z_i) \right\rangle_{g_s}^{\mathbb{D}_\pm} = 
            \left[\prod\limits_{i=1}^{n_o}\frac{\delta}{\delta J_{\omega_i}^{A_iB_i}(t_i)} \prod\limits_{i=1}^{n_c}\frac{\delta}{\delta J_{w_i}(z_i)} \right] 
            \left.\frac{Z_{g_s}^{\mathbb{D}_\pm}[J]}{Z_{g_s}^{\mathbb{D}_\pm}[0]} \right|_{J=0}
        \end{align}
        of the grand-canonical theory can be approximated by fixing $N_\pm$ and $P$ to their expectation values.
        If we saddle-point approximate\footnote{
        Approximating the sum
        \begin{align}
            Z^{\mathbb{D}_\pm}_{g_s}[0] 
            =&~ \sum_{P=0}^{\infty}\sum_{N_\pm-P=0}^{\infty} \frac{(\mu z)^P (\nu_- g_-)^{N_--P} (\nu_+ g_+)^{N_+-P}}{P!(N_--P)!(N_+-P)!} 
        \end{align}
        by the largest summand, which occurs at $P\sim \mu z, N_\pm-P \sim \nu_\pm g_\pm$, produces 
        \begin{align}
            Z^{\mathbb{D}_\pm}_{g_s}[0]  \sim \prod_{\Lambda\in\{ \mu z, \nu_- g_-, \nu_+ g_+\}} \frac{e^\Lambda}{\sqrt{2\pi \Lambda}},
        \end{align}
        which underestimates the full sum by the Gaussian fluctuation prefactor $\sqrt{2\pi \Lambda}$. 
        However, this does not affect our leading-order estimation since the numerator and denominator have the same width and the prefactor cancels in the normalised ratio.
        } the numerator $\delta_J \left.Z^{\mathbb{D}_\pm}_{g_s}[J]\right|_{J=0}$ and the denominator $Z^{\mathbb{D}_\pm}_{g_s}[0]$, correlators of twist fields in the normalisation of \eqref{eq:sourcebulktwist} and \eqref{eq:sourceinterfacetwist} reproduce the results of Sec.~\ref{sec:CanonicalLargeN} to leading order in $g_s \sim 1/\sqrt{N}$.
    
\section{Conclusion}\label{sec:conclusion}

    In this work, we have developed the necessary covering map machinery to efficiently analyse interface correlation functions of symmetric product orbifolds on arbitrary Riemann surfaces. 
    These methods open up the possibility to answer several interesting questions about the nature of interfaces in symmetric product orbifolds and the holography of AdS$_2$ branes in pure NSNS AdS$_3$ string backgrounds.
    We would like to conclude by briefly sketching some of the potential directions to pursue in future research.
    
    \paragraph{Deriving AdS$_3$/CFT$_2$ for open strings.}
    In Section \ref{sec:string_theory_&_GCE}, we established a structural match between the genus expansion of interface correlators in the symmetric product orbifold and string perturbation theory. 
    To complete the derivation of the duality between our interfaces and AdS$_2$ branes in minimal pure NSNS AdS$_3$, it is necessary to furthermore show that open string scattering amplitudes also quantitatively match the proposed dual correlators at each order in the string genus expansion.
    The analogous derivation of the duality for closed string scattering amplitudes provided in the seminal paper \cite{Eberhardt:2019ywk} was based on a localisation of the string world sheet to ramified coverings of the holographic boundary.
    It was already conjectured in \cite{Gaberdiel:2021kkp} that this localisation property should extend to the open string case.
    Now that we identified the class of interface covering maps to which the world sheet is expected to localise in that case, this conjecture can be fully established.

    A derivation of the bulk correspondence that directly matches CFT correlations functions with string scattering amplitudes, without the need to compute either of them was proposed in \cite{Hikida:2023jyc}. It passes through a deformation of the string background that involves 
    AdS$_3$ backgrounds with NSNS flux $k>1$. The dual 
    CFT is a non-rational symmetric product orbifold that involves a Liouville-like direction \cite{Eberhardt:2021vsx}. In \cite{Hikida:2023jyc},
    the H$_3^+$-Liouville correspondence of 
    \cite{Ribault:2005wp,Hikida:2007tq} was run backwards to reconstruct string theory on AdS$_3$ from the non-rational symmetric product orbifold, see also \cite{Knighton:2024qxd} for 
    related work based on path integral representations. In the limit $k \rightarrow 1$, a $\mathcal{N}=4$ supersymmetric Liouville field theory decouples from the CFT$_2$, leaving a rational symmetric product orbifold for the supersymmetric 4-torus behind and hence one recovers the holographic relation of \cite{Eberhardt:2019ywk}. It would be interesting to extend this approach to interfaces and open strings using previous work on a boundary version of the 
    H$_3^+$-Liouville relation, see  \cite{Fateev:2007wk,Creutzig:2010bt}.

     
    \paragraph{Thermal correlators and superficial holography.}
    Reference \cite{Belin:2025nqd} reports universal behaviour of thermal correlation functions in symmetric product orbifolds, compatible with a BTZ black hole as semi-classical gravity dual.
    The result implies that conditions on the analysed thermal correlators of local operators cannot be sufficient criteria to determine whether a given CFT has a geometric bulk dual.
    It would be interesting to investigate how the inclusion of extended operators such as $\mathcal{I}^{(P)}_{|a_\pm\rangle}$ fits into this picture -- presumably leading to the construction of a superficial semi-classical gravity dual of our interfaces.
    
    \paragraph{SUSY interface covering maps.}
    Recently, a supersymmetric extension of the covering map approach to symmetric product orbifolds was formulated \cite{Nairz:2025kou}. 
    We expect that our construction of interface covering maps naturally extends to coverings of super Riemann surface, using the formalism that Reference \cite{Nairz:2025kou} developed for the bulk case.
    
    \paragraph{Interface deformations of symmetric product orbifolds.}
    Reference \cite{Benjamin:2025knd} provides a detailed description of a broad class of generalised symmetries in symmetric product orbifolds, especially emphasizing constraints on deformations by operators from twisted sectors of the theory. 
    It would be interesting to generalise these results to our interface setup and systematically analyse deformations by interface twist fields. Another direction that is now accessible would be to study 
    the effect of marginal bulk deformations on the spectrum of interface operators. Of particular relevance in this context is the deformation that 
    drives the system towards the geometric regime in 
    which the interacting CFT$_2$ possesses a dual 
    supergravity description. The effect of such 
    deformations on the bulk spectrum has been addressed
    e.g. in \cite{Pakman:2009mi,Gaberdiel:2023lco,Fabri:2025rok}. The interfaces we have studied here seem to be ideal new probes for this deformation. 

    \paragraph{Partial permutations and the interface grand-canonical ensemble.}
    From the definition of grand-canonical correlations function (Section \ref{sec:bulkgce}), neither the existence of a state operator correspondence nor the consistency of the operator product expansion are manifest.
    Reference \cite{Benizri:2025bok} solves these problems by working with an enlarged Hilbert space graded by conjugacy classes of partial permutations.
    The construction should straightforwardly generalise to the interface grand-canonical ensemble proposed in Sec.~\ref{sec:interfacegce}, providing a more complete picture of the interface spectrum and OPE.
    In the context of gauge gravity dualities for topological symmetric product orbifolds \cite{Benizri:2024mpx}, such a boundary generalisation was recently investigated in Reference
    \cite{Troost:2025eqm}.
    
\acknowledgments
We thank Lior Benizri, Andrea Dei, Gaston Giribet, Michael Gutperle, Yasuaki Hikida, Emil Martinec, Sylvain Ribault and Yifan Wang for useful comments and discussions. 
This project received funding from the German Research Foundation DFG under Germany's Excellence Strategy - EXC 2121 Quantum Universe - 390833306 and the Collaborative Research Center - SFB 1624 “Higher structures, moduli spaces and integrability” - 506632645. 
SH is further supported by the Studienstiftung des Deutschen Volkes. 
TT is supported by JSPS KAKENHI Grant Number JP24KJ1374. 

\appendix

\section{Symmetric product orbifolds as discrete gauge theories}\label{app:usual_gauge_theory}

    Symmetric product orbifolds are discrete gauge theories and therefore we often use gauge theory language in the main text to describe objects in these theories.
    At a first glance, the formulation in terms of covering maps provided in Section \ref{sec:symm_orb} however might seem very different from the usual construction of a gauge theory.
    The purpose of this appendix is to bridge this gap.
    We first formulate symmetric product orbifolds in the usual language of gauge theory i.e.~using sections of associated bundles of principal bundles.
    We then discuss how to recover the covering map description for vacuum partition functions.
    Finally, we extend the match to also include twist fields.
    The overall dictionary that we end up with is summarised in the box below.

    \begin{tcolorbox}[left=0pt,right=0pt,top=-17pt,bottom=0pt]
        \begin{align*}
            \text{$N$ sheeted covering maps $\Gamma$} &\leftrightarrow \text{$S_N$ principal bundles $P$} \\
            \text{Group of Deck transformations $\text{Deck}[\Gamma]$} &\leftrightarrow \text{Gauge group $\mathcal{G}(P)$} \\
            \text{Measure monodromy of coverings} &\leftrightarrow \text{Insert Wilson lines} \\
            \text{Change covering via cutting and glueing} &\leftrightarrow \text{Insert 't Hooft lines} \\
            \text{$S_N$ twist fields} &\leftrightarrow \text{$S_N$ monopoles}
        \end{align*}
    \end{tcolorbox}
    \paragraph{Symmetric product orbifolds and $S_N$ principal bundles.} For simplicity, let us assume that the seed theory can be formulated in terms of some real scalar field $\phi$ with Lagrangian density $\mathcal{L}[\phi]$ (consider for instance $\mathcal{M}$ to be the free boson).
    For every $S_N$ principal bundle $\pi: P \rightarrow X$, we can define an associated fibre bundle
    \begin{align}
        \Gamma_P: \Sigma_P \rightarrow X \quad \text{with} \quad \Sigma_P:=(P \times \underline{N}) / S_N
    \end{align}
    whose fibres are sets of $N$ discrete points,
    \begin{align}
         \underline{N} = \{1, 2 , \dots , N\}.
    \end{align}
    In the construction, we let $S_N$ act on $\underline{N}$ in the canonical way by permutations.
    By considering the elements of $\underline{N}$ as basis vectors of an $\mathbb{R}$ vector-space, we can also define an $\mathbb{R}^N$ vector-bundle $\mathbb{R}^N_P$ over $X$.
    In local trivialisations, sections of $\mathbb{R}^N_P$ are maps
    \begin{align}
        \Phi: U \subseteq X \rightarrow U \times \mathbb{R}^N, \quad z \mapsto (z,\Phi^{(1)}(z),\Phi^{(2)}(z), \dots ,\Phi^{(N)}(z)).
    \end{align}
    The action of $\text{Sym}^N(\mathcal{M})$ is a functional on the space of such sections, which we define in local coordinates by using the Lagrangian $\mathcal{L}$ of $\mathcal{M}$:
    \begin{align}\label{eq:action_Phi}
        S_X[\Phi] = \int\limits_X d^2z \sum\limits_{i=1}^N \mathcal{L}[\Phi^{(i)}].
    \end{align}
    Note that the action is well defined, and in particular not dependent on the choice of trivialisation, since the coordinate changes by construction act as permutations on the components of $\phi$ and the sum $\sum\limits_{i=1}^N \mathcal{L}[\phi^{(i)}]$ is permutation invariant.

    \paragraph{Vacuum partition functions.} As usual in gauge theory, the vacuum partition function is a sum over all gauge bundles $P$ weighted by the volumes of their gauge groups $\mathcal{G}(P)$.
    For the symmetric orbifold on a space time $X$, it is defined as the path integral
    \begin{align}\label{eq:partition_function_sections}
        Z^X_N := \sum\limits_{P} 
        \tfrac{1}{|\mathcal{G}(P)|} \hspace{-0.7 cm}
        \int\limits_{ \text{Sections of }\mathbb{R}^N_P} \hspace{-0.7 cm} D\Phi e^{-S_X[\Phi]}.
    \end{align}
    To translate this into the formulation in terms of covering maps, note that there is a natural identification of sections $\Phi$ of $\mathbb{R}^N_P$ with functions $\phi : \Sigma_P \rightarrow \mathbb{R}$, namely
    \begin{align}\label{eq:partition_function_functions_on_Sigma_P}
        \Phi(z) = (z,\Phi^{(1)}(z),\dots,\Phi^{(n)}(z)) \quad \leftrightarrow \quad \phi(z,i) = \Phi^{(i)}(z).
    \end{align}
    Defining this map does not require a choice of trivialisation. Instead we are using that, by construction, there is a natural basis of the fibre $(\mathbb{R}^N_P)_z$ of $\mathbb{R}^N_P$ over $z$ given by the fibre $(\Sigma_P)_z$ of $\Sigma_P$ over $z$.
    Through this identification, we can rewrite eq.~\eqref{eq:partition_function_sections} as
    \begin{align}
        Z_X^N = \sum\limits_{P}\tfrac{1}{|\mathcal{G}(P)|}\hspace{-0.35 cm} \int\limits_{\phi: \Sigma_{P} \rightarrow \mathbb{R} } \hspace{-0.5 cm} D\phi e^{-S_{\Sigma_P}[\phi]},
    \end{align}
    where $S_{\Sigma_P}$ is just the action of the seed theory $\mathcal{M}$ on $\Sigma_P$.
    This already almost looks like the sum over covering spaces in e.g.~equation \eqref{eq:torus_partition}.
    To complete the match, we need to replace the sum over $S_N$ principal bundles by a sum over unramified covering maps.
    For later convenience, let us directly write down an extended version of this correspondence that also holds for branched covering maps.

    \paragraph{Singular $S_N$ principal bundles and branched covering maps.}There is a natural one-to-one correspondence between the $N$ sheeted branched coverings of $X$ whose set of branch points is contained in a finite subset $B$ of $X$ and singular $S_N$ principal bundles $\pi : P \rightarrow X$ over $X$ whose singular locus\footnote{For our purposes, a singular principal bundle over $X$ is a principle bundle over $X \setminus S$, where $S$ is a finite subset of $X$, which we call the singular locus of the bundle.} is a subset of $B$.

    To show this, let us first construct a covering map from the principal bundle $P$.
    By construction, the projection map $\Gamma_P$ of the associated fibre bundle $(P \times \underline{N}) / S_N$ to the base space $X \setminus B$ is a $N$ sheeted covering map.
    We can induce a complex structure of $P$ by pulling back the complex structure on $X$ along $\Gamma_P$.
    Then, $\Gamma_P$ is trivially holomorphic.
    There is a unique analytic continuation of $\Gamma_P$ to a ramified covering map of $X$, whose branch points are a subset of $B$.

    Clearly, this construction is reversible.
    Starting with a branched covering map $\Gamma$, we can remove the ramification points from its domain to obtain a $\underline{N}$ bundle over $X \setminus B$.
    The structure group of this fibre bundle is $S_N$ and the associated $S_N$ principal bundle is the bundle $P_\Gamma$ we are after.

    Since furthermore $\text{Deck}[\Gamma_P]$ is identical to the group of bundle automorphisms of $\Sigma_P$ which is isomorphic to $\mathcal{G}(P)$, we can use the above bijection to rewrite eq.~\eqref{eq:partition_function_functions_on_Sigma_P} as a sum
    \begin{align}
        Z_X^N = \sum\limits_{P}\tfrac{1}{|\mathcal{G}(P)|}\hspace{-0.35 cm} \int\limits_{\phi: \Sigma_{P} \rightarrow \mathbb{R} } \hspace{-0.5 cm} D\phi e^{-S_{\Sigma_P}[\phi]} = \sum\limits_{\Gamma} \frac{\langle \mathds{1} \rangle_{\mathcal{M}}^\Sigma}{|\text{Deck}[\Gamma]|} ,
    \end{align}
    over $N$ sheeted covering maps $\Gamma$ and thus reproduce the prescription of Section \ref{sec:symm_orb}.
 
    \paragraph{Gauge transformations.} In gauge theory, the name ``gauge transformation'' is used for several types of operations, which are distinguished in the language of principal bundles as
    \begin{enumerate}
        \item \textbf{Small gauge transformations} i.e.~the action of isomorphisms of the gauge bundle that are homotopic to the identity map.
        \item \textbf{Large gauge transformation} i.e.~the action of topologically non trivial isomorphisms of the gauge bundle.
        \item \textbf{Coordinate changes} of a trivialisation of the bundles, which are sometimes also considered as ``small gauge transformations''.
        \item \textbf{Changes of the global structure of the bundles}, which are sometimes also considered ``large gauge transformations''.
    \end{enumerate}
    
    Since the gauge group $S_N$ is discrete, the only small gauge transformation is the identity.
    Relatedly, discreteness of the fibre implies that every automorphism of the gauge bundle is already fixed over each connected component of $X$ by the action on a single fibre. 
    Thus, if $X$ has $n$ connected components, then the group of large gauge transformations $\mathcal{G}(P)$ is isomorphic to a subgroup of $S_N^{n}$.
    
    If $(U_i,s_i)_{i \in I}$ is a trivialisation of $P$ (i.e.~a collection of local sections) then the coordinate changes are described by transition functions
    \begin{align}
        t_{ij} : U_i \cap U_j \rightarrow S_N,
    \end{align}
    such that $s_j(z) = s_i(z) t_{ij}(z)$.
    The transition functions satisfy the cocycle condition
    \begin{align}
         t_{ik}(z) = t_{ij}(z) t_{jk}(z) && \text{for all} && z \in U_i \cap U_j \cap U_k
    \end{align}
    and fully encode the structure of the bundle $P$.

    We can change the global structure of the gauge bundle by altering the trivialisation through a cut and glue procedure that allows us to act with oriented closed curves on $X$ labelled by permutations $g \in S_N$ on the set of $S_N$ principal bundles.
    Physically, this corresponds to the insertion of a 't Hooft line of the product theory.
    The procedure consists of the following three steps.
    \begin{enumerate}
        \item \textbf{Setup.} Given an oriented non-self-intersecting closed curve $\gamma$ on $X$, a permutation $g \in S_N$ and an $S_N$ principal bundle $\pi:P\rightarrow X$, we define a new bundle $P^\gamma_g$.
        \item \textbf{Trivialisation of $P$.} Since $\gamma([0,1]) \cong \mathbb{S}^1$, we can pick a trivialisation of $P$ such that only two charts $U$ and $V$ have a non trivial intersection with $\gamma([0,1])$. 
        Also, we can ensure that $U,V$ are simply connected, the intersection of $U$ and $V$ with $\gamma([0,1])$ is connected and that $U \cap V$ has exactly two connected components $UV$ and $\widetilde{UV}$.
        The transition function $t_{UV}$ is specified by two permutations: Its values on $UV$ and $\widetilde{UV}$ respectively.
        W.l.o.g.~let the value on $\widetilde{UV}$ be $1$ and denote the value on $UV$ by $h$.
        
        \item \textbf{Trivialisation of $P^\gamma_g$.} The curve $\gamma$ splits both $U$ and $V$ in two connected components $U_\pm'$ and $V_\pm'$ such that $U_\pm' \cap V_\mp' = \emptyset$. 
        Thicken $U_\pm'$ and $V_\pm'$ slightly to obtain open sets $U_\pm$ and $V_\pm$ that cover $\gamma$ but whose intersection $U_\pm \cap V_\mp$ is still disjoint from all charts of $P$ except for $U$ and $V$. 
        If $g$ commutes with $h$, then replacing $(U,t_U)$ and $(V,t_V)$ by the four local sections
        \begin{align}\label{eq:new_sections}
            (U_-,s_U|_{U_-}) && (V_-,s_V|_{V_-}) &&(U_+, (s_U|_{U_+})\cdot g)  && \text{and} && (V_+, (s_V|_{V_+}) \cdot g), 
        \end{align}
        leads to 
        \begin{align}
            t_{U_+ V_+} = g^{-1} \cdot (t_{UV}|_{U_+ \cap V_+}) \cdot g = t_{UV}|_{U_+ \cap V_+} && \text{and} && t_{U_- V_-} = t_{UV}|_{U_- \cap V_-} \,.
        \end{align}
        Therefore, we can use the transition functions of $P$ to consistently construct local sections over the entire connected component of $X$ containing $\gamma$ from the four new local sections defined in eq.~\eqref{eq:new_sections}.
        Complementing this with the local sections of $P$ over the rest of $X$ yields the bundle $P^\gamma_g$. 
    \end{enumerate}
    Note that, with the choice of trivialisation of $P$ made above, the permutation $h$ is exactly the monodromy of $P$ along $\gamma$. 
    The condition that $h$ and $g$ need to commute can therefore be phrased as the statement that we can only act with $g$ along $\gamma$ on $P$ if the $S_N$ monodromy of $P$ along $\gamma$ commutes with $g$.

    \paragraph{Monopoles.} The field $\Phi$ whose action we defined in eq.~\eqref{eq:action_Phi} is the lift of $\phi$ to the untwisted sector of the symmetric orbifold.
    In addition, the spectrum of local excitations also includes monopoles i.e.~operators whose insertion in the path integral restricts the sum over gauge bundles by fixing a certain monodromy around their insertion point. 
    Under the identification between $S_N$ gauge bundles and degree $N$ covering maps, this translates to the statements that the insertion of monopoles creates ramification points with a fixed index. 
    Thus, the $S_N$ monopoles are exactly the twist fields as introduced in Sec.~\ref{sec:symm_orb}.
        
\section{Wilson and 't Hooft lines in symmetric orbifolds}\label{App:Wilson_tHooft}

    In this appendix, we first use Wilson loops to obtain an alternative definition of gauge fixed twist fields. 
    We then discuss the 't Hooft lines that implement an $S_N$ action on the global structure of covering maps.
    These provide us with a different perspective on gauge choices and gauge invariance in symmetric orbifolds and also allow us to quite conveniently describe other permutation orbifolds.

        \paragraph{Wilson lines.} 
        The data defining a Wilson line $\mathcal{W}[\gamma]$ is a curve $\gamma : [0,1] \rightarrow X$.
        The insertion of such a line does not affect the set of covering maps associated with an operator $\mathcal{O} \in \mathcal{A}_X[\text{Sym}^N(\mathcal{M})]$, that is
        \begin{align}
            C[\mathcal{O} \mathcal{W}[\gamma] ]  = C[\mathcal{O}] \quad \text{for all }\mathcal{O}\in\mathcal{A}_X[\text{Sym}^N(\mathcal{M})].
        \end{align}
        It only acts through its lift, which is a bijection from $\Gamma^{-1}(\gamma(0))$ to $\Gamma^{-1}(\gamma(1))$, i.e.
        \begin{align}
            \mathcal{W}[\gamma]^\Gamma \in \text{Bij}[\Gamma^{-1}(\gamma(0)),\Gamma^{-1}(\gamma(1))].
        \end{align}
        Concretely, the value of $\mathcal{W}[\gamma]^\Gamma$ is the map defined by parallel transport along $\Gamma^*\gamma$.
        Through local gauge choices of $C[\mathcal{O}]$ at $\gamma(0)$ and $\gamma(1)$, we can identify $\mathcal{W}[\gamma]^\Gamma$ with an element $\mathcal{W}[\gamma]^\Gamma_{\text{gf}}$ of $S_N$.
        If we change the local gauge at $\gamma(0)$ by acting with a permutation $g_0$ and that at $\gamma(1)$ by acting with a permutation $g_1$, then $\mathcal{W}[\gamma]^\Gamma_{\text{gf}}$ transforms as
        \begin{align}\label{eq:W_transformation}
            \mathcal{W}[\gamma]^\Gamma_{\text{gf}} \rightarrow g_1 \mathcal{W}[\gamma]^\Gamma_{\text{gf}} g^{-1}_0.
        \end{align}
        \paragraph{Wilson loops.} 
        For closed Wilson lines i.e.~Wilson loops, $\gamma(0) = \gamma(1)$ and we only make a single local gauge choice.
        Thus, Wilson loops transform by conjugation under a change of the gauge.
        We can hence define a $\mathbb{C}$ valued operator
        \begin{align}
            \mathcal{W}_R[\gamma]:= \chi_R(\mathcal{W}[\gamma])
        \end{align}
        for every closed loop $\gamma$ and every representation $R$ of $S_N$ with character $\chi_R$ by making an arbitrary gauge choice to evaluate $\chi_R(\mathcal{W}[\gamma]^\Gamma)$.
        As usual in gauge theory, these topological loops are not the only gauge invariant extended objects that can be built from $\mathcal{W}[\gamma]^\Gamma_{\text{gf}}$.
        Another option is to consider Wilson lines whose end-points are dressed with operators charged under $S_N$.

    \paragraph{Gauge fixed twist fields.}
    Using Wilson lines, we can define gauge fixed versions $\sigma_g$ of the twist fields $\sigma_{[g]}$:
    \begin{align}
        \sigma_g(z) := \lim\limits_{\epsilon \rightarrow 0} \delta_{\mathcal{W}[\gamma_\epsilon]_{\text{gf}},\,g}  \, \sigma_{[g]}(z),
    \end{align}
    where $\gamma_\epsilon$ is a small loop winding once around $z$ clockwise at distance $\epsilon$.
    The gauge dependence of $\sigma_g$ comes from the choice of how to identify $\mathcal{W}[\gamma_\epsilon]$ with an element of $S_N$.
    In particular, the transformation behaviour \eqref{eq:W_transformation} of the Wilson loop implies that under a change $h$ of the local gauge choice, $\sigma_g$ transforms as
    \begin{align}\label{eq:sigma_g_transform}
        \sigma_g \mapsto  \lim\limits_{\epsilon \rightarrow 0} \delta_{h\mathcal{W}[\gamma_\epsilon]_{\text{gf}}h^{-1},\, g}  \, \sigma_{[g]} =  \lim\limits_{\epsilon \rightarrow 0} \delta_{\mathcal{W}[\gamma_\epsilon]_{\text{gf}}, \,h^{-1}gh}  \, \sigma_{[g]} = \sigma_{h^{-1} g h}.
    \end{align}
    Note that the linear combination
    \begin{align}
        \hat{\sigma}_{[g]} = \sum\limits_{\tilde g \in [g] } \sigma_{\tilde g}
    \end{align}
    is gauge invariant. But of course, $\hat{\sigma}_{[g]}$ is simply ${\sigma}_{[g]}$
    \begin{align}
        \hat{\sigma}_{[g]} = \sum\limits_{\tilde g \in [g]}\lim\limits_{\epsilon \rightarrow 0} \delta_{\mathcal{W}[\gamma_\epsilon], \tilde{g}}  \, \sigma_{[g]} = \sigma_{[g]}.
    \end{align}
    
    \paragraph{'t Hooft lines.}
    The insertion of Wilson lines does not affect the set of covering maps associated with a correlation function and only acts through the lift of the lines along the covering maps.
    Let us now construct a dual set of lines that only act by changing the structure of covering maps but lift to the trivial defect in the seed theory. These oriented topological lines $\mathcal{L}_g[\gamma]$ are labeled by a permutation $g\in S_N$ and a non-self intersecting curve $\gamma$.
    They depend on a local gauge choice at $\gamma(0)$ only.

    The set $C[\mathcal{O}\mathcal{L}_g[\gamma]]$ of covering maps computing correlators in the presence of $\mathcal{L}_g[\gamma]$ is given by applying a cutting and glueing procedure that maps each  $\Gamma \in C[\mathcal{O}]$ to a new covering which we shall denote by $\mathcal{L}_g[\gamma]\cdot\Gamma$. 
    To construct $\mathcal{L}_g[\gamma]\cdot\Gamma$, first cut out the pre-image $\Gamma^{-1}(\gamma([0,1]))$ of the image of $\gamma$ under $\Gamma$ from the covering space $\Sigma$ to obtain
    \begin{align}
        \tilde \Sigma:= \Sigma \setminus \Gamma^{-1}(\gamma([0,1])).
    \end{align}
    By parallel transport of the local gauge choice at $\gamma(0)$ along $\gamma$, every point on $\Gamma^{-1}(\gamma([0,1]))$ has a label $i\in \{1,\dots,N\}$.
    We now glue $\tilde \Sigma$ back together along its cut by identifying the $i^\text{th}$ sheet with the $gi^{\text{th}}$ sheet upon travelling along the cut from below (w.r.t.~the orientation of $\gamma$).
    In particular, this identifies several of the $N$ points in $\Gamma^{-1}(\gamma(0))$ (and $\Gamma^{-1}(\gamma(1))$) among each other, namely all those that are elements of the same $g$ orbit.
    Let the Riemann surface constructed in this manner be denoted by $\mathcal{L}_g[\gamma] \cdot \Sigma$.
    We define $\mathcal{L}_g[\gamma] \cdot \Gamma$ as the analytic extension of $\Gamma$ from $\tilde \Sigma$ to $\mathcal{L}_g[\gamma] \cdot \Sigma$.

    Changing the local gauge at $\gamma(0)$ by acting with a permutation $h$ transforms $\mathcal{L}_g[\gamma]$ as
    \begin{align}
        \mathcal{L}_g[\gamma] \rightarrow \mathcal{L}_{hgh^{-1} }[\gamma]
    \end{align}
    since glueing the $i^{\text{th}}$ sheet to that labelled by $gi$ is the same as first relabelling $i \rightarrow i' = hi$ and then glueing the sheet with the new label $i'$ to that carrying the new label $h (g i) = h g h^{-1} i'$.

    Thus, a simple way to construct gauge invariant lines from $\mathcal{L}_g[\gamma]$ is to average over characters $\chi_R$ of $S_N$. This leads to the lines
    \begin{align}\label{eq:L_rep_lines}
        \mathcal{L}_R[\gamma]:= \sum\limits_{g \in S_N} \chi_R(g) \mathcal{L}_g[\gamma].
    \end{align}
    Note that if we average over a particular conjugacy class,
    \begin{align}
        \mathcal{L}_{[g]}[\gamma]:=  \sum\limits_{\tilde g \in [g]} \mathcal{L}_{\tilde g}[\gamma],
    \end{align}
    the resulting non-local operators are in effect pairs of $[g]$-twist fields inserted at $\gamma(0)$ and $\gamma(1)$ that are coupled to each other in a non-local fashion.

    There is an obstruction to inserting $\mathcal{L}_g$ along a closed curve $\gamma$.
    The operator $\mathcal{L}_g[\gamma]$ is only well defined for closed $\gamma$ if the monodromy along $\gamma$ as measured by $\mathcal{W}[\gamma]$ commutes with $g$.
    Thus, we define $\mathcal{L}_g[\gamma] \cdot \Gamma$ for closed curves $\gamma$ as
    \begin{align}
        \mathcal{L}_g[\gamma] \cdot \Gamma =
        \begin{cases}
            \text{as in the $\gamma(0) \neq \gamma(1)$ case} & \text{if } \mathcal{W}[\gamma]^\Gamma \in \mathcal{C}_g \\
            \Gamma & \text{else.}
        \end{cases}
    \end{align}
    
    \paragraph{Global gauge choices and gauge invariance.}
    As made precise in Appendix \ref{app:usual_gauge_theory}, the sums over covering maps in symmetric product orbifolds are sums over gauge bundles of $S_N$ gauge theories.
    The closed permutation lines introduced at the end of the previous paragraphs can in this sense be thought of as implementing large gauge transformations.
    Gauge invariance of the symmetric product orbifold is the statement that for every closed curve $\gamma$, every $g\in S_N$ and every $\mathcal{O} \in \mathcal{A}_X[\text{Sym}^N(\mathcal{M})]$,
    \begin{align}
            C[\mathcal{O}] = C[\mathcal{O} \mathcal{L}_g].
    \end{align}
    As an example, let us verify that the vacuum of the symmetric product orbifold as defined in eq.~\eqref{eq:C[1]} is gauge invariant. Given any holomorphic covering map $\Gamma: \Sigma \rightarrow X$ and a closed curve $\gamma$ on $X$, there are two cases to consider. The first case is that the monodromy of $\Gamma$ along $\gamma$ does not commute with $g$. In this case, we have defined the permutation line as acting trivially. In the second case $\mathcal{L}_g[\gamma]$ maps $\Gamma$ into another holomorphic covering map without changing the monodromy along $\gamma$.
    Thus, $\mathcal{L}_g[\gamma]$ simply permutes the different holomorphic covering maps among each other, leaving the set $C[\mathds{1}]$ invariant.

    \paragraph{Other permutation orbifolds.} As a side comment, let us briefly note that it is also possible to consider a variation of the above procedure where one only imposes invariance under a certain subset of the operators $\mathcal{L}_g$.
    This, together with an apropriate analogue of eq.~\eqref{eq:C[1]}, leads to more general permutation orbifolds.

    For example, we could choose to impose no invariance at all. Together with
    \begin{align}
        C_{\mathcal{M}^{\otimes N}}[\mathds{1}] := \{\text{trivial covering of $X$ by $X^N$}\},
    \end{align}
    this leads to the product theory. While the lines $\mathcal{L}_R[\gamma]$ defined in eq.~\eqref{eq:L_rep_lines} are trivial in the symmetric product orbifold if $\gamma$ is closed, they do have interesting correlation functions in the product theory.
    
\section{Comments on the normalisation of the interfaces}\label{app:comments_on_normalisation}
    
    The purpose of this appendix is to compare the normalisation of the interfaces as defined in eq.~\eqref{eq:Interfaces_unnormalised_corr} to the normalisation fixed through the explicit expressions for boundary states of the folded theory given in Reference \cite{Harris:2025wak}.
    
    In \cite{Harris:2025wak}, the normalisation of the boundary states $|p^{L/R}\rangle'$ was fixed by demanding that  
    \begin{align}
        Z_{N_\pm,p_{L/R}}^{\text{torus}} :=  \langle p^{L}|' \exp(\pi i \hat t (L_0 + \bar L_0 - \tfrac{c}{12}) |p^{R}\rangle'
    \end{align}
    leads to formula \eqref{eq:torus_gc_result} for the grand-canonical torus partition function \eqref{eq:torus_definition}.
    Since we derived the very same formula using the normalisation of interfaces fixed by eq.~\eqref{eq:Interfaces_unnormalised_corr}, the approach presented in this paper clearly agrees with the approach of \cite{Harris:2025wak} for the two interface torus partition function. 
    
    However, to obtain a fully consistent picture, the boundary states should be rescaled  
    \begin{align}
        |p\rangle := \sqrt{N_-!}\sqrt{N_+!}|p\rangle' 
    \end{align}
    and the overlap should be divided by $N_-!N_+!$, so that the torus partition function
    \begin{align}
        Z_{N_\pm,p_{L/R}}^{\text{torus}} := \frac{1}{N_-! N_+!}  \langle p^{L}| \exp(2\pi i \hat t (L_0 + \bar L_0 - \tfrac{c}{12}) |p^{R}\rangle && \text{with} && c = (N_- + N_+)c_\mathcal{M}
    \end{align}
    remains unchanged. 
    As long as we are only interested in $Z_{N_\pm,p_{L/R}}^{\text{torus}}$, the question of whether we work with $ |p\rangle'$ or $|p\rangle$ is of course irrelevant and purely a matter of taste. 
    But once we move on to other partition functions, we should work with $|p\rangle$. 
    Let us for instance consider the sphere partition function with a single interface insertion, splitting the sphere into the two disks $\mathbb{D}_\pm$. 
    The fixed $N_\pm,p$ partition function in terms of $|p\rangle$ is
    \begin{align}
    \label{eq:spherevacuum}
        Z_{N_\pm,p}^{\mathbb{D}_-,\mathbb{D}_+} = \frac{\langle p | 0 \rangle}{N_-! N_+!} = \frac{g_-^{N_--p}g_+^{N_+-p}}{(N_--p)!(N_+-p)!p!} ,
    \end{align}
    where we obtained the r.h.s.~from an expression for $\langle 0 | p \rangle'$ given in eq.~(2.40) of \cite{Harris:2025wak}.
    This result agrees with the expression derived from the interface covering map perspective in the main text, see eq.~\eqref{eq:D-D+_main_text}.
    As another indication towards the fact that \eqref{eq:spherevacuum} is the correct prescription, observe that the grand-canonical partition function
    \begin{align}
        Z^{\mathbb{D}_-,\mathbb{D}_+}_{g_s} := \sum\limits_{N_\pm=0}^\infty\sum\limits_{p=0}^{\text{min}(N_-,N_+)} \mu^{p} \nu^{N_- + N_+ - 2p} Z_{N_\pm,p}^{\mathbb{D}_-,\mathbb{D}_+}
    \end{align}
    with $\mu := g_s^{-2}$ and $\nu:= g_s^{-1}$ exponentiates as\footnote{As a consistency check, note that setting $\nu = 0$ reproduces $Z_{g_s}^{\text{sphere}}[0]$ in a normalisation where the sphere partition function of the seed theory is set to 1.}
    \begin{align}\label{eq:exponentiated_sphere}
        Z^{\mathbb{D}_-,\mathbb{D}_+}_{g_s} = \exp(\mu + \nu g_- + \nu g_+)
    \end{align}
    only if we work with $|p\rangle$ and not $|p'\rangle$.
    
\section{Generalised Lunin-Mathur construction -- computational details}\label{app:LM_computations}
    
    This appendix contains detailed derivations and computations underlying the results discussed in Section \ref{sec:Lunin_Mathur}.
    
    \subsection{Details on the regularisation procedure}\label{app:details_regularisation}
        In this appendix, we derive eq.~\eqref{eq:S_L_boundary_integral}.
        In the first short paragraph, we briefly describe the setup of the derivation to make the appendix self-contained. 
        We then begin with the derivation by dividing the covering space $\Sigma$ into three regions. 
        One of the three regions contains a pathological subregion, which we cut out in order to define the regularised Liouville action \eqref{eq:Liouville_reg}.
        The next three paragraphs evaluate the Liouville action on each individual region.
        After this, we add the three contributions to obtain the regularised vacuum partition function \eqref{eq:ZSigma1eps}.
        The regularised partition function still depends on an unphysical regulator, which we remove in the final paragraph of the section by rescaling the twist fields. 
        This final step directly produces the result \eqref{eq:S_L_boundary_integral} that we want to show.
        \paragraph{The setup.}
        In Section \ref{sec:bulk_LM}, we introduced the family
        \begin{align}\label{eq:def_delta}
            ds^2_{\delta} = 
            \begin{cases}
                 dzd\bar{z} & \text{if } |\delta z| < 1 \\
                 \frac{dzd\bar{z}}{|z\delta |^4} & \text{else}
           \end{cases}
        \end{align}
        of continuous metrics on the Riemann sphere parametrised by $\delta>0$.
        While $ds^2_{\delta}$ is continuous, it is not smooth. 
        In particular, its scalar curvature is the distribution 
        \begin{align}
            R_\delta = 4 \delta\left(|z|-\tfrac{1}{\delta}\right)
        \end{align}
         whose support is the circle $|z\delta|=1$.
         We noted in the main text that the partition function of a CFT with central charge $c$ on $\overline{\mathbb{C}}$ equipped with $ds^2_\delta$ is
        \begin{align}
            Z^{\overline{\mathbb{C}}_\delta}_1 = Q \delta^{- \tfrac{c}{3}},
        \end{align}
        where $Q$ is an arbitrary constant, which we fixed to $Q = \eta^{\frac{c}{3}}$ so that $X = \overline{\mathbb{C}}_{\eta}$ has $Z_1^X = 1$.
        In this appendix, we show in detail how to evaluate $Z_1^\Sigma$ for a sphere $\Sigma$ with metric induced via a covering map $\Gamma: \Sigma \rightarrow X$, which is unramified at $\Gamma^{-1}(\infty)$ and satisfies $\Gamma(\infty) = \infty$.
        As explained in the main text, we can do so (for an appropriate $\delta>0$ specified in the next paragraph) via the relation
        \begin{align}
             Z^\Sigma_1 = e^{S_L[\phi]} Z^{\overline{\mathbb{C}}_\delta}_{1}\quad \text{with} \quad S_L[\phi] = \frac{c}{96 \pi} \int\limits_\Sigma d^2 t \sqrt{g_{\delta}} \left( \partial_\mu \phi \partial_\nu \phi g^{\mu \nu}_{\delta} + 2 R_\delta\phi\right),
        \end{align}
        where $\phi$ is the Weyl factor relating $\Sigma$ to $\overline{\mathbb{C}}_\delta$. 
        
        \paragraph{Divide and cover.} 
        To organise the computation, we divide $\Sigma$ into three disjoint regions and evaluate the Liouville action separately on each of these.  
        The first of the regions is
         \begin{align}
            \mathcal{B} = \{t \in \mathbb{C} : |\eta \Gamma(t)| < 1\} \subseteq \Sigma.
        \end{align}
        We pick $\delta$ sufficiently small to ensure that 
        \begin{align}
            \mathcal{B} \subseteq \Sigma^0:= \{t \in \mathbb{C}: |t \delta| < 1\}.
        \end{align}
        Furthermore, we denote the complement of $\Sigma^0$ by
        $\Sigma^\infty$. Our decomposition of $\Sigma$ is
        \begin{align}
                \Sigma = \mathcal{B} \sqcup (\Sigma^0\setminus \mathcal{B}) \sqcup \Sigma^\infty\, .
        \end{align}
        We use this decomposition to evaluate the Liouville action 
        \begin{align}\label{eq:Liouville_decomp}
            S_L[\phi] = S_L^{\mathcal{B}}[\phi] + S_L^{\Sigma^0\setminus \mathcal{B}}[\phi] + S_L^{\Sigma^\infty}[\phi]
        \end{align}
        for the Weyl factor $\phi$ relating metric $ds^2$, induced on $\Sigma$ via $\Gamma$, to the metric $ds^2_\delta$.
        
        \paragraph{Regularisation of $S_L[\phi]$.}
        The region $\mathcal{B}$ contains all ramification points of $\Gamma$. 
        At the ramification points, the induced metric degenerates, the Weyl factor diverges and the Liouville action needs to be regularised. 
        To do so, we introduce a regulator $\varepsilon>0$ that is sufficiently small to ensure that 
        \begin{align}
            R_\varepsilon := \bigcup\limits_{r \in R[\Gamma]} \{t:|\Gamma(t)-\Gamma(r)|<\varepsilon\} \subseteq \mathcal{B}.
        \end{align}
        We regularise the Liouville action by removing $R_\varepsilon$ from $\mathcal{B}$. 
        That is, we define $\mathcal{B}_\varepsilon := \mathcal{B} \setminus R_\varepsilon $ and use $S_L^{\mathcal{B}_\varepsilon}[\phi]$ as a regularised version of $S_L^{\mathcal{B}}[\phi]$, leading to
        \begin{align}\label{eq:Liouville_reg}
            S_L^\varepsilon[\phi] := S_L^{\mathcal{B}_\varepsilon}[\phi] + S_L^{\Sigma^0\setminus \mathcal{B}}[\phi] + S_L^{\Sigma^\infty}[\phi]
        \end{align}
        as a regularised version of \eqref{eq:Liouville_decomp}.
        \paragraph{Evaluation of $S_L^{\mathcal{B}_\varepsilon}[\phi]$}
        On $\mathcal{B}_\varepsilon$ the metric $ds_\delta^2 = dtd\bar t$ is flat. Thus, only the kinetic term of the Liouville action contributes. The latter evaluates to
        \begin{align}\label{eq:Sbepsilon}
            S_L^{\mathcal{B}_\epsilon}[\phi] = \frac{c}{96 \pi} \int\limits_{\mathcal{B}_\epsilon} d^2 t 4 \partial \phi \bar \partial \phi = \frac{c}{96 \pi} \int\limits_{\mathcal{B}_\epsilon} d^2 t 4\partial( \phi \bar \partial \phi) = \frac{c}{96 \pi}\int\limits_{\partial \mathcal{B}_\epsilon} d s   \phi \partial_n \phi.
        \end{align}
        via Stokes theorem, where $n$ is the normal vector of $\partial \mathcal{B}_\epsilon$. In the second step, we used that
        \begin{align}
            \phi = \log(\partial \Gamma) + \log(\bar \partial \bar \Gamma) \quad \text{and hence} \quad \partial \bar \partial \phi = 0. 
        \end{align}
        To further evaluate the integral, note that each connected component of $\partial \mathcal{B}$ either belongs to a pole $p \in P[\Gamma]$ of $\Gamma$ in the complex plane, the point $\infty$, or to a ramification point $r \in R[\Gamma]$.
        Around these three different types of point, $\Gamma$ behaves as
        \begin{align}
            -\Gamma(p+t) = \frac{b_p}{t} + O(1), && \Gamma(t) = b_\infty t + O(1) && \text{and} && \Gamma(r+t) = \Gamma(r) + a_r t^w + O(t^{w+1}).
        \end{align}
        Thus, a pole $p$ in the complex plane contributes
        \begin{align}
            -\int\limits_{0}^{2\pi} ds \, b_p \eta \, (-2\log(b_p \eta^2))(-4b_p^{-1}\eta^{-1}) = -16 \pi  \log(b_p \eta^2)
        \end{align}
        to \eqref{eq:Sbepsilon}, the point $\infty$ contributes nothing and a ramification point $r$ contributes 
        \begin{align}
             -\int\limits_{0}^{2\pi} ds \, \varepsilon_r \, (-2\log(wa_r \varepsilon_r^{w-1}))(-2(w-1)\varepsilon_r^{-1}) = -8 \pi (w-1) \log(w a_r \varepsilon_r^{w-1})
        \end{align}
        to leading order in $\varepsilon$ and $\eta$. Here, $\varepsilon_r := \left(\frac{\varepsilon}{a_r}\right)^{\frac{1}{w}}$. In total, we conclude
        \begin{align}\label{eq:Bepsilon}
            S_L^{\mathcal{B}_\epsilon}[\phi] = -\frac{c}{12}\sum\limits_{r \in R[\Gamma]} (w-1)\log\left(w \varepsilon \left(\tfrac{a_r}{\varepsilon}\right)^{\frac{1}{w}}\right) - \frac{c}{6}\sum\limits_{p \in P[\Gamma]} \log(b_p \eta^2) 
        \end{align}
        \paragraph{Evaluation of $S_L^{\Sigma^0\setminus \mathcal{B}}[\phi]$.}
        On $\Sigma^0\setminus \mathcal{B}$ the metric $ds_\delta^2 = \frac{dtd\bar t}{|\delta t|^4}$ is flat. 
        Thus, only the kinetic term is  non-vanishing. 
        Two regions contribute to the corresponding integral. 
        The first consists of disks around the poles of $\Gamma$. 
        Their contribution is suppressed by $\delta^2$.
        To leading order, $S_L^{\Sigma^0\setminus \mathcal{B}}[\phi]$ is thus fully determined by the integral over the second region: The annulus around $\infty$. There, $\phi = -\log(b_\infty^2 |\delta t|^4)$ and we get 
        \begin{align}\label{eq:Sigma0/Bepsilon}
            S_L^{\Sigma^0\setminus \mathcal{B}}[\phi] =\frac{c}{96 \pi}\int\limits_{(\Sigma^0 \setminus \mathcal{B})_\infty} d^2t \, 4 \partial \phi \bar \partial \phi= \frac{c}{24 \pi}\int\limits_{\frac{1}{b_\infty \eta}}^{\frac{1}{\delta}} dr \int\limits_{0}^{2\pi} d\theta \, {r b_\infty}\frac{4}{r^2} = -\frac{c}{3} \log\left(\frac{\delta}{b_\infty \eta}\right)
        \end{align}
        
        \paragraph{Evaluation of $S_L^{\Sigma^\infty}[\phi]$.} On $\Sigma^\infty$ the Weyl factor $\phi = \log(\tfrac{\delta^4}{b_\infty^2 \eta^4})$ is constant.
        Thus, the kinetic term vanishes and $S_L^{\Sigma^\infty}[\phi]$ is fully determined by the curvature term, which yields
        \begin{align}\label{eq:Sigmainf}
            S_L^{\Sigma^\infty}[\phi] = \frac{c}{96 \pi}  16\pi \phi = \frac{c}{3} \log \delta^2 - \frac{c}{3} \log(b_\infty \eta^2)
        \end{align}
        
        \paragraph{Putting everything together.} Combining eqs.~\eqref{eq:Bepsilon}, \eqref{eq:Sigma0/Bepsilon} and \eqref{eq:Sigmainf}, we conclude
        \begin{align}
            S_L^\varepsilon[\phi] = -\frac{c}{12}\sum\limits_{r \in R[\Gamma]} (w-1)\log\left(w \varepsilon \left(\tfrac{a_r}{\varepsilon}\right)^{\frac{1}{w}}\right) - \frac{c}{6}\sum\limits_{p \in P[\Gamma]} \log(b_p \eta^2) . 
        \end{align}
        Therefore, the regularised version of $Z_{1}^\Sigma$ is
        \begin{align}
            Z_{1;\varepsilon}^\Sigma := e^{S_L^\varepsilon[\phi]} Z_1^{\overline{ \mathbb{C}}_\delta} =\left( \prod\limits_{r \in R[\Gamma]} w^{w-1} \varepsilon^{\frac{(1-w)^2}{w}} a_r^{1-\frac{1}{w}}\prod\limits_{p \in P[\Gamma]} b_p^{2} \eta^{4}\right)^{-\frac{c}{12}}  \, .
        \end{align}
        Since $\infty$ is not a branch point of $\Gamma$, we have $|P[\Gamma]|=\text{Deg}[\Gamma]-1$. Thus,
        \begin{align}\label{eq:ZSigma1eps}
            Z_{1;\varepsilon}^\Sigma = \eta^{-\frac{c}{3}(\text{Deg}[\Gamma]-1)} \left( \prod\limits_{r \in R[\Gamma]} w^{w-1} \varepsilon^{\frac{(1-w)^2}{w}} a_r^{1-\frac{1}{w}}\prod\limits_{p \in P[\Gamma]} b_p^{2}\right)^{-\frac{c}{12}}  \, .
        \end{align}
        \paragraph{Rescaling the twist fields.}
        The regularisation scheme introduced an explicit dependence of $Z_{1;\varepsilon}^\Sigma$ on the unphysical parameter $\varepsilon$.
        To remove the regulator dependence, we need to rescale the twist fields i.e.~replace $\sigma_w$ by $\mathcal{N}_w \sigma_w$.
        We make the minimal choice
        \begin{align}\label{eq:bulk_normalisation}
            \mathcal{N}_w = \varepsilon^{\frac{c}{12} \frac{(1-w)^2}{w}}
        \end{align}
        so that, after rescaling, $Z_{1;\varepsilon}^\Sigma$ becomes the physical quantity 
        \begin{align}\label{eq:Z1Sigma_appendix_result}
            Z_{1}^\Sigma = \eta^{-\frac{c}{3}(\text{Deg}[\Gamma]-1)} \prod\limits_{r \in R[\Gamma]} w^{-\frac{c}{12}(w-1)}a_r^{-\frac{c}{12}(1-\frac{1}{w})}\prod\limits_{p \in P[\Gamma]} b_p^{-\frac{c}{6}}
        \end{align}
        which is independent of $\varepsilon$.
        
    \subsection{Regularisation for the upper half plane}\label{app:details_regularisation_disk}
    In this appendix, we regulate the Liouville action for interface coverings of the upper half complex plane by the upper half complex plane. 

    \paragraph{The setup.}
    The setup treated in this appendix is almost identical to that described in the beginning of Appendix \ref{app:details_regularisation}.
    In particular, the reference metric that we consider is the restriction of the metric $ds^2_\delta$ defined in eq.~\eqref{eq:def_delta} to the upper half plane.
    Generically, we would have to evaluate the extrinsic curvature contribution in addition to the bulk Liouville action in the interface case. 
    For the upper half plane equipped with $ds^2_\delta$, the derivative of the normal vector w.r.t.~the tangent vector of the boundary is however always proportional to the normal vector and hence the extrinsic curvature vanishes.
    Accordingly, only the bulk Liouville action is relevant. 
    As we argue in the next paragraph, the latter can be evaluated using the method of images.
    
    \paragraph{Method of images.}
    Given an interface covering map $\Gamma$ that covers the upper half complex plane by itself, we can readily construct a covering map
    \begin{align}
        \Gamma': \mathbb{C} \rightarrow \mathbb{C}, \quad z \mapsto 
        \begin{cases}
            \Gamma(z) \quad \text{if Im}(z)>0  \\
            \overline{\Gamma(\bar{z})} \quad \text{else}.
        \end{cases}
    \end{align}
    For each bulk ramification point of $\Gamma$, the doubled map $\Gamma'$ has two ramification points with the same index. 
    For each interface ramification point of index $\omega$, it has a bulk ramification point with index $w = \omega$ on the real line. 
    The bulk part of the regulated Liouville action for the conformal factor $\phi$ relating the upper half plane with metric induced by $\Gamma'$ to the upper half plane with metric $ds^2_\delta$ is given by
    \begin{align}
        S_{L}^\varepsilon[\phi]_{\text{Bulk}} = \frac{1}{2}S_{L}^\varepsilon[\phi']
    \end{align}
    in terms of the Liouville action for the conformal factor $\phi'$ of the doubled covering map.
    We can therefore directly generalise eq.~\eqref{eq:ZSigma1eps} to
    \begin{align}\label{eq:InterfaceZSigma1eps}
        Z_{1;\varepsilon}^\Sigma = \Omega(\eta) \left(\prod\limits_{r \in R_B[\Gamma]} \hspace{-0.2 cm} w^{w-1} \varepsilon^{\frac{(1-w)^2}{w}} a_r^{1-\frac{1}{w}}\hspace{-0.2 cm}\prod\limits_{p \in P_B[\Gamma]} \hspace{-0.2 cm} b_p^{2} \right)^{-\frac{c}{12}} \left( \prod\limits_{r \in R_I[\Gamma]} \hspace{-0.2 cm} \omega^{\omega-1} \varepsilon^{\frac{(1-\omega)^2}{\omega}} a_r^{1-\frac{1}{\omega}}\hspace{-0.2 cm}\prod\limits_{p \in P_I[\Gamma]} \hspace{-0.2 cm} b_p^{2}\right)^{-\frac{c}{24}}  \hspace{-0.5 cm} ,
    \end{align}
    where $\Omega(\eta) := \eta^{-\frac{c}{6}(N_-+N_+-1)}$ and $N_\pm = \text{Deg}_\pm[\Gamma]$. Furthermore, $R_B[\Gamma]$ and $R_I[\Gamma]$ are the sets of bulk and interface ramification points of $\Gamma$ respectively. 
    Likewise, $P_B[\Gamma]$ and $P_I[\Gamma]$ are the sets of poles in the bulk and on the interface, not including $\infty$.
    
    \paragraph{Rescaling the twist fields.}
    The result \eqref{eq:InterfaceZSigma1eps} is consistent with the bulk twist field normalisation found in eq.~\eqref{eq:bulk_normalisation}. 
    To fully remove the regulator dependence, we need to also redefine the normalisation of the interface twist fields. 
    We work with the minimal bulk and interface normalisations
    \begin{align}
        \mathcal{N}_w^B = \varepsilon^{\frac{c}{12} \frac{(1-w)^2}{w}} && \text{and} && \mathcal{N}_\omega^I = \varepsilon^{\frac{c}{24} \frac{(1-\omega)^2}{\omega}}.
    \end{align}
    In this normalisation, we obtain the physical partition function 
    \begin{align}
        Z_1^\Sigma = \eta^{-\frac{c}{6}(N_-+N_+-1)}\left( \prod\limits_{r \in R_B[\Gamma]} \hspace{-0.2 cm} w^{w-1} a_r^{1-\frac{1}{w}}\hspace{-0.2 cm}\prod\limits_{p \in P_B[\Gamma]} \hspace{-0.2 cm} b_p^{2}\right)^{-\frac{c}{12}} \left( \prod\limits_{r \in R_I[\Gamma]} \hspace{-0.2 cm} \omega^{\omega-1} a_r^{1-\frac{1}{\omega}}\hspace{-0.2 cm}\prod\limits_{p \in P_I[\Gamma]} \hspace{-0.2 cm} b_p^{2}\right)^{-\frac{c}{24}}  \hspace{-0.2 cm} .
    \end{align}
    
    \subsection{Computing covering maps}\label{app:compute_cover}

    This appendix discusses how covering maps are computed in the first section of the mathematica notebook \texttt{CoveringMaps.nb} uploaded in the ancillary files associated with the arXiv page of this publication. 
    We focus on coverings $\Gamma: \mathbb{C} \rightarrow \mathbb{C}$ that have branch points 
    \begin{align}
        x_1 = 0, \quad x_2, \quad x_3 = 1, \quad \text{and} \quad x_4 = \infty
    \end{align}
    which correspond to ramification points
    \begin{align}
        s_1 = 0, \quad s_2, \quad s_3 = 1, \quad \text{and} \quad s_4 = \infty
    \end{align}
    with ramification indices $w_1, \dots w_4$. The covering map $\Gamma$ satisfies
    \begin{align}
        \Gamma(s_i + s) = x_i + a_i s^{w_i}.
    \end{align}
    The choices $s_1 = x_1 = 0$ and $x_4 = s_4 = \infty$ ensure that we can directly satisfy two of these conditions by making the Ansatz
    \begin{align}
        \Gamma(s) = \frac{q(s)}{p(s)} 
    \end{align}
    with polynomials $q$ and $p$ that satisfy
    \begin{align}
        q(s) = a_1 s^{w_1} + O(s^{w_1+1}), \quad p(s) = 1 + O(s) \quad \text{and} \quad \deg(p)-\deg(q) = w_4. 
    \end{align}
    Since the degree of the covering map coincides with the degree of the polynomial $p$, we can fix it through the Riemann-Hurwitz formula. This leads to
    \begin{align}
        \text{deg}(p) = \frac{1}{2}(w_1 + w_2 + w_3 + w_4) - 1 \quad \text{and} \quad \text{deg}(q)=  \frac{1}{2}(w_1 + w_2 + w_3 - w_4) - 1.
    \end{align}
    Fixing the polynomials $p$ and $q$ hence amounts to computing $w_2+w_3-1$ coefficients and the additional unknown $s_2$. 
    This is achieved in the notebook by solving the $w_2+w_3$ equations 
    \begin{align}
        \partial^n_s(\Gamma(s)-x_i)|_{s=s_i} = 0 \quad \text{for } \quad i\in\{2,3\} \quad \text{ and } \quad n=0,1,\dots w_i-1. 
    \end{align}

\section{Proving the diagrammatic rules}\label{app:proof_rules}
    In this appendix, we sketch a proof for the validity of interface PRR diagrammatics. 
    In the first paragraph, we introduce sharpened terminology that allows us to distinguish between embeddings of the diagrams and the intrinsic data of the diagrams more clearly than in the main text.
    With this preparation, we then show that the map from equivalence classes of covering maps to abstract PRR diagrams is injective i.e.~that two covering maps producing the same diagram are equivalent. 
    Next, we show that the map is also surjective i.e.~that every abstract PRR diagram indeed arises from a covering map. 
    Together, these two steps imply that there is indeed a one-to-one correspondence between PRR diagrams and equivalence classes of covering maps.
   
    \paragraph{Embedded and abstract PRR diagrams.} \, \\
    \noindent \underline{Definition.} \emph{Embedded (interface) PRR diagrams} are the graphs embedded in Riemann surfaces that can be constructed from the vertices in Sec.~\ref{sec:PRR_diagrams} and Sec.~\ref{sec:Interface_PRR} with the constraints outlined in those sections.
    Consider two embedded diagrams $\mathcal{D}$ and $\mathcal{D}'$ on Riemann surfaces $\Sigma$ and $\Sigma'$ whose vertices are labelled by the same branch points.
    We say that the diagrams are isomorphic if there is an isomorphism between $\Sigma$ and $\Sigma'$ that restrict to an isomorphism of graphs for the diagrams $\mathcal{D}$ and $\mathcal{D}'$ preserving the labels of the vertices. 
    \smallskip
    
    \noindent \underline{Definition.} \emph{Abstract (interface) PRR diagrams} are equivalence classes of embedded diagrams w.r.t.~the notion of isomorphism introduced in the previous paragraph. 
    
    \paragraph{Injectivity of PRR diagrammatics.}\, \\
    \underline{Claim.} \emph{If taking the preimage of the contour \ref{fig:interface_Gamma_to_D} for two covering maps $\Gamma$ and $\Gamma'$ with the same branch points produces the same abstract PRR diagram, then $\Gamma$ and $\Gamma'$ are equivalent i.e.~identical up to an isomorphism of the covering space.}
    \smallskip
    
    \noindent \underline{Proof.} If the two embedded PRR diagrams $\mathcal{D}$ and $\mathcal{D}'$ are isomorphic, then there is an isomorphism $\phi: \Sigma \rightarrow \Sigma'$ such that $\phi(\mathcal{D}) = \mathcal{D}'$. 
    We now show that $\phi$ is also an isomorphism of the associated covering maps i.e.~$\Gamma = \tilde \Gamma := \Gamma' \circ \phi$.
    By definition, $\phi$ identifies ramification points that are labelled by the same branch-point.
    Thus, $\Gamma(r) = \tilde \Gamma (r)$ for every ramification point $r \in \Sigma$ of $\Gamma$. 
    Furthermore, $\Gamma(x) = \tilde\Gamma(x)$ for every point $x$ on an edge of $\mathcal{D}$, since the value at the vertices and the fact that $\Gamma$ and $\tilde \Gamma$ have to be both length and orientation preserving already uniquely characterises these functions on the edges. 
    Finally, the restrictions of $\Gamma$ and $\tilde \Gamma$ to a face of $\mathcal{D}$ are restrictions of two holomorphic maps between disks that coincide on the boundary of the disk. 
    Therefore, $\Gamma$ and $\tilde \Gamma$ also coincide on the faces of $\mathcal{D}$.
    
    \paragraph{Surjectivity of PRR diagrammatics.}\,\\
    \underline{Claim.} \emph{Every abstract PRR diagram that can be constructed using the rules of Section \ref{sec:PRR} has a representative that is the preimage of the contour \ref{fig:interface_Gamma_to_D} w.r.t.~an interface covering map.}

    \noindent \underline{Proof.} Take an arbitrary realisation of the PRR diagram on a surface $\Sigma$.
    We now define a map $\Gamma$ by first setting its value at each ramification-point to the branch-point by which it is labelled and then picking an arbitrary continuous, bijective, orientation preserving identification between each edge of the diagram and the interval of the contour on the sphere that connects the branch-points corresponding to the ramification points at the ends of the edge. 
    The PRR rules enforce that the restriction of the map $\Gamma$ constructed in this way to any face of the diagram is a homeomorphism that identifies that face with the boundary of a disk on the sphere. 
    For each face, extend $\Gamma$ to the interior such that it becomes a homeomorphism between the face and a disk on the sphere. 
    Away from the ramification points, $\Gamma$ is a local homeomorphism and therefore, we can use it to induce a metric and complex structure on $\Sigma$, which makes the restriction of $\Gamma$ to the complement of the ramification points a holomorphic local isometry by construction. 
    Clearly, the PRR diagram corresponding to $\Gamma$ is the diagram that we used to construct $\Gamma$.

\bibliographystyle{JHEP}
\bibliography{biblio.bib}
\end{document}